\documentclass[prx, notitlepage, twocolumn]{revtex4-1}
\usepackage{amsmath,amsfonts, amssymb, amsthm, dsfont}
\usepackage{yfonts}
\usepackage{bm}
\usepackage{mathrsfs}
\usepackage{graphicx}
\usepackage{verbatim}
\usepackage{hyperref}
\usepackage[normalem]{ulem}
\usepackage{float}
\usepackage{empheq}
\usepackage{tensor}
\usepackage{mathtools}
\usepackage{adjustbox}
\usepackage{bbold}
\usepackage{natbib}
\usepackage{lipsum}
\usepackage{flushend}
\usepackage{tikz}

\newcommand\tikzmark[2]{%
\tikz[remember picture,baseline=(#1.base)]  
\node[inner sep=0,outer sep=3pt] (#1) {$#2$};%
}

\usetikzlibrary{positioning}
\usetikzlibrary { decorations.pathmorphing, decorations.pathreplacing, decorations.shapes,decorations.text}
\usetikzlibrary{shapes.geometric}
\usetikzlibrary{arrows.meta}
\usetikzlibrary{calc}
\tikzstyle ->-=[postaction={decorate,decoration={markings,
    mark=at position .6 with {\arrow[scale=0.7]{>}}}}]
\tikzstyle -<-=[postaction={decorate,decoration={markings,
    mark=at position .4 with {\arrowreversed[scale=0.7]{>};}}}]
    
\tikzstyle ->>-=[postaction={decorate,decoration={markings,
    mark=at position .54 with {\arrow[scale=0.7]{>}}}}]
\tikzstyle -<<-=[postaction={decorate,decoration={markings,
    mark=at position .46 with {\arrowreversed[scale=0.7]{>};}}}]

\newcommand{\io}[3]{\iota^{(#1)}_{\beta_#2}({#3})}

\DeclareMathOperator{\imag}{imag}

\newcommand{\dd}[1]{\operatorname{d}\!#1\,} 
\newcommand{\ee}{\mathrm{e}} 
\newcommand{\ii}{\operatorname{i}} 
\newcommand{\TT}{\mathcal{T}} 
\newcommand{\ZZ}{\mathbb{Z}} 
\newcommand{\HH}{\mathbb{H}} 
\newcommand{\WW}{\mathbb{W}}
\newcommand{\I}{\hat{\mathbf{1}}} 
\newcommand{\pa}{{\partial A}}
\newcommand{\pe}{P_{edge}}

\newcommand{\Tr}{\mathrm{Tr}}
\newcommand{\dg}{\dagger}
\newcommand{\tp}{\mathrm{t}}

\newcommand{\wt}[1]{\widetilde{#1}}
\newcommand{\ket}[1]{\lvert#1\rangle}
\newcommand{\bra}[1]{\langle#1\rvert}
\newcommand{\lrangle}[1]{\langle#1\rangle}
\newcommand{\abs}[1]{\left\lvert#1\right\rvert}
\newcommand{\subf}{\mathop{_{\mkern-4mu f}}}

\newcommand{\vket}[1]{\left| #1\right)}
\newcommand{\vbra}[1]{\left( #1\right|}

\newcommand{\IGG}{\mathrm{IGG}}
\newcommand{\IGA}{\mathrm{IGA}}

\renewcommand{\vec}[1]{\boldsymbol{\mathbf{#1}}}
\newcommand{\act}[2]{\tensor*[^{#1}]{#2}{}}
\newcommand{\fTr}{{\rm{fTr}}}
\newcommand{\BOF}[2]{\bigotimes_{#1}^{#2}\mathop{}_{\mkern-4mu f}}

\usetikzlibrary{decorations.pathreplacing,decorations.markings}
\tikzset{middlearrow/.style n args={2}{
        decoration={markings,
            mark= at position #1 with {\arrow{#2}} ,
        },
        postaction={decorate}
    }
}

\hypersetup{colorlinks=true}
\begin{document}

\title{Unveiling Correlated Two-dimensional Topological Insulators through Fermionic Tensor Network States -- Classification, Edge Theories and Variational Wavefunctions}
\author{Chao Xu}
\author{Yixin Ma}
\email{mayixin21@mails.ucas.ac.cn}
\author{Shenghan Jiang}
\email{jiangsh@ucas.ac.cn}
\affiliation{Kavli Institute for Theoretical Sciences, University of Chinese Academy of Sciences, Beijing 100190, China}

\begin{abstract}
    The study of topological band insulators has revealed fascinating phases characterized by band topology indices and anomalous boundary modes protected by global symmetries.
    In strongly correlated systems, where the traditional notion of electronic bands becomes obsolete, it has been established that the topological insulator phases persist as stable phases, separate from the trivial insulators.
    However, due to the inability to express the ground states of such systems as Slater determinants, the formulation of generic variational wavefunctions for numerical simulations is highly desirable.

    In this paper, we tackle this challenge for two-dimensional topological insulators by developing a comprehensive framework for fermionic tensor network states.
    Starting from simple assumptions, we obtain possible sets of tensor equations for any given symmetry group, capturing consistent relations governing symmetry transformation rules on tensor legs. 
    We then examine the connections between these tensor equations and \emph{non-chiral} topological insulators by construing edge theories and extracting quantum anomaly data from each set of tensor equations.
    By exhaustively exploring all possible sets of equations, we achieve a systematic classification of non-chiral topological insulator phases.
    Imposing the solutions of a given set of equations onto local tensors, we obtain generic variational wavefunctions for corresponding topological insulator phases. 
    Our methodology provides an important step towards simulating topological insulators in strongly correlated systems.
    We discuss the limitations and potential generalizations of our results, paving the way for further advancements in this field.
\end{abstract}

\maketitle
\tableofcontents

\section{Introduction}\label{sec:introduction}
Since the groundbreaking discovery of topological insulators\cite{KaneMele2005,konig2007quantum,fu2007topological,HasanKane2010,QiZhang2011}, the interplay between topology and symmetry has emerged as an exciting research frontier in both non-interacting and interacting systems
\footnote{When first proposed, topological insulators means 2D or 3D band insulators with helical Dirac edge mode protected by charge conservation and time reversal symmetry. 
In this work, we use a broader definition, referring to electronic insulators with anomalous edge modes protected by any symmetry group.}.
In weakly interacting systems, topological insulators are classified based on their band topology, with symmetry indicators providing valuable insights\cite{schnyder2008classification,kitaev2009periodic,ChiuTeoSchnyderRyu2016,kane2005z,fu2007topologicalInversion,kruthoff2017topological,po2017symmetry,song2018quantitative,tang2019comprehensive,po2020symmetry}.
The generalization of topological insulators, along with topological superconductors, to interacting systems has given rise to the fascinating realm of fermionic symmetry-protected topological (FSPT) phases\cite{GuWen2014symmetry,WangSenthil2014interacting,Kapustin2015fermionic,WangGu2018,WangGu2020construction,aasen2021characterization,barkeshli2022classification}.

 In contrast to the fractional topological insulators\cite{LevinStern2009fractional,neupert2011frac,chen2012interaction,repellin2014z2}, which host fractionalized quasiparticles within the bulk, the elementary bulk excitations of FSPT phases are gapped fermions.
 Nevertheless, they host anomalous edge excitations. 

In strongly interacting systems, the conventional concept of electronic bands loses relevance.
However, FSPT phases can still be characterized by a rich array of experimental observables, including quantized bulk response functions\cite{QiHughesZhang2008,YeWang2013}, and the emergence of intriguing anomalous boundary modes\cite{LuVishwanath2012,VishwanathSenthil2013}.

In the realm of strongly correlated systems, the classification of topological insulators and superconductors takes on a distinct character compared to their non-interacting counterparts.
Interactions can give rise to entirely new topological phases\cite{WangPotterSenthil2014classification}, while also leading to the adiabatic connection of certain band topological phases to trivial ones upon their introduction\cite{FidkowskiKitaev2010effects,FidkowskiKitaev2011topological,turner2011topological}.
For instance, in one-dimensional time-reversal invariant spinless superconductors~(class BDI\cite{schnyder2008classification}), the number of edge Majorana zero modes characterizes $\mathbb{Z}$ free fermion classes.
However, the addition of interactions can generate an energy gap via eight Majorana modes, resulting in a classification of eight distinct classes\cite{FidkowskiKitaev2010effects,FidkowskiKitaev2011topological,turner2011topological}.
Thus, the development of new tools for detecting and classifying interacting fermionic symmetry-protected topological (FSPT) phases becomes imperative.

In recent years, significant progress has been made in this direction.
Field theories have been proposed to explore the quantized response functions of systems coupled to external gauge fields, revealing their stability against interactions\cite{WangPotterSenthil2014classification,cheng2014topological}.
While these approaches provide valuable insights into the universal properties of FSPT phases, achieving complete classifications remains a formidable task.
To address this challenge, sophisticated mathematical tools such as group supercohomology \cite{GuWen2014symmetry} and topological field theories incorporating spin structures \cite{Kapustin2015fermionic,gaiotto2016spin} have been developed for the classification of FSPT phases.
However, most of these methods focus on systems with finite symmetry groups and are inadequate for capturing the full richness of topological insulators with $U_f(1)$ charge conservation symmetry.

Notably, exact solvable models have been constructed to realize interacting topological superconductors in two dimensions \cite{TarantinoFidkowski2016,WangNingChen2018}, which were subsequently generalized to two-dimensional topological insulators \cite{SonAlicea2019,Metlitski20191d}.
These exact solvable models not only provide fixed-point wavefunctions for topological phases but also offer a comprehensive classification of FSPT phases with or without $U_f(1)$ symmetry \cite{WangGu2020construction,WangQiFangGu2021exactly}.
Such advancements provide invaluable insights into the intricate landscape of interacting topological insulators and highlight the remarkable interplay between topology and strong electronic correlations.

Apart from the important task of classifying FSPT phases, it is also crucial to search for these phases within realistic models using numerical techniques.
In this regard, variational methods serve as widely employed tools in the study of strongly correlated systems.
By parameterizing a subset of the total Hilbert space as the space of variational wavefunctions, ground state wavefunctions can be obtained through energy minimization algorithms within this variational space.
Various observables can then be extracted from the optimized wavefunction, enabling the identification of the quantum phase.
Notable examples of variational methods in strongly correlated models include the variational Monte Carlo method based on Gutzwiller projective wavefunctions, and the density matrix renormalization group (DMRG) method based on matrix product states (MPS) \cite{white1992density,white1993density,Ostlund1995thermodynamic,Dukelsky1998equivalence}.

In this work, our focus lies on providing variational wavefunctions for non-chiral fermionic topological insulators in two-dimensions.
In particular, we will use fermionic tensor network states \cite{barthel2009contraction,kraus2010fermionic,CorbozEvenblyVerstraeteVidal2010,gu2010grassmann,SchuchGarciaCirac2011,Bultinck2017fermionic,BultinckWilliamsonHaegemanVerstraete2017fermionicmps,wille2017fermionic,cirac2021matrix} as tools to provide variational wavefunctions.
One of the main advantages of tensor networks lies in their ability to capture complicated quantum entanglement patterns, a key feature of topological phases\cite{zeng2019quantum,wen2019choreographed}.
By representing the many-body wavefunction as a network of interconnected tensors, fermionic tensor networks provide an efficient and compact description of topological phases in electronic systems.
Moreover, the variational nature of tensor network methods allows for systematic optimization of wavefunctions within the variational space, enabling the exploration of a wide range of topological phases\cite{levin2007tensor,GuLevinWen2008,orus2009simulation,xie2012coarse,evenbly2015tensor,yang2017loop,liao2019differentiable}.
This combination of entanglement representation and variational flexibility makes fermionic tensor networks an ideal choice for studying and characterizing the rich landscape of interacting topological insulators and superconductors.
It is worth noting that the unique properties of topological phases are intricately encoded in their complex entanglement patterns, which are often challenging to extract solely from local observable measurements within the optimized wavefunction.
To overcome this difficulty, one can leverage the classification results of topological phases.
We introduce an innovative approach distinct from the direct energy optimization within the full variational space of fermionic tensors.
Our proposed strategy involves constraining the variational space into distinct subspaces, each associated with a specific topological phase. 
By applying optimization algorithms exclusively within each subspace, we can ascertain the quantum phase by comparing the optimized energies across all these subspaces.
This approach allows for a more targeted exploration of the phase diagram, taking advantage of the known classification of topological phases.

Before delving into our specific approach, let us briefly review how tensor network states are employed in the context of bosonic symmetry-protected topological (SPT) phases\cite{ChenGuWen2011,ChenGuWen2011complete,Chen:2012ctz,LevinGu2012,SenthilLevin2013}.
As the name suggests, tensor network states are constructed from small tensors that possess both physical and internal legs.
Physical wavefunctions are obtained by contracting the internal legs while leaving the physical legs untouched. 
Although not directly manifest in the physical states, the internal legs encode crucial entanglement information regarding the interplay between topology and symmetry\cite{schuch2010peps,SchuchGarciaCirac2011,turner2011topological,haegeman2012order,PollmannTurner2012}.
By studying the transformation rules of the symmetry group on the internal legs, one can construct and classify SPT phases in various dimensions. 
In one-dimensional SPT phases, MPS were initially employed for their classification, with distinct classes corresponding to different projective representations of the symmetry group on the internal legs\cite{ChenGuWen2011, PollmannTurner2012, SchuchGarciaCirac2011}.
In two-dimensional SPT phases, projected entangled pair states (PEPS) with symmetry actions on the internal legs represented by inequivalent matrix product operators (MPO) or gauge transformations were shown to capture the classification\cite{williamson2017symmetry, Bultinck2017fermionic, schuch2010peps, JiangRan2017}.
Given the symmetry action on the internal legs, one can then solve for a subspace of local tensors that satisfy the symmetry constraints, thereby obtaining variational tensor wavefunctions.

To construct variational wavefunctions for topological insulators, fermionic tensor network states, characterized by their fermionic legs, are essential\cite{kraus2010fermionic,BultinckWilliamsonHaegemanVerstraete2017fermionicmps,Bultinck2017fermionic}
In particular, we will employ a specific type of fermionic tensor network state known as fermionic projected entangled-pair states (FPEPS) as our variational ansatz.
In our previous work \cite{ma2023variational}, FPEPS have demonstrated their efficacy in accurately capturing the behavior of interacting quantum spin Hall phases \cite{WangQiFangGu2021exactly}.
In this study, instead of focusing on specific examples, we adopt a more general approach.
We start by making certain assumptions about the symmetries associated with the internal legs and derive consistent conditions, which we refer to as tensor equations.
Notably, these tensor equations capture the algebraic data that characterize the quantum anomalies of the edge theories in topological insulators.
By exploring the various sets of tensor equations, we achieve a comprehensive classification of topological insulators, aligning with previous investigations \cite{WangQiFangGu2021exactly}.
Additionally, we obtain the symmetry actions on the internal legs by solving the tensor equations.
Subsequently, by imposing these symmetry constraints on the local tensors, we construct the variational spaces corresponding to different topological insulator phases.

Tensor equations offer a compelling avenue for solving variational wavefunctions for topological insulators, serving as a crucial initial step towards their simulation.
Moreover, they provide a novel and insightful description of anomalous edge theories associated with these systems.
This framework allows for the systematic construction of edge Hilbert spaces and the characterization of symmetry actions on edges, shedding light on the intriguing interplay between symmetry and topology in the edge theory. 

This paper is organized as follows.
In Section~\ref{sec:example}, we warm up with the FPEPS construction of a topological insulator protected by $U_f(1) \times \ZZ_2^{\text{Ising}}$ in 2D.
Motivated by exact solvable models, we present the FPEPS representation of the fixed-point wavefunction and extract the tensor equations for symmetry transformation rules on the internal legs of this topological phase.
In Section~\ref{sec:classification}, we introduce a classification scheme for topological insulators based on anomalous symmetry actions on the edge\cite{ElseNayak2014classifying}, which yields consistent results with recent classification proposals based on exactly solvable models \cite{WangQiFangGu2021exactly}. 
We then develop a general framework for the FPEPS construction of any topological insulator phases in Section~\ref{sec:tensor_equation}.
By assuming simple forms of symmetries on the internal legs, we derive different sets of tensor equations. 
Exhaustively exploring all possible sets of tensor equations allows us to reproduce the previous classification results of topological insulators.
For each set of tensor equations, we provide at least one representative solution, demonstrating the existence of non-empty solution spaces for all topological insulators.
In Section~\ref{sec:edge_theory}, we extract the edge Hilbert spaces and anomalous edge symmetry actions from these equations, providing concrete realizations of the proposals introduced in Section~\ref{sec:classification}.
We discuss subtle issues in Section~\ref{sec:kasteleyn}, where symmetries on internal legs are not parity-even, leading to Kasteleyn orientations, a discrete version of spin structures.
In Section~\ref{sec:solution}, we re-examine the examples proposed in Section~\ref{sec:example} by solving the constraints on local tensors and thus provide variational wavefunctions beyond the fixed-point.
Finally, in Section~\ref{sec:discussion}, we summarize the main findings and discuss future directions.
Technical details and additional mathematical formulations can be found in the Supplementary Material (SM)\footnotemark[10]
\footnotetext[10]{See Supplementary Material.}.

\section{A simple example -- electronic insulators with $U_f(1)\times \ZZ_2^{Ising}$ symmetry}\label{sec:example}
In this section, we investigate a specific example of a topological insulator protected by the $U_f(1)\times\ZZ_2^{\text{Ising}}$ symmetry, characterized by the group relation $g^2=e$ and $g\cdot n_f=n_f\cdot g$, where $n_f$ is the generator for $U_f(1)$ and $g$ is the generator for $\ZZ_2^{\text{Ising}}$.
This symmetry group gives rise to $\ZZ_4$ classes of FSPT phases, and our focus is on the root phase.

We begin by deriving the interacting edge theory using bosonization techniques.
This approach leads us to a decorated-vortex picture for the correlated topological insulator wavefunction, providing insights into the fixed-point wavefunction proposed in previous literature.
Next, we translate the fixed-point wavefunction into a representation using FPEPS, enabling us to extract the symmetry actions on internal legs and derive tensor equations that describe these actions.
It is worth noting that in our previous work\cite{ma2023variational}, we have already explored the FPEPS representation of the quantum spin Hall phase.

For readers interested in the general framework, they can skip this section without affecting their understanding of the subsequent discussions.
We will come back to this example in Section~\ref{sec:solution}, where we apply our general framework to analyze this specific example.

\subsection{From edge theory to decorated vortices}\label{subsec:ll_edge}
Similar to the quantum spin Hall phase\cite{WuBernevigZhang2006,XuMoore2006}, the non-interacting edge modes of this phase are described by helical Dirac fermions, whose low-energy Hamiltonian reads
\begin{align}
    H_{\text{edge}} = \int \dd{x} \ii v_F \, \Psi^{\dagger} \partial_x \tau_z \Psi,
    \label{eq:ti_non-interacting_edge}
\end{align}
where $\Psi = (\psi_L, \psi_R)^\tp$ describes fermion modes with strong spin-orbit coupling, and the right/left movers $\psi_{R/L}$ correspond to spin up/down, respectively.
Consequently, the symmetry actions on $\Psi$ are given by
\begin{align}
    U_f(\varphi): \Psi \to \mathrm{e}^{-\ii \varphi} \Psi, \quad
    g: \Psi \to \tau^z \Psi,
    \label{eq:ti_sym_on_psi}
\end{align}
which forbid any possible mass terms.

To observe the effect of interactions, we bosonize \cite{haldane1981luttinger} Eq.~(\ref{eq:ti_non-interacting_edge}) using the fields $\phi(x)$ and $\theta(x)$, both having a periodicity of $2\pi$.
Here, we define $\psi_{R/L} = \exp[-\ii(\theta \pm \phi/2)]$, and the commutation relation is given by $[\phi(x), \theta(x')] = -2\pi \ii \Theta(x - x')$, where $\Theta(x)$ is a step function.
By comparing with Eq.~(\ref{eq:ti_sym_on_psi}), we determine the symmetry actions on the bosonic fields as
\begin{align}
    U_f(\varphi): \theta \to \theta + \varphi, \quad \phi \to \phi, \notag \\
    g: \theta \to \theta+\frac{\pi}{2}, \quad \phi \to \phi + \pi,
    \label{eq:ti_sym_on_theta_phi}
\end{align}
respectively.
The Lagrangian density for this interacting edge system is
\begin{align}
    \mathcal{L} = \frac{1}{4\pi K} \left[ \frac{(\partial_t \phi)^2}{v_F} - v_F (\partial_x \phi)^2 \right] + \gamma \cos (2\phi - \phi_0),
\end{align}
where $\gamma$ and $\phi_0$ depend on microscopic details.
In the absence of interactions, we have $K = 2$ for free fermions.
Without loss of generality, we can set $\gamma > 0$ and $\phi_0 = 0$.
The $\gamma$-term represents the dominant scattering term allowed by symmetry, with scaling dimension $2K$.
For $K > 1$, this term is irrelevant, resulting in a gapless system.
However, for $K < 1$, the term becomes relevant, driving the system into a gapped phase.
In the limit of large $\gamma$, this gapped phase exhibits two degenerate ground states characterized by $\langle \phi \rangle = 0$ or $\langle \phi \rangle = \pi$.
According to Eq.~(\ref{eq:ti_sym_on_theta_phi}), these ground states are transformed into each other by the action of $g$, leading to the identification of this gapped phase as a spontaneous $\ZZ_2^{\text{Ising}}$ symmetry-breaking phase.

\begin{figure}
    \includegraphics[scale=0.33]{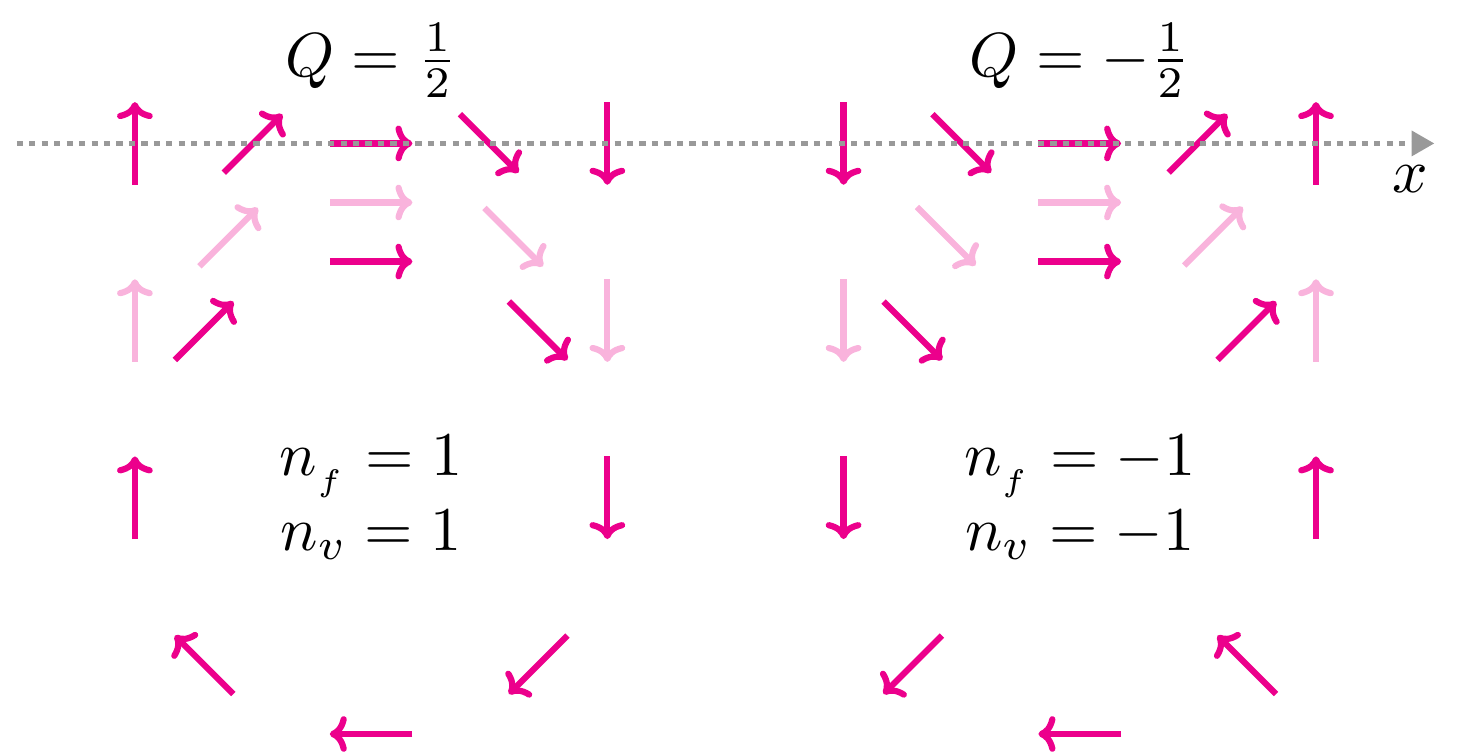}
    \caption{Schematic plot of the edge soliton and bulk vortex. 
        Gray dashed line denotes the edge of a 2D system.
        Configurations of $Q=\pm \frac{1}{2}$ on the edge correspond to clockwise/counter-clockwise rotating $\phi$ along the edge.
        We extend edge field $\phi$ to the bulk, where a solition now is interpreted as ``half'' vortex($n_v=1$)/anti-vortex($n_v=-1$).
    So, a vortex/anti-vortex contains $\pm 1$ charge.} 
    \label{fig:bulk_edge_soft_conf}
\end{figure}

Furthermore, this symmetry-breaking edge phase exhibits anomalous behavior, which can be observed in the properties of its domain walls.
To illustrate this, let us consider a $\ZZ_2^{Ising}$ domain wall located at $x_0$, where $\langle\phi\rangle=0$ for $x<x_0-\varepsilon$ and $\langle\phi\rangle=\pi$ for $x>x_0+\varepsilon$, as depicted in Fig.~\ref{fig:bulk_edge_soft_conf}.
In the language of bosonization, the charge density operator is given by $\rho = -\partial_x \phi/2\pi$, and thus the total charge $Q = \pm\frac{1}{2}$ is enclosed around the domain wall, known as the Goldstone-Wilczek half soliton \cite{GoldstoneWilczek1981}:
\begin{align}
Q = \int_{x_0-\varepsilon}^{x_0+\varepsilon} -\dd{x} \frac{\partial_x\phi}{2\pi}=\pm \frac{1}{2},
\end{align}
where the $\pm$ signs correspond to counterclockwise and clockwise rotations, respectively.

To establish a connection with spin degrees of freedom, we can map the compact bosonic field $\phi$ to spin using $S^+ = \exp(\ii\phi)$.
In this mapping, $\phi=0$ corresponds to "spin up" and $\phi=\pi$ corresponds to "spin down".

As illustrated in Fig.~\ref{fig:bulk_edge_soft_conf}, we extend the edge field $\phi$ into the bulk, where an edge domain wall can be viewed as ``half'' of a bulk vortex ($n_v=1$) or anti-vortex ($n_v=-1$), depending on its rotation direction \cite{ChenLuVishwanath2014,LiuGuWen2014}.
As an edge soliton carries $Q=\pm\frac{1}{2}$ charge, it is natural to associate vortices with $U_f(1)$ charge, with $n_f=n_v$ where $n_v$ labels the number of vortices.

The anomalous nature of the edge theory can also be understood through the fusion rules of symmetry flux.
Acting $g$ on a subregion $M$ of the edge, where $\phi\to \phi+\pi$ within $M$, creates $g$-flux at the ends of $M$.
If we apply the $g$ action twice within $M$, treating it as a continuous process in time, the current operator $j=\Psi^{\dagger}\tau_z\Psi = \partial_t \phi/2\pi$ indicates the presence of a current within $M$, resulting in the accumulation of unit charges at the two ends of $M$.
In other words, the fusion of two $g$-flux leads to unit fermions, which is impossible in a conventional 1D system.

\subsection{The fixed-point wavefunction and FPEPS construction}
\subsubsection{The fixed-point wavefunction on a honeycomb lattice}
\begin{figure}[htpb]
    \includegraphics[scale=1]{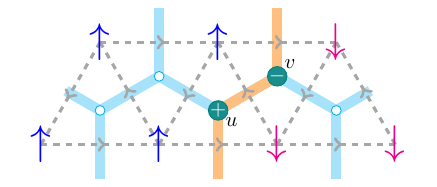}
    \caption{A typical configuration for the fixed-point wavefunction of the topological insulator phase.
        Spinless fermion $f$ lives at a honeycomb lattice, while $\ket{\tau}$ resides at the dual triangular lattice.
        Spin up/down is colored blue/red. 
        When crossing an Ising domain wall (colored yellow) along/against oriented bonds of the dual lattice, One adds $\pm\pi$ .
        The arrows indicate the convention: when a domain wall crosses a site with sublattice index $u/v$, a vortex/anti-vortex decorated with a hole/electron appears at this site.
    }
    \label{fig:bulk_conf}
\end{figure}

Motivated by the decorated vortex picture derived in the above subsection, we introduce the fixed-point wavefunction of the interacting topological insulator phase protected by $\ZZ_2^{Ising}\times U_f(1)$.
As shown in Fig.~\ref{fig:bulk_conf}, the system consists of two types of local degree of freedom: spinless fermions $f$ at a honeycomb lattice and Ising spins $\ket{\tau}$ at the plaquette centers of this lattice, forming a dual triangular lattice.
$\ZZ_2^{Ising}$ rotates Ising spins, while acts trivially on fermions:
\begin{align}
    g:f\to f\,,~\ket{\uparrow}\leftrightarrow\ket{\downarrow}\,.
    \label{eq:ti_ising_action}
\end{align}
In other words, the Ising spin mimics $\phi$-field, which rotates by $\pm\pi$ when crossing an Ising domain wall.
To distinguish between $\pm\pi$ rotation, arrows are added on the bonds of the dual lattice, and $\phi$ rotate $\pm\pi$ when crossing an Ising domain wall along/against the arrow.
In the arrow convention in Fig.~\ref{fig:bulk_conf}, $n_v=\pm1$ lives at site with sublattice index $u/v$ whenever an Ising domain wall passes through this site.
To match the decorated vortex picture where $n_f=n_v$, fermions at sublattice $u/v$ are set to be electrons/holes:
\begin{align}
    [n_{f;s},f_{s}]=-(-1)^{s}f_{s}~,
    \label{eq:ti_u1f_charge_phys}
\end{align}
where $(-1)^s=\pm1$ for $s=u/v$.
Namely, $n_{f;s}=(-1)^s f_s^\dg f_s$.

Fig.~\ref{fig:bulk_conf} shows a typical decorated vortex configurations with fermions.
The fixed-point wavefunction can be expressed as superposition of all such configurations: 
\begin{align}
    \ket{\Psi}=\sum_{c} \Psi(c)\ket{c}\otimes\ket{\psi_c}~,
    \label{eq:ti_wf_superposition}
\end{align}
where $\ket{c}$ is a product state of Ising spins and $\ket{\psi_c}$ is the fermion decoration.
The sum is taken over all Ising spin configurations.
Here, $\abs{\Psi(c)}=1$, and fermion ordering are fixed to ensure that there is no sign ambiguity for $\Psi(c)$.
It is important to note that on any domain wall loop, there are always the same number of $u-$ and $v-$sites.
Thus, according to Eq.~(\ref{eq:ti_u1f_charge_phys}), $\ket{\psi_c}$ has zero $U_f(1)$ charge, making $\ket{\Psi}$ $U_f(1)$ symmetric. 
Additionally, to ensure $\ZZ_2^{Ising}$ symmetry of $\ket{\Psi}$, we require $\Psi(c)=\Psi(g\circ c)$
\footnote{The specific form of $\Phi(c)$ depend on the ordering of fermions on sites and can be quite complicated. 
    However, for the purpose of this discussion here, we do not need to consider the detailed forms of $\Phi(c)$'s.
}.

\subsubsection{FPEPS representation}
We now proceed to construct the FPEPS representation of the fixed-point wavefunction on the honeycomb lattice.
FPEPS are formed using fermionic tensors, which are quantum states in the fermionic tensor product ($\otimes_f$) of physical and internal legs.
Legs with inward/outward arrows represent fermionic Hilbert spaces for ket/bra states.
The connection between outward and inward internal legs involves contracting states using the $\fTr$ operation, defined as
\begin{align}
    \fTr[\bra{i}\otimes_f\ket{j}]=(-)^{\abs{i}\abs{j}}\fTr[\ket{j}\otimes_f\bra{i}]=\delta_{ij}
    \label{}
\end{align}
where $(-1)^{\abs{i}}$ denotes the fermion parity of the state $\ket{i}/\bra{i}$, with $\abs{i}\in{0,1}$.
Physical wavefunctions for FPEPS are obtained by contracting all internal legs.
Throughout this paper, all local tensors are assumed to be \emph{parity even}.
More details on fermionic tensor networks are presented in Sec.I of SM\footnotemark[10].

\begin{figure}[htpb]
    \includegraphics[scale=0.7]{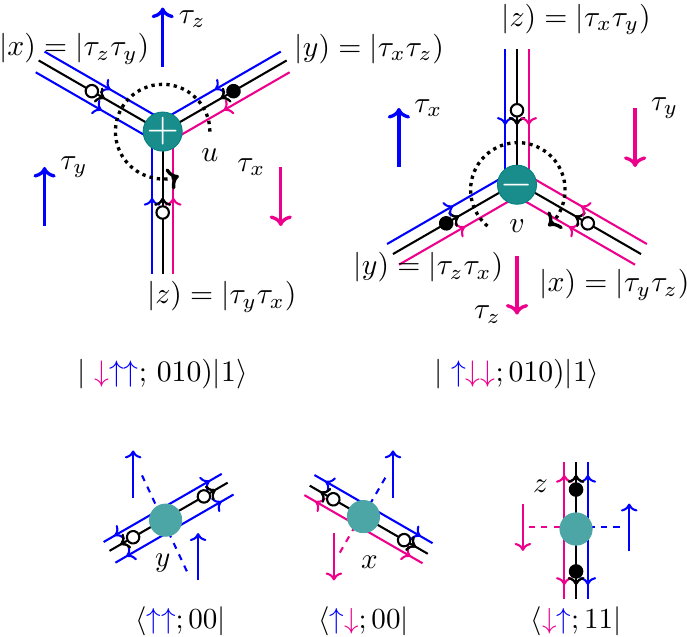}
    \caption{Illustration of basis states for site tensors.
        Physical fermions locate at the site center~(green circle), where the $\pm$ signs label $U_f(1)$ charges.
        Each internal leg is represented by three lines, where the two side lines are plaquette Ising spins, and the middle line gives an internal fermion.
    }
    \label{fig:uvtensor_hilbert}
\end{figure}

By imposing translation symmetry on the honeycomb lattice, we introduce two types of site tensors ($\hat{T}_{u,v}$) and three types of bond tensors ($\hat{B}_{x,y,z}$), as illustrated in Fig.~\ref{fig:uvtensor_hilbert}.
Each site tensor supports a physical fermion, while each bond tensor accommodates two physical Ising spins.
The Ising spins within a plaquette are set to be the same, effectively representing the plaquette spins.

The internal leg $(s\alpha)$ extends from $\hat{B}_\alpha$ to $\hat{T}_s$ and is represented as a triple line in Fig.~\ref{fig:uvtensor_hilbert}.
The middle line corresponds to a spinless fermion $c_{(s\alpha)}$ with charge $(-1)^{1-s}$, while the lines on the sides carry two internal Ising spins.
Thus, the basis states for the internal leg $(s\alpha)$ are $(c_{(s\alpha)}^\dg)^{n_\alpha}\vket{0}{(s\alpha)}\vket{\tau\beta\tau_\gamma}{(s\alpha)}$, where the ordering of $\tau{\beta}\tau_\gamma$ follows the direction of the dashed arcs in Fig.~\ref{fig:uvtensor_hilbert}.

As all spins within a plaquette are identical, a typical state for site tensor $\hat{T}_u$ is given by 
\begin{align}
    &(c_{(ux)}^\dg)^{n_x}(c_{(uy)}^\dg)^{n_y}(c_{(uz)}^\dg)^{n_z}\vket{0}\vket{\tau_z\tau_y}_{(ux)}\vket{\tau_x\tau_z}_{(uy)}\vket{\tau_y\tau_x}_{(uz)}\notag\\
    &\otimes_f(f_{u}^\dg)^{n}\ket{0}
    \label{}
\end{align}
which can be conveniently expressed as $\vket{\tau_x\tau_y\tau_z;\,n_xn_yn_z}\ket{n}$.
Similar notation applies to typical states in $\hat{T}_v$.

By employing the decorated vortex construction discussed earlier, a site accommodates one fermion whenever an Ising domain wall intersects the site.
Additionally, in order to ensure that the physical wavefunction carries zero $U_f(1)$ charge, all local tensors are designed to be charge-neutral.
To fulfill these requirements, for the internal leg $(s\alpha)$, we can simply introduce a fermion mode $c^{\dagger}{(s\alpha)}\vket{0}$ for the state $\vket{\downarrow\uparrow}$, while keeping zero fermions for the other three spin states $\vket{\uparrow\uparrow}$, $\vket{\uparrow\downarrow}$, and $\vket{\downarrow\downarrow}$.
Here, $c{(s\alpha)}$ carries the opposite charge compared to $f_s$.
Thus, in comparison with Eq.~(\ref{eq:ti_u1f_charge_phys}), we have
\begin{align}
    [n_{f;(s\alpha)},c_{(s\alpha)}]=(-1)^s c_{(s\alpha)},
    \label{eq:ti_u1f_charge_internal}
\end{align}
which implies that $n_{f;(s\alpha)}=(-1)^{s+1}c_{(s\alpha)}^\dg c_{(s\alpha)}$.

Using the concise notation, we assert that the site tensors for the fixed-point wavefunction can be expressed as:
\begin{align}
    \hat{T}_{u}={}&\big[\vket{\uparrow\uparrow\uparrow}+\vket{\downarrow\downarrow\downarrow}\big]\ket{0}+\big[\ii \vket{\downarrow\uparrow\downarrow;001}+ \vket{\uparrow\downarrow\uparrow;100}+\vket{\downarrow\uparrow\uparrow;001}\notag\\
    &+\ii\vket{\uparrow\downarrow\downarrow;010} +\ii\vket{\downarrow\downarrow\uparrow;100}+\vket{\uparrow\uparrow\downarrow;010} \big]\ket{1}\notag\\
    \hat{T}_{v}={}&\big[\vket{\uparrow\uparrow\uparrow}+\vket{\downarrow\downarrow\downarrow} \big]\ket{0}+\big[\vket{\downarrow\uparrow\downarrow;001}+\ii \vket{\uparrow\downarrow\uparrow;100}+\ii\vket{\downarrow\uparrow\uparrow;001}\notag\\
        &+\vket{\uparrow\downarrow\downarrow;010} +\vket{\downarrow\downarrow\uparrow;100}+\ii\vket{\uparrow\uparrow\downarrow;010} \big]\ket{1}
    \label{eq:ti_site_tensor}
\end{align}
which can be represented graphically as Fig.~\ref{fig:T_uv}:
\begin{figure}[H]
    \includegraphics{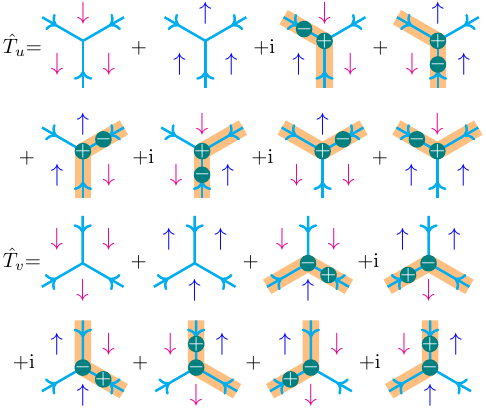}
    \caption{Illustration of site tensors.}
    \label{fig:T_uv}
\end{figure}
Coefficients of site tensors are chosen according to $\ZZ_2^{Ising}$ symmetry, as will be discussed in Section~\ref{subsubsec:fpeps_ft_wf_equiv}.

Due to the constraints where all Ising spins within a plaquette are identical, a typical state for bond tensor $\hat{B}_\alpha$ reads
\begin{align}
    \bra{\tau_0\tau_1}_{\alpha}\otimes_f\vbra{\tau_0\tau_1}_{(u\alpha)}\vbra{\tau_0\tau_1}_{(v\alpha)}(c_{(u\alpha)})^{n_u}(c_{(v\alpha)})^{n_v}~,
    \label{eq:bond_tensor_B}
\end{align}
which can be shorten as $\bra{\tau_0\tau_1;n_un_v}_\alpha$.
Here, the order of $\tau$'s on leg $(s\alpha)$ follows arrows in dashed arcs in Fig.~\ref{fig:uvtensor_hilbert}.
In this work, bond tensors are set to be maximal entangled states as
\begin{align}
    \hat{B}_\alpha=\bra{\uparrow\uparrow;00}_\alpha+\bra{\downarrow\downarrow;00}_\alpha+\bra{\uparrow\downarrow;00}_\alpha-\bra{\downarrow\uparrow;11}_\alpha
    \label{eq:ti_bond_tensor}
\end{align}
Such tensors are apparently charge neutral.
Note that to match internal states in site tensors, only $\bra{\downarrow\uparrow}$ is accompanied with a fermion mode.
The $\pm$ signs of bond tensor entries are important for $\ZZ_2^{Ising}$ symmetry, and will be discussed in Section~\ref{subsubsec:fpeps_ft_wf_equiv}.

\subsubsection{Equivalence between the fixed-point wavefunction and the FPEPS representation}\label{subsubsec:fpeps_ft_wf_equiv}
In this part, we demonstrate that by contracting all internal legs, the FPEPS construction presented in the previous part yields the fixed-point wavefunction.
The physical wavefunction reads $\ket{\Psi} = \fTr \left[ \mathbb{T}\otimes_f \mathbb{B} \right]$ with
\begin{align}
    \mathbb{T} = \bigotimes_{\vec{r}s}\subf \hat{T}_{(\vec{r}s)}~,\quad
    \mathbb{B}=\bigotimes_{\vec{r}\alpha}\subf \hat{B}_{\vec{r}\alpha}~,
    \label{}
\end{align}
where $\vec{r}$ is coordinate for unit cell, and entries of $\hat{T}_{(\vec{r}s)}$ and $\hat{B}_{\vec{r}\alpha}$ follow Eq.~\eqref{eq:ti_site_tensor} and Eq.~\eqref{eq:ti_bond_tensor} respectively.
The ordering of tensors is not important as all local tensors are parity even.

$\ket{\Psi}$ can be organized according to plaquette Ising spin configurations $c$, as in Eq.~(\ref{eq:ti_wf_superposition}).
In the following, we will show that such tensor wavefunction shares the same properties as the fixed-point wavefunction: for all $c$, $\abs{\Psi(c)}=1$ and $\Psi(c)=\Psi(g\circ c)$.

Given an Ising spin configuration $c$, by overlapping $\ket{\Psi}$ with $\ket{c}$, we obtain $\Psi(c)\ket{\psi(c)}$, which is a new tensor network state formed by site tensors $\hat{T}_{(\vec{r}s)}^c$ and bond tensors $\hat{B}^c_{\vec{r}\alpha}$.
For such a new tensor network, the internal legs are all one-dimensional Hilbert spaces with either bosonic or fermionic internal states.
Since the entries of site and bond tensors have unit modulus, we conclude that $\abs{\Psi(c)}=1$.

To show $\Psi(c)=\Psi(g\circ c)$ for arbitrary spin configurations, we discuss three cases separately.
\begin{enumerate}
    \item For the trivial case where all Ising spins points up/down, it is easy to see that $\Psi(c)=\Psi(g\circ c)=1$.

    \item Consider a configuration $c$ characterized by a single domain wall loop, where the inside and outside of the loop are labeled as $\downarrow$ and $\uparrow$, respectively.
        In this case, the configuration $g\circ c$ also results in a single domain wall loop, but with the inside and outside regions interchanged, i.e., $\uparrow$ inside and $\downarrow$ outside.
        The honeycomb lattice's bipartite nature ensures that the loop contains an equal number of $u$-sites and $v$-sites, which we label as $N$.
        Consequently, the loop consists of $2N$ bonds.
        Tensors away from the loop correspond to pure spin states with a coefficient of $1$, so all the subtleties arise from the tensors on the loop.
        We assign labels $\hat{T}_j^c$ to the site tensors along the loop and $\hat{B}^c_{j,j+1}$ to the bond tensors connecting sites $j$ and $j+1$, where $j=\{1,\cdots,2N\}$ with an arbitrary starting point, proceeding clockwise along the loop.
        Without loss of generality, we choose site $j=1$ to be a $u$-site.
        The tensor network state $\ket{\psi(c)}$ is depicted schematically as follows:
        \begin{widetext}
            \begin{align}
                \Psi(c)\ket{\psi(c)}=\adjincludegraphics[scale=1,valign=c]{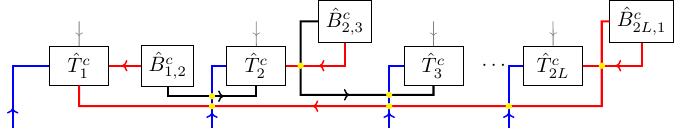}~,
                \label{eq:loop_contraction}
            \end{align}
        \end{widetext}
        where only tensors in the domain wall loop are presented.
        The blue legs are physical fermions, the light gray legs are internal spins connecting to tensors within domains, and the yellow dots represent fermion swapping gates. 
        The internal legs traveling from right to left are colored red and introduce an additional factor of $-1$ when contracting fermionic states according to Eq.~(3) in Sec.I of SM\footnotemark[10].
        The tensor network state $\Psi(g\circ c)\ket{\psi(g\circ c)}$ can be graphically represented in a similar manner.
        The coefficients for these two configurations come from two contributions: fermion swapping and tensor entries.

        From Eq.~(\ref{eq:ti_bond_tensor}), the tensor $\hat{B}_{2k-1,2k}^c$ contains fermion modes $\vbra{0}c_{2k-1}$ and $\vbra{0}c_{2k}$, while $\hat{B}_{2k,2k+1}^c$ carries zero fermion charge.
        Contracting $\hat{B}_{2k-1,2k}^c$ with $\hat{T}_{2k-1}^c$ and $\hat{T}_{2k}^c$, and $\hat{B}_{2k,2k+1}^c$ with $\hat{T}_{2k}^c$ and $\hat{T}_{2k+1}^c$ both yield a factor of $-1$, resulting in no sign factor overall.
        For the configuration $g\circ c$ with an $\uparrow$ domain inside the loop, $\hat{B}_{2k,2k+1}^{g\circ c}$ contains fermion modes $\vbra{0}c_{2k}$ and $\vbra{0}c_{2k+1}$, while $\hat{B}_{2k-1,2k}^{g\circ c}$ is a pure spin state.
        Contracting $\hat{B}_{2k,2k+1}^{g\circ c}$ with $\hat{T}_{2k}^{g\circ c}$ and $\hat{T}_{2k+1}^{g\circ c}$ for $k<N$ yields a factor of $-1$, and contracting $\hat{B}_{2N,1}^{g\circ c}$ with $\hat{T}_1^{g\circ c}$ and $\hat{T}_{2N}^{g\circ c}$ does not introduce a phase factor.
        Thus, the phase difference between $\Psi(c)$ and $\Psi(g\circ c)$ due to fermion swapping for a single loop with $2N$ sites is $(-)^{N-1}$.

        We now analyze the contributions from tensor entries.
        According to Eq.~(\ref{eq:ti_bond_tensor}), entries of $\hat{B}_{j,j+1}^c$ and $\hat{B}_{j,j+1}^{g\circ c}$ have opposite signs for all $j$, implying that the $2N$ bonds contribute no phase difference.
        We consider the sites on a domain wall loop, which can be classified into two types based on their angles: $\pm\frac{2\pi}{3}$-sites.
        For loops with $2N$ sites, the number of $\pm\frac{2\pi}{3}$-sites is equal to $N\pm3$.
        To see this, let us consider a domain wall loop with $2N$ sites and $N_+$ $\frac{2\pi}{3}$-sites.
        Expanding such a loop by adding a unit honeycomb loop, the two $\frac{2\pi}{3}$-sites on the original loop transform into $-\frac{2\pi}{3}$-sites.
        We assume that $n_-$ $-\frac{2\pi}{3}$-sites (where $0\leq n_-\leq 4$) on the original loop become sites in the domain.
        After expansion, the half number of total sites ($N$) as well as the number of $\frac{2\pi}{3}$-sites both increase by $(2-n_-)$.
        Considering that a unit honeycomb loop has $N=3$ and $N_+=N+3=6$, and any loop can be expanded from a unit honeycomb loop, we conclude that there are $(N+3)$ $\frac{2\pi}{3}$-sites and $(N-3)$ $-\frac{2\pi}{3}$-sites.
        An example is shown in Fig.~\ref{fig:scheme_add_one_plaquette}.
        
        \begin{figure}
            \includegraphics[scale=1]{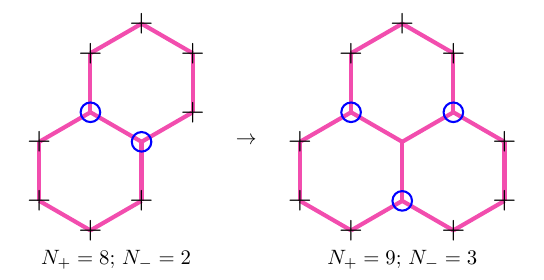}
            \caption{An example of the inductive proof.
            Before adding a new plaquette to the loop in the left figure, there are 8 $N_+$ sites and 2 $N_-$ sites, and after adding a new plaquette to the loop, one site is packed into the domain ($n_-=1$), then both $N_+$ and $N_-$ increase by $1 = 2-n_-$.
        }
        \label{fig:scheme_add_one_plaquette}
        \end{figure}

        Let the number of $\pm\frac{2\pi}{3}$-$s$-sites be $N_{s\pm}$.
        Since the number of $u$-sites is equal to the number of $v$-sites on the loop, we have $N_{u-}=N-N_{u+}$, $N_{v+}=N+3-N_{u+}$, and $N_{v-}=N_{u+}-3$.
        From Eq.~(\ref{eq:ti_site_tensor}), we conclude that the phase difference between configurations $c$ and $g\circ c$ due to entries of site tensors is given by
        \begin{align}
            \ii^{N_{u+}+N_{v-}}\cdot(-\ii)^{N_{u-}+N_{v+}}=(-1)^{N-1}.
        \end{align}

        By considering the contributions from fermion swapping and tensor entries, we can conclude that $\Psi(c)=\Psi(g\circ c)$ for $c$ with a single domain wall.

    \item For configurations with multiple domain wall loops, one can move tensors belonging to each individual loop together.
        The analysis presented above can then be applied to each loop separately, leading to the conclusion that $\Phi(g\circ c) = \Phi(c)$ is always satisfied.
\end{enumerate}
In conclusion, for any configuration $c$ with $N_{dw}(c)$ domain wall loops, we have $\Psi(c)= \Psi(g\circ c)$, and thus $\ZZ_2^{Ising}$ symmetric.

\subsection{Tensor equations from the fixed-point wavefunction}\label{subsec:tensor_eq_ft_wf}
In this work, we make the assumption that symmetries act as gauge transformations on internal legs, which are canceled out during tensor contraction, resulting in symmetric physical wavefunctions.
These gauge transformations, in general, do not satisfy the same group relations as physical symmetries in topological phases.
The relations between these gauge transformations are characterized by tensor equations.
Please see more details in Section~\ref{subsec:sym_fpeps} and Sec.I of SM\footnotemark[10].

Tensor equations play a crucial role in our work.
They serve two main purposes: First, they provide algebraic data that characterizes the anomalous edge theory of FSPT phases.
Second, solutions to tensor equations yield symmetry actions on internal legs of local tensors, which can be imposed to obtain variational tensor wavefunctions for FSPT phases.
In this subsection, we introduce tensor equations by examining the symmetries of local fermionic tensors constructed in the previous subsection.
The tensor equations for generic topological insulators will be discussed in Section~\ref{sec:tensor_equation}.

Let's begin by considering the $U_f(1)$ symmetry with the charge operator $n_f$.
From Eq.(\ref{eq:ti_u1f_charge_phys}) and Eq.(\ref{eq:ti_u1f_charge_internal}), it is easy to verify that the local tensors in Eq.(\ref{eq:ti_site_tensor}) and Eq.(\ref{eq:ti_bond_tensor}) are charge neutral:
\begin{align}
    \left( n_{f;s}+\sum_\alpha n_{f;(s\alpha)} \right)\cdot \hat{T}_s&=0\,,\notag\\
    \hat{B}_\alpha\cdot \left( n_{f;(u\alpha)}+n_{f;(v\alpha)} \right)&=0~,.
    \label{eq:ti_u1f_sym}
\end{align}
Thus, the physical wavefunction obtained by contracting internal legs is also charge neutral.

Next, let's consider the $\mathbb{Z}_2^{\text{Ising}}$ symmetry acting on local tensors.
It can be observed that the local tensors in Eq.~\eqref{eq:ti_site_tensor} and \eqref{eq:ti_bond_tensor} satisfy the following equations:
\begin{eqnarray}
    &W_{(sx)}(g)\otimes_f W_{(sy)}(g) \otimes_f W_{(sz)}(g) \cdot \hat{T}_{s} = \hat{T}_{s}\,,
    \label{eq:ti_z2_sym}\\
    &V_{(\alpha1)}(g)\otimes V_{(\alpha2)}(g)\cdot \hat{B}_{\alpha}\cdot W^{-1}_{(u\alpha)}(g)\otimes_f W^{-1}_{(v\alpha)}(g)  = \hat{B}_{\alpha}\,,\notag
\end{eqnarray} 
where $V_{(\alpha1/2)}(g)$ are $g$ action on physical Ising spin $\tau_{0}/\tau_{1}$ defined in Eq.~(\ref{eq:ti_ising_action}). 
$W(g)$'s are $g$-action on internal legs, which read
\begin{widetext}
    \begin{align}
        W_{(sx)}(g)=W_{(sz)}(g)&=\vket{\uparrow\uparrow}\vbra{\downarrow\downarrow} + \vket{\downarrow\uparrow,1}\vbra{\uparrow\downarrow} + \ii \vket{\uparrow\downarrow}\vbra{\downarrow\uparrow,1}+ \vket{\downarrow\downarrow}\vbra{\uparrow\uparrow}\,,\notag\\
        W_{(sy)}(g) &= \vket{\uparrow\uparrow}\vbra{\downarrow\downarrow} - \vket{\downarrow\uparrow,1}\vbra{\uparrow\downarrow}- \ii \vket{\uparrow\downarrow}\vbra{\downarrow\uparrow,1}+ \vket{\downarrow\downarrow}\vbra{\uparrow\uparrow}\,.
        \label{eq:ti_z2_internal}
    \end{align}
\end{widetext}
It is important to note that $W(g)$'s do not have fixed parity. 
Eq.~(\ref{eq:ti_z2_sym}) does not imply a $\mathbb{Z}_2^{\text{Ising}}$-symmetric physical wavefunction, as swapping $W(g)$'s introduces additional fermion swapping gates that may break the cancellation between $W(g)$ and $W^{-1}(g)$.
However, as we will show later in Section \ref{sec:kasteleyn}, Eq.~(\ref{eq:ti_z2_sym}) leads to the Kasteleyn orientation, which in turn yields a global $\mathbb{Z}_2^{\text{Ising}}$ symmetry.

We now explore the group relation between $n_f$ and $g$ on internal legs.
While they commute with each other on physical legs, $W(g)$'s do not commute with $n_f$, instead satisfying
\begin{align}
    W_{(s\alpha)}(g)\circ n_{f;(s\alpha)}=n_{f;(s\alpha)}+n_{D;(s\alpha)}
    \label{eq:ti_z2_u1f_commutator}
\end{align}
Here, we use the notation `` $A\circ B$ '' to represent $A\cdot B\cdot A^{-1}$.
The operator $n_D$ can be expressed as
\begin{align}
    n_{D;(s\alpha)}\equiv(-1)^{s+1}\vket{\uparrow\downarrow}\vbra{\uparrow\downarrow}+(-1)^s\vket{\downarrow\uparrow}\vbra{\downarrow\uparrow}~,
    \label{eq:ti_nd}
\end{align}
$n_D$'s are also ``symmetry charges'' for local tensors, but they act trivially on physical legs:
\begin{align}
    &\left( \sum_{\alpha} n_{D;(s\alpha)} \right)\cdot \hat{T}_{s} = 0,,\notag\\
    &\hat{B}_\alpha\cdot\left( n_{D;(u\alpha)}+n_{D;(v\alpha)} \right)=0,.
    \label{eq:ti_nd_tensor}
\end{align}
These internal leg symmetries also form a group, referred to as the invariant gauge group (IGG)\cite{Wen2002,JiangRan2015symmetric,JiangRan2017anyon}.
In cases where the IGG is a Lie group, its Lie algebra is called the invariant gauge algebra (IGA).
To ensure a symmetric physical wavefunction, the group relations on internal legs may differ from those on physical legs by an arbitrary IGG element.
Roughly speaking, the IGG can be seen as identity elements on internal legs.

Physically, $n_{D;(s\alpha)}$ contains information about domain walls on the internal leg $(s\alpha)$: it returns $0$ for configurations without domain walls, while returning $\pm 1$ when domain walls are present.
It is worth mentioning that $n_D$'s can be further decomposed into "plaquette IGA" $n_\lambda$'s:
\begin{align}
    n_{D;(s\alpha)}=n_{\lambda;(s\alpha1)}+n_{\lambda;(s\alpha2)}
    \label{eq:ti_nd_to_nlambda}
\end{align}
where
\begin{align}
    n_{\lambda;(s\alpha a)}=(-1)^{a+1}\vket{\downarrow}_{(s\alpha a)} \vbra{\downarrow}_{(s\alpha a)},,~a\in{1,2}
    \label{eq:ti_nlambda}
\end{align}
Here, $\vket{\uparrow\downarrow}_{(s\alpha a)}$ denotes the $a$th Ising spin on the internal leg $(s\alpha)$, with $a=1,2$ ordered counter-clockwise for both sites:
\begin{align}
    \adjincludegraphics[valign=c,scale=0.5]{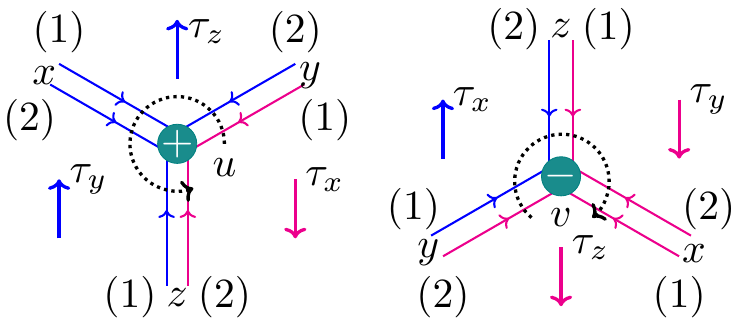}\,,
\end{align}
As mentioned, $n_{\lambda}$'s form a ``plaquette IGA'', meaning that local tensors vanish when acting $n_\lambda$'s on legs within a plaquette:
\begin{align}
    &(n_{\lambda;(s,\alpha,1)}+n_{\lambda;(s,\alpha+1,2)})\cdot \hat{T}_s=0\,,\notag\\
    &\hat{B}_\alpha\cdot (n_{\lambda;(s\alpha1)}+n_{\lambda;(s\alpha2)})=0\,.
    \label{eq:ti_nlambda_tensor}
\end{align}

We point out that another group relation $g^2=e$ also no longer holds on internal legs.
Instead, we have
\begin{align}
    \exp\left[ \ii\frac{\pi}{2}n_{D;(s\alpha)}^2 \right]\cdot\left[ W_{(s\alpha)}(g) \right]^2=\hat{1}
    \label{eq:ti_wg_wg}
\end{align}
Such complicated group relation will be explained in Section~\ref{sec:tensor_equation} and Section~\ref{sec:solution}.

To complete the tensor equations, we also consider the nontrivial symmetry action on IGG elements.
It is easy to observe that $W(g)$ anticommutes with $n_{D}$:
\begin{align}
    \{ W_{(s\alpha)}(g),n_{D;(s\alpha)} \}=0.
    \label{eq:ti_wg_nd}
\end{align}
Regarding $W(g)$'s action on $n_{\lambda}^{(a)}$, we have
\begin{align}
    W_{(s\alpha)}(g)\circ n_{\lambda;(s\alpha a)} = -n_{\lambda;(s\alpha a)} + (-1)^a.
    \label{eq:ti_wg_nlambda}
\end{align}

In summary, we can group the equations obtained from the FPEPS construction of the fixed-point wavefunction into three categories:
\begin{enumerate}
    \item Eq.~\eqref{eq:ti_z2_u1f_commutator}, \eqref{eq:ti_nd_to_nlambda}, \eqref{eq:ti_wg_wg}, \eqref{eq:ti_wg_nd}, and \eqref{eq:ti_wg_nlambda} are referred to as \emph{tensor equations}, representing the ``group relations'' of the symmetry group and IGG on the internal legs of FPEPS.
    \item Eq.~\eqref{eq:ti_z2_internal}, \eqref{eq:ti_nd}, and \eqref{eq:ti_nlambda} provide one solution to these tensor equations, forming a representation of the symmetry group and IGG on internal legs.
    \item Once obtaining a representation, we impose constraints on the local tensors as given in Eq.~\eqref{eq:ti_u1f_sym}, \eqref{eq:ti_z2_sym}, \eqref{eq:ti_nd_tensor}, and \eqref{eq:ti_nlambda_tensor}.
        By solving these constraints, we obtain a variational space of tensor wavefunctions for the given topological insulator phase.
        \label{item:ti_constraints}
\end{enumerate}

While these equations are derived from the fixed-point wavefunction, it is important to verify whether they truly characterize the topological insulator phase.
In Section~\ref{sec:edge_theory}, we address this concern by analyzing the anomalous edge theory using these tensor equations.
We demonstrate that they indeed characterize the topological insulator phase, so solutions obtained by methods in Item~\ref{item:ti_constraints} indeed give variational wavefunctions beyond the fixed-point.

\section{Classification of topological insulators through symmetry action on edge}\label{sec:classification}
Topological insulators are renowned for their metallic edge states, which are protected by the anomalous symmetry actions when confined to the edge.
This anomalous symmetry action can be related to the anomalous fusion rules of symmetry defects, as discussed in Ref.~\onlinecite{ElseNayak2014classifying} and \onlinecite{Metlitski20191d}.
In this section, by exploring the potential anomalous fusion rules of symmetry defects, we extract the algebraic data for a complete classification 2+1D topological insulators\cite{WangQiFangGu2021exactly}.

\subsection{Symmetry group for electronic insulators}\label{subsec:sym_grp}
Let us first give a general description for symmetry group of insulators.
For an electronic insulator, its symmetry group $G_f$ can generally be expressed as an extended group:
\begin{equation}
    G_f = U_f(1)\rtimes_{\tilde{\rho},\alpha}G_b,
\end{equation}
where $\tilde{\rho}:G\to\ZZ_2$ and $\alpha:G\times G\to U_f(1)$ characterize the group extension.
A group element is represented as $(g,\theta)$, with $g\in G_b$, and $\theta\in[0,2\pi)$.
The group multiplication rule is given by
\begin{equation}
    (g_1,\theta_1)\cdot (g_2,\theta_2) = (g_1g_2,\theta_1 + (-)^{\tilde{\rho}(g_1)}\theta_2 + \alpha(g_1,g_2)\pmod 2\pi)\,.
    \label{eq:group_mul_rule}
\end{equation}
From associativity, we derive the following relations:
\begin{align}
    \tilde{\rho}(g_1)+\tilde{\rho}(g_2)&=\tilde{\rho}(g_1g_2)\,,
    \label{eq:group_rho_alpha}\\
    \alpha(g_1,g_2)+\alpha(g_1g_2,g_3)&=(-)^{\tilde{\rho}(g_1)}\cdot\alpha(g_2,g_3)+\alpha(g_1,g_2g_3)\,.\notag
\end{align}
In other words, $\tilde{\rho}\in H^1(G_b,\mathbb{Z}_2)$ and $\alpha\in H^2_{\wt{\rho}}(G_b,U_f(1))$.
One can choose gauge such that $\alpha(e,g)=\alpha(g,e)=0$, and then from Eq.~(\ref{eq:group_mul_rule}), we have
\begin{align}
    (g,0)\cdot(e,\theta)=(e,(-)^{\tilde{\rho}(g)}\theta)\cdot(g,0)
    \label{eq:group_g_on_u1f}
\end{align}
We introduce another index $s\in H^1(G_b,\ZZ_2)$: $s(g)=0/1$ if $g\in G_b$ is a unitary/anti-unitary symmetry action.
Let $n_f$ be the generator of $U_f(1)$, Eq.~(\ref{eq:group_g_on_u1f}) is equivalent to
\begin{equation}
    g\cdot n_f \cdot g^{-1} = (-)^{\rho(g)} n_f\,,
    \label{eq:group_g_on_nf}
\end{equation}
where $\rho(g) \equiv \tilde{\rho}(g) + s(g)$.
Namely, $g$ is a particle-hole symmetry if $\rho(g)=1$.

By setting $\theta_{1,2}=0$ in Eq.~(\ref{eq:group_mul_rule}), we have
\begin{align}
    (g_1,0)\cdot (g_2,0) = (g_1g_2,\alpha(g_1,g_2))\,.
    \label{eq:group_g12}
\end{align}
It is easy to verify that Eq.~(\ref{eq:group_g_on_nf})~(or Eq.~(\ref{eq:group_g_on_u1f})) and (\ref{eq:group_g12}) give Eq.~(\ref{eq:group_mul_rule}).

\subsection{Fermionic three-cocycle from edge theory}\label{subsec:anomaly_cocycle}
For an electronic system with Hilbert space $\HH = \bigotimes_{f,i} \HH_i$, the action of $n_f$ and $g \in G_b$ on $\HH$ is given by
\begin{align}
    n_f \equiv \sum_i n_{f;i}\,,\quad
    U(g)K^{s(g)} \equiv \bigotimes_i U_i(g)K^{s(g)}\,,
    \label{eq:sym_on_bulk}
\end{align}
where $n_{f;i}$ and $U_i(g)$ act on the local Hilbert space $\HH_i$.
Note that in this work, we focus on onsite symmetries.

For a short-range entangled (SRE) phase on a finite-size sample $A$, its low-energy eigenstates are identified as edge states.
Following Ref.~\cite{ElseNayak2014classifying}, we assume that these edge states can be constructed from a superposition of certain Wannier orbitals localized near the boundary.
In other words, the edge Hilbert space $\HH_{edge}$ has a tensor product structure:
\begin{align}
    \HH_{edge}\equiv\bigotimes_{j\in \partial A} \HH_j
    \label{eq:edge_hilbert_tensor_prod}
\end{align}
We mention that edge Wannier orbitals are, in general, different from the original local Hilbert spaces.
However, this difference does not affect our analysis in the following.

Now, we examine the symmetry actions projected on $\HH_{edge}$, which may be anomalous when the edge is regarded as a (1+1)D system.
These actions are not necessarily onsite.
Here, we consider the case where elements in $U_f(1)$ act as onsite symmetries on the edge, namely,
\begin{align}
    n_{f;edge}=\sum_{j\in \partial A} n_{f;j}
    \label{eq:edge_nf_onsite}
\end{align}
For $g\in G_b$, its action on the edge is labeled as $U_{edge}(g)K^{s(g)}$, where $U_{edge}(g)$ is assumed to be a finite-depth local unitary.
Note that such action satisfies the group relation defined in Eq.~(\ref{eq:group_mul_rule}) and Eq.~(\ref{eq:group_g_on_nf}):
\begin{align}
    &U_{edge}(g)K^{s(g)}\cdot n_{f}\cdot [U_{edge}(g)K^{s(g)}]^{-1}=(-)^{\rho(g)}\cdot n_{f}\,,\notag\\
    &U_{edge}(g_1)K^{s(g_1)}\cdot U_{edge}(g_2)K^{s(g_2)}\notag\\
    &=\exp\left[ \ii\alpha(g_1,g_2)\cdot n_{f} \right]\cdot U_{edge}(g_1g_2)K^{s(g_1g_2)}\,.
    \label{eq:uedge_relation}
\end{align}
where we use $n_{f}$ as a short-hand for $n_{f;edge}$.

Since $U_{\text{edge}}(g)$ is of finite depth, it can be truncated to a subregion $M\subset \partial A$, denoted as $U_M(g)$.
We require the truncation result to be charge neutral, satisfying
\begin{align}
    U_{M}(g_a)K^{s(g_a)}\cdot n_f\cdot [U_{M}(g_a)K^{s(g_a)}]^{-1}=&(-)^{\rho(g_a)}\cdot n_f
    \label{eq:um_charge_neutral}
\end{align}
When restricted to $M$, the group relation becomes:
\begin{align}
    &U_M(g_1)K^{s(g_1)}\cdot U_M(g_2)K^{s(g_2)}=\nonumber\\
    &\Omega(g_1,g_2)\cdot \exp\left[ \ii\alpha(g_1,g_2)\cdot n_f \right]\cdot U_M(g_1g_2)K^{s(g_1g_2)}
    \label{eq:u1f_um_condition}
\end{align}
where $\Omega(g_1,g_2)=\Omega_a(g_1,g_2)\cdot \Omega_b(g_1,g_2)$, and $\Omega_{a/b}(g_1,g_2)$ are unitary operators located at $\partial M=\{a,b\}$, which are interpreted as local excitations at ends of $M$.
Note that such decomposition is defined up to a $U(1)$ phase factor, as $\Omega(g_1,g_2)=(\ee^{\ii\theta}\Omega_a(g_1,g_2))\cdot (\ee^{-\ii\theta}\Omega_b(g_1,g_2))$.

From Eq.~\eqref{eq:um_charge_neutral} and \eqref{eq:u1f_um_condition}, we have
\begin{align}
    [n_f,\Omega(g_1,g_2)]=0
    \label{}
\end{align}
Thus, $\Omega_{a/b}$ can carry opposite $U_f(1)$ quantum number: 
\begin{equation}
    [n_f,\Omega_{a/b}(g_1,g_2)] = \mp \sigma(g_1,g_2) \Omega_{a/b}(g_1,g_2)\,,
    \label{eq:Omega_ab_fermion_number}
\end{equation}
where $\sigma(g_1,g_2)\in\ZZ$. 
By definition, $\Omega_{a/b}(g_1,g_2)$ are operators with even/odd fermion parity iff $\sigma(g_1,g_2)$ is even/odd.

From associativity and Eq.~(\ref{eq:group_rho_alpha}), we have
\begin{align}
    \Omega(g_1,g_2)\Omega(g_1g_2,g_3)={}^{U_M(g_1)K^{s(g_1)}}\Omega(g_2,g_3)\Omega(g_1,g_2g_3)\,.
    \label{eq:omega_two-cocycle}
\end{align}
where ${}^{U_M(g_1)K^{s(g_1)}}\Omega\equiv U_M(g)K^{s(g)}\cdot \Omega\cdot K^{s(g)} U_M^{-1}(g)$.
As the decomposition of $\Omega$ to $\Omega_a\cdot \Omega_b$ is unique up to a $U(1)$ phase, we conclude
\begin{widetext}
\begin{align}
    \Omega_a(g_1,g_2)\Omega_a(g_1g_2,g_3)&=(-)^{\sigma(g_1,g_2)\sigma(g_1g_2,g_3)+\sigma(g_2,g_3)\sigma(g_1,g_2g_3)}\cdot \omega(g_1,g_2,g_3)\cdot {}^{U_M(g_1)K^{s(g_1)}}\Omega_a(g_2,g_3)\Omega_a(g_1,g_2g_3)\nonumber\\
    \Omega_b(g_1,g_2)\Omega_b(g_1g_2,g_3)&=\omega^{-1}(g_1,g_2,g_3)\cdot {}^{U_M(g_1)K^{s(g_1)}}\Omega_b(g_2,g_3)\Omega_b(g_1,g_2g_3)\,,
    \label{eq:omegaab_twist_two-cocycle}
\end{align}
\end{widetext}
where $(-)^{\sigma\sigma}$ factors come from exchanging $\Omega_a$ and $\Omega_b$.
Thus, the anomalous symmetry action are characterized by two pieces of data $\sigma$ and $\omega$, which constitutes the fermionic 3-cocycle in the presence of $U_f(1)$ symmetry.
By calculating commutators between $n_f$ and two sides of the equations in Eq.~\eqref{eq:omegaab_twist_two-cocycle}, we conclude that $\sigma\in H^2_{\rho}(G_b,\mathbb{Z})$.
Namely,
\begin{align}
    &\sigma(g_1,g_2)+\sigma(g_1g_2,g_3)=(-)^{\rho(g_1)}\sigma(g_2,g_3)+\sigma(g_1,g_2g_3)
    \label{eq:u1f_sigma_2-cocycle}
\end{align}
Consistent conditions for $\omega$'s are given by
\begin{align}
    &\frac{\omega(g_1,g_2,g_3)\omega(g_1,g_2g_3,g_4)[\omega(g_2,g_3,g_4)]^{1-2s(g_1)}}{\omega(g_1g_2,g_3,g_4)\omega(g_1,g_2,g_3g_4)}\notag\\
    ={}&\exp\left[ \ii(-)^{\rho(g_1g_2)}\alpha(g_1,g_2)\sigma(g_3,g_4) \right]\cdot (-)^{\sigma(g_1,g_2)\cdot \sigma(g_3,g_4)}\,.
    \label{eq:u1f_fermionic_3-cocycle}
\end{align}
which can be derived by calculating $\Omega_a(g_1,g_2)\Omega_a(g_1g_2,g_3)\Omega_a(g_1g_2g_3,g_4)$ in two different ways, where details are shown in Sec.II of SM\footnotemark[10].

One may initially expect that distinct solutions of Eq.~\eqref{eq:u1f_sigma_2-cocycle} and \eqref{eq:u1f_fermionic_3-cocycle} correspond to different edge anomalies.
However, as we will demonstrate in the following subsection, there exist redundancies arising from gauge choices of $U_M(g)$ and $\Omega_{a/b}(g)$.
To obtain the correct classification, these redundancies must be factored out.

\subsection{The coboundary condition}
It is important to note that the restriction of $U_{edge}(g)K^{s(g)}$ to region $M$ is not unique: $U_M(g)$ are defined up to operators $\Sigma(g)\equiv\Sigma_a(g)\cdot\Sigma_b(g)$ that acts on $\partial M=\{a,b\}$. 
Under the action of $\Sigma(g)$, we have $U_M(g)\to\wt{U}_M(g)=\Sigma(g)\cdot U_M(g)$.
In particular, $\Sigma_{a/b}$ can carry opposite $U_f(1)$ quantum number, where 
\begin{align}
    [n_f,\Sigma_{a/b}]=\mp\mu(g)\Sigma_{a/b}(g)
    \label{}
\end{align}
with $\mu(g)\in\ZZ$.
Due to this ambiguity, the $U_f(1)$ quantum number $\sigma(g_1,g_2)$ of $\Omega_{a/b}$ is defined up to a 2-coboundary:
\begin{align}
    \sigma(g_1,g_2)\to\wt{\sigma}(g_1,g_2)=\sigma(g_1,g_2)+\mu(g_1)+\mu(g_2)-\mu(g_1g_2)
    \label{eq:u1f_fermionic_2-coboundary}
\end{align}

Another ambiguity comes from the $U(1)$ phase ambiguity when decomposing $\Omega$ to $\Omega_a\cdot\Omega_b$.
Combined with the above $\Sigma$ ambiguity, we have
\begin{align}
    \tilde{\Omega}_a(g_1,g_2) ={}& \beta(g_1,g_2)\cdot \Sigma_a(g_1)\cdot{}^{U_M(g_1)K^{s(g_1)}}\Sigma_a(g_2)\notag\\
    &\cdot\Omega_a(g_1,g_2)\cdot\Sigma^{-1}(g_1g_2)\,.
\end{align}
$\wt{\Omega}_{a/b}$ also satisfy similar condition as Eq.~(\ref{eq:omegaab_twist_two-cocycle}):
\begin{align}
    &\tilde{\Omega}_a(g_1,g_2)\cdot\tilde{\Omega}_a(g_1g_2,g_3)
    \label{eq:tildew}\\
    ={}& \tilde{\omega}(g_1,g_2,g_3){}^{\tilde{U}_M(g_1)K^{s(g_1)}}\tilde{\Omega}_a(g_2,g_3)\cdot\tilde{\Omega}_a(g_1,g_2g_3)\notag
\end{align}
The detailed derivation of the relation between $\wt{\omega}$ and $\omega$ is presented in Sec.II of SM\footnotemark[10], where we conclude
\begin{widetext}
\begin{align}
    \tilde{\omega}(g_1,g_2,g_3)
    = \omega(g_1,g_2,g_3)\cdot \frac{\beta(g_1,g_2)\beta(g_1g_2,g_3)}{\beta(g_1g_2,g_3)(\beta(g_2,g_3))^{1-2s(g_1)}}\cdot (-)^{\mu(g_1)\tilde{\sigma}(g_2,g_3) + \sigma(g_1,g_2)\mu(g_3)}\cdot \ee^ {-\ii(-)^{\rho(g_1g_2)+s(g_1g_2)}\alpha(g_1,g_2)\mu(g_3)}
    \label{eq:u1f_fermionic_coboundary}
\end{align}
\end{widetext}
Eq.~(\ref{eq:u1f_fermionic_2-coboundary}) and Eq.~(\ref{eq:u1f_fermionic_coboundary}) gives the coboundary conditions for the anomalous symmetry action at edge.

In summary, the classification of topological insulators with symmetry group $G_f$ is given by the fermionic three-cohomology, which are constructed using fermionic three-cocycles $(\sigma,\omega)$ that satisfy Eq.~\eqref{eq:u1f_sigma_2-cocycle} and \eqref{eq:u1f_fermionic_3-cocycle} modulo coboundaries whose definitions are presented in Eq.~\eqref{eq:u1f_fermionic_2-coboundary} and \eqref{eq:u1f_fermionic_coboundary}.

\section{Tensor equations for topological insulators}\label{sec:tensor_equation}
In this section, we delve into the exploration of the symmetry actions on the internal legs of FPEPS while making specific assumptions.
Building upon these assumptions, we systematically formulate an exhaustive set of tensor equations that encompass a wide range of possible scenarios.
From each distinct set of tensor equations, we extract crucial algebraic information, offering insights that align with the classification outcomes for topological insulators discussed in Section~\ref{sec:classification}.
Of paramount importance, these algebraic insights establish a profound link to the anomalous edge theories of topological insulators, which we meticulously elaborate upon in Section~\ref{sec:edge_theory}.

\subsection{Symmetric FPEPS}\label{subsec:sym_fpeps}
In this subsection, we provide a brief introduction to the symmetric conditions on local tensors for FPEPS, with a focus on the symmetry groups discussed in Section~\ref{subsec:sym_grp}.
We begin with FPEPS composed of site tensors $\hat{T}_s$ and bond tensors $\hat{B}_{ss'}$, where $s$ and $s'$ denote site coordinates.
The physical wavefunction is obtained by contracting the internal legs, which can be represented as
\begin{align}
    \ket{\Psi}=\fTr\left[ \Bigg( \bigotimes_{\lrangle{ss'}}\subf\hat{B}_{ss'} \Bigg)\otimes_f \Bigg( \bigotimes_{s}\subf \hat{T}_s \Bigg) \right]
    \label{}
\end{align}
For simplicity, we adopt the convention that physical legs reside exclusively on site tensors, while all internal legs of site/bond tensors correspond to ket/bra spaces, pointing inward/outward.
For the purpose of illustration, we consider the FPEPS on a square lattice.
Nonetheless, it is important to note that the framework developed in this work can be readily extended to other lattice geometries without any loss of generality.

As discussed in Sec.I of SM\footnotemark[10], $\forall g\in G_f$, we assume that the action of $g$ on physical legs can be pushed to gauge transformations on internal legs.
Since our focus is on onsite symmetries in this work, we can express the symmetric condition as a constraint on each single local tensor as
\begin{align}
    U_{s}(g)\otimes_f\left( \bigotimes_{\alpha}\subf W_{(s\alpha)}(g) \right)K^{s(g)}\cdot \hat{T}_{s}=\hat{T}_{s}~,\notag\\
    K^{s(g)}\cdot\hat{B}_{ss'}\cdot K^{s(g)} \cdot\left( W^{-1}_{(s\alpha)}(g)\otimes_f W^{-1}_{(s'\alpha')}(g) \right)=\hat{B}_{ss'}
    \label{eq:sym_on_local_tensors}
\end{align}
where $U_s(g)$ are representation of $g$ on physical leg $s$. 
$(s\alpha)$ and $(s'\alpha')$ label internal legs, and 
\begin{align}
    W_{(s\alpha)}(g) \vket{i}_{(s\alpha)}=\sum_{j}[W_{(s\alpha)}(g)]_{ji}\vket{j}_{(s\alpha)}~,\notag\\
    \vbra{i}_{(s\alpha)} W^{-1}_{(s\alpha)}(g)=\sum_{j}\vbra{j}_{(s\alpha)}\left[ W_{(s\alpha)}^{-1}(g) \right]_{ij}
\end{align}
We note that $W(g)$'s may not be even parity, and in such cases, the order of $W(g)$ (and $W^{-1}(g)$) must be carefully chosen, which ensures that $W_{(s\alpha)}(g)$ and $W_{(s\alpha)}^{-1}(g)$ can be moved together without additional fermion swapping gates.
As we will demonstrate in Section~\ref{sec:kasteleyn}, such specific ordering gives rise to Kasteleyn orientation.

As an illustration, for a site tensor on square lattice, we have
\begin{equation*}
    \adjincludegraphics[valign=c]{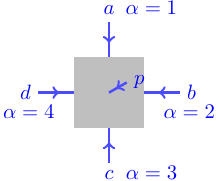}=T_{abcd,p}|a)_1|b)_2|c)_3|d)_4\ket{p}\,,
\end{equation*}
and the above symmetric condition is represented graphically as
\begin{equation*}
    \adjincludegraphics[valign=c]{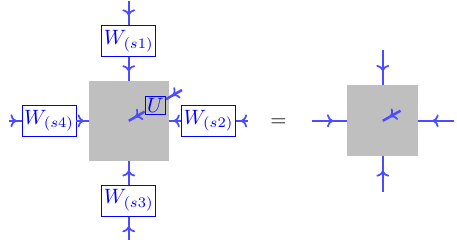}\,.
\end{equation*}
An example of bond tensors is illustrated as 
\begin{equation*}
    \adjincludegraphics[valign=c]{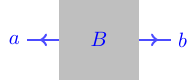}=B_{ss^{\prime},ab} \vbra{a}_{\alpha}\vbra{b}_{\alpha^{\prime}}\,,
\end{equation*}
with corresponding symmetry condition
\begin{equation*}
    \hspace{-0.3cm}\adjincludegraphics[valign=c]{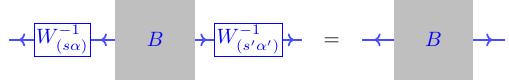}\,.
\end{equation*}

For $U_f(1)$ symmetry, the above symmetric condition are translated to zero fermion number of local tensors
\footnote{In general, the physical wavefunction preserves $U_f(1)$ symmetry if local tensors carry constant fermion number.
    If the fermion numbers of local tensors are odd, such tensors have odd fermion parity.
However, in this work, we will not consider such cases.}: 
\begin{align}
    \left(n_{f;s}+\sum_{\alpha}n_{f;(s\alpha)} \right)\cdot \hat{T}^s = 0\,,\notag\\
    \hat{B}_{ss'}\cdot (n_{f;(s\alpha)}+n_{f;(s'\alpha')})=0\,.
\end{align}

As mentioned in Section~\ref{sec:example}, we have identified a unique type of local tensor constraint on FPEPS.
This constraint corresponds to a gauge transformation on the internal legs while leaving each local tensor unchanged:
\begin{align}
    \bigotimes_{\alpha}\subf D_{(s\alpha)}\cdot\hat{T}_s=\hat{T}_s\,,\quad
    \hat{B}_{ss'}\cdot D_{(s\alpha)}\otimes\subf D_{(s'\alpha')}=\hat{B}_{ss'}
    \label{eq:global_igg_cond}
\end{align}
In this work, we direct our attention to the scenario where all IGG elements exhibit \emph{even parity}.
As demonstrated in Sec.I of SM\footnotemark[10], the IGG elements are intricately connected to the entanglement characteristics of the wavefunction.
Specifically, the presence of "global" IGG generally induces low-energy gauge dynamics, resulting in long-range entanglement.
To circumvent this, we make further assumption that all IGG elements can be decomposed into products of plaquette IGG elements\cite{JiangRan2017} which solely act on the internal legs within a plaquette.
\begin{align}
    &\hspace{1.2cm}D_{(s\alpha)}=\lambda_{(s\alpha)}^{p_2}\cdot \lambda_{(s\alpha)}^{p_1}\label{eq:global_igg_decomp}\\
    &\adjincludegraphics[valign=c]{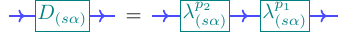}\nonumber
\end{align} 
where $p_{1/2}$ label plaquette coordinate.
For later convenience, we also use $\lambda^{(1/2)}_{(s\alpha)}$ to denotes action on different plaquette.

Constraints on local tensors from the plaquette IGG are
\begin{align}
    \lambda_{(s\alpha)}^{(1)}\otimes\lambda_{(s\alpha+1)}^{(2)}\cdot \hat{T}_s{}&=\hat{T}_s\,,\notag\\
    \hat{B}_{ss'}\cdot \left[ \lambda_{(s\alpha)}^{(a)} \right]^{-1} \otimes \left[ \lambda_{(s'\alpha')}^{(\bar{a})} \right]^{-1}{}&=\hat{B}_{ss'}\,,
    \label{}
\end{align}
with $a$ taking value 1 or 2 and $\bar{a}=3-a$.
\begin{equation*}
    \adjincludegraphics[valign=c]{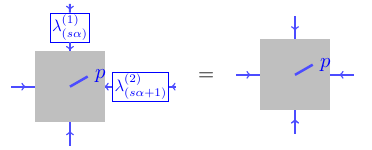}\,.
\end{equation*}

In the following, we demonstrate that the plaquette IGG elements belonging to different plaquettes, denoted as $p_1$ and $p_2$, can always be chosen to commute with each other.
This statement is evidently true when $p_1$ and $p_2$ have no shared leg.
Therefore, we focus on the scenario where $p_1$ and $p_2$ are neighbouring plaquettes with $(s\alpha)$ representing the shared leg.
The commutation relation between $\lambda^{p_1}$ and $\lambda^{p_2}$ on leg $(s\alpha)$ equals that of $\lambda^{p_1}$ and $\lambda^{p_2}$ on all legs:
\begin{align}
    &\lambda_{(s\alpha)}^{p_1}\cdot\lambda^{p_2}_{(s\alpha)}\cdot \left[ \lambda^{p_1}_{(s\alpha)} \right]^{-1} \cdot \left[ \lambda^{p_2}_{(s\alpha)} \right]^{-1}\notag\\
    ={}&\left( \bigotimes_{(s_1\alpha_1)\in p_1}\lambda_{(s_1\alpha_1)}^{p_1} \right) \cdot \left( \bigotimes_{(s_2\alpha_2)\in p_2}\lambda^{p_2}_{(s_2\alpha_2)} \right) \cdot\notag\\
    &\left( \bigotimes_{(s_1\alpha_1)\in p_1} \left[ \lambda^{p_1}_{(s_1\alpha_1)} \right]^{-1} \right) \cdot \left( \bigotimes_{(s_2\alpha_2)\in p_2}\left[ \lambda^{p_2}_{(s_2\alpha_2)} \right]^{-1} \right)
    \label{eq:plq_igg_comm}
\end{align}
It is important to note that, by definition, the commutation between IGG elements should also result in an IGG element.
Furthermore, an IGG element that only acts on a single leg $(s\alpha)$ can be simplified to $\hat{1}$ by discarding internal states that are not invariant under such a single-leg IGG operation.
Thus, Eq.~\eqref{eq:plq_igg_comm} can always be trivialized.

We would like to emphasize the presence of a special $U(1)$ subgroup in the plaquette IGG, which corresponds to the decomposition of the action of $\hat{1}$ on all internal legs.
Specifically, $\lambda_{(s\alpha)}^p=\exp[\pm\ii\theta]$, where the $\pm$ signs are chosen such that $\lambda_{(s\alpha)}^{p_1}=[\lambda_{(s\alpha)}^{p_2}]^*$ for two neighboring plaquettes $p_{1,2}$ sharing the leg $(s\alpha)$.
Due to this redundancy, the decomposition of $D_{(s\alpha)}$ in Eq.~(\ref{eq:global_igg_decomp}) is not unique:
\begin{align}
    D_{(s\alpha)}=\lambda_{(s\alpha)}^{p_2}\cdot \lambda_{(s\alpha)}^{p_1}
    =\left( \ee^{\ii\theta}\cdot\lambda_{(s\alpha)}^{p_2} \right) \cdot \left( \ee^{-\ii\theta}\cdot \lambda_{(s\alpha)}^{p_1} \right)
    \label{eq:global_igg_decomp_ambiguity}
\end{align} 

In our investigation, we will also encounter nontrivial $U(1)$ IGG in the subsequent analysis.
In such cases, the condition for "global" IGG in Eq.~(\ref{eq:global_igg_cond}) can be expressed as follows:
\begin{align}
    \sum_{\alpha} n_{D;(s\alpha)}\cdot\hat{T}_s=0\,,\quad
    \hat{B}_{ss'}\cdot \left( n_{D;(s\alpha)}+ n_{D;(s'\alpha')} \right)=0
    \label{eq:global_iga_cond}
\end{align}
where $\exp[\ii\theta n_D]$ is an $U(1)$ IGG element, and the generator $n_D$ is named as an IGA element.
Such nontrivial $n_D$ gives rise to $U(1)$ gauge dynamics, which we aim to eliminate to study the SRE phase.
To achieve this, we assume the decomposition into plaquette IGA as follows:
\begin{align}
    n_{D;(s\alpha)}=n_{\lambda;(s\alpha)}^{(2)}+n_{\lambda;(s\alpha)}^{(1)}
    \label{eq:global_iga_decomp}
\end{align}
where $\exp[\ii\theta n_\lambda^p]$ is an plaquette IGG, giving constraints on local tensors as
\begin{align}
    \left(n_{\lambda;(s\alpha)}^{(1)}+n_{\lambda;(s\alpha+1)}^{(2)}  \right)\cdot \hat{T}^s &= 0\,,\notag\\
    \hat{B}_{ss'}\cdot\left( n_{\lambda;(s\alpha)}^{(a)}+n_{\lambda;(s'\alpha')}^{(\bar{a})} \right)&=0\,,~a=1/2
\end{align}
And Eq.~\eqref{eq:plq_igg_comm} leads to the following commutation relation
\begin{align}
    [n_{\lambda;(s\alpha)}^{p_1},~n_{\lambda;(s\alpha)}^{p_2}]=0 
    \label{}
\end{align}

Similar as Eq.~\eqref{eq:global_igg_decomp_ambiguity}, the decomposition in Eq.~(\ref{eq:global_iga_decomp}) is also not unique:
\begin{align}
    n_{\lambda;(s\alpha)}^{(2)}+n_{\lambda;(s\alpha)}^{(1)}
    =(n_{\lambda;(s\alpha)}^{(2)}+\varepsilon)+(n_{\lambda;(s\alpha)}^{(1)}-\varepsilon)
    \label{eq:global_iga_decomp_ambiguity}
\end{align}
where $\varepsilon$ is a constant.

\subsection{Tensor equations for symmetric FPEPS}
As in the example discussed in Section~\ref{sec:example}, the presence of IGG leads to distinct multiplication rules for the symmetry action on internal legs compared to those on physical legs (Eq.~(\ref{eq:group_mul_rule})).
Namely, symmetries act as projective representations on internal legs, with IGG serving as the coefficient.
A well-known example is the MPS representation of AKLT states, where the physical legs are spin-1 states, while the internal legs support a spin-$\frac{1}{2}$, representing a projective representation of the $SO(3)$ symmetry with the coefficient in the form of a $U(1)$ phase IGG.
The characterization of such projective representation on internal legs is described by equations known as \emph{tensor equations}.
In the following, we will discuss the tensor equations for the symmetry group $G_f$ with multiplication rules given by Eq.~\eqref{eq:group_g_on_nf} and \eqref{eq:group_g12}.

\subsubsection{Commutators between $n_f$ and elements in $G_b$}
By pushing Eq.~(\ref{eq:group_g_on_u1f}) to internal legs, we obtain:
\begin{align}
    W(g)K^{s(g)}\circ U_{f}(\theta)= U_{D;g}\left( (-)^{s(g)}\theta \right) \cdot U_{f}\left( (-)^{\rho(g)+s(g)}\theta  \right)
    \label{eq:u1f_g_act_u1f}
\end{align}
Here, $U_{D;g}(\theta)$ is an IGG element, and can be expressed using its generator $n_D(g)$ as $\exp[\ii\theta n_D(g)]$.
In the following discussion, we will occasionally omit the internal leg index $(s\alpha)$ when there is no ambiguity.
Expressed in terms of $n_D$'s, we find from Eq.~(\ref{eq:group_g_on_nf}) that Eq.~\eqref{eq:u1f_g_act_u1f} becomes
\begin{align}
    W(g)\circ n_{f}= (-)^{\rho(g)}n_{f} + n_{D}(g)\,,
    \label{eq:u1f_g_act_nf}
\end{align}
As $U_f(0)=U_f(2\pi)=1$, we conclude eigenvalues of $n_{D}$'s are all integers. 

As mentioned earlier, the $U(1)$ IGG symmetry leads to long-range entanglement, which can be eliminated through the plaquette IGG decomposition given by Eq.~\eqref{eq:global_igg_decomp}:
\begin{align}
    U_{D;g}(\theta)=U_{\lambda;g}^{(2)}(\theta)\cdot U_{\lambda;g}^{(1)}(\theta)
    \label{}
\end{align}
By expressing $U_{\lambda;g}^{(a)}(\theta)$ as $\exp[\ii\theta n_\lambda^{(a)}(g)]$, we get decomposition of $n_D(g)$ to $n_\lambda(g)$ as
\begin{align}
    n_{D}(g)=n_{\lambda}^{(2)}(g)+n_{\lambda}^{(1)}(g)\,.
    \label{eq:nd_decomp}
\end{align}
Here, $n_D(g)$'s and $n_\lambda^p$'s carry zero fermion charge.

In the following, we establish consistent conditions for $n_D(g)$'s and $n_{\lambda}^{(a)}(g)$'s.
There are two ways to act $g_1g_2$ on $n_f$ on internal legs, resulting in the same outcome:
\begin{align}
    &W_{(s\alpha)}(g_1)K^{s(g_1)}\cdot W_{(s\alpha)}(g_2)K^{s(g_2)}\circ n_{f;(s\alpha)}\nonumber\\
    =&W_{(s\alpha)}(g_1g_2)K^{s(g_1g_2)}\circ n_{f;(s\alpha)}\,,
    \label{eq:g1g2_on_nf}
\end{align}
The derivation of this equation involves intricate details, and we relegate it to Sec.III of SM\footnotemark[10] for a comprehensive treatment.
We can introduce the notation $\act{g}{n}_f\equiv W(g)K^{s(g)}\circ n_f$, which is unambiguous based on the aforementioned equation.
Combining this with Eq.~(\ref{eq:u1f_g_act_nf}), we obtain:
\begin{equation}
    (-)^{\rho(g_2)}\cdot n_{D}(g_1)+\act{g_1}{n_{D}(g_2)}=n_{D}(g_1g_2)\,,
    \label{eq:u1f_nD_fusion}
\end{equation}
where $\act{g}{n}_D\equiv W(g)K^{s(g)}\circ n_D$.

Similarly, by acting $g_1g_2$ on $U_f(1)$ and utilizing Eq.~(\ref{eq:u1f_g_act_u1f}), we acquire
\begin{align}
    &\exp\left[ \ii(-)^{s(g_1g_2)+\rho(g_2)}\theta \cdot n_D(g_1) \right]\cdot \exp\left[ \ii(-)^{s(g_1g_2)}\theta\cdot \act{g_1}{n_D(g_2)} \right]\notag\\
    ={}&\exp\left[ \ii(-)^{s(g_1g_2)}\theta \cdot n_D(g_1g_2) \right]
    \label{eq:u1f_u1D_fusion}
\end{align}
Through a comparison between Eq.~\eqref{eq:u1f_nD_fusion} and \eqref{eq:u1f_u1D_fusion}, it is reasonable to assume that $[n_D(g_1),\act{g_1}{n_D(g_2)}]=0$.
After performing straightforward calculations, as detailed in Sec.III of SM\footnotemark[10], we arrive at the following simple commutation relation:
\begin{equation}
    [\act{g_1}{n}_{_{D,j}}(g_2),\act{g_3}{n}_{_{D,j}}(g_4)]= 0,~~\forall g_{1,2,3,4}\in G_b\,.
    \label{eq:nd_comm}
\end{equation}

Based on the above discussion on $n_D$'s, we come to derive relations for $n_\lambda$'s.
Due to the ambiguity presented in Eq.~\eqref{eq:global_iga_decomp_ambiguity}, $n_{\lambda}$'s satisfy similar relations as in Eq.~\eqref{eq:u1f_nD_fusion} up to an integer function $\sigma:G_b\times G_b\to \ZZ$
\begin{align}
    &(-)^{\rho(g_2)}\cdot n_{\lambda;(s\alpha)}^{(a)}(g_1)+\act{g_1}{n}_{\lambda;(s\alpha)}^{(a)}(g_2)\nonumber\\
    ={}&n_{\lambda;(s\alpha)}^{(a)}(g_1g_2)+(-)^{a+\rho(g_1g_2)}\sigma(g_1,g_2)\,,~a=1,2
    \label{eq:u1f_n_lam_fusion}
\end{align}
Here, we add $(-)^{a+\rho(g_1g_2)}$ to match results in Section~\ref{sec:classification}~(see details in Sec.VI of SM\footnotemark[10]).
Similar as $n_D$'s, we also assume $n_\lambda$'s satisfy the following simple commutation relations:
\begin{equation}
    [\act{g_1}{n}^{(a)}_{\lambda;(s\alpha)}(g_2),\act{g_3}{n}_{\lambda;(s\alpha)}^{(a)}(g_4)]= 0,~~\forall g_{1,2,3,4}\in G_b\,.
    \label{eq:nlambda_comm}
\end{equation}

\begin{widetext}
Here, $\sigma$'s are two-cocycles, which satisfy Eq.~\eqref{eq:u1f_sigma_2-cocycle}.
To see this, let us decompose $n_{\lambda}(g_1g_2g_3)$ in two different ways using Eq.~\eqref{eq:u1f_n_lam_fusion}:
\begin{align}
    n_{\lambda}^{(a)}(g_1g_2g_3) 
    ={}&(-)^{\rho(g_3)}n_{\lambda}^{(a)}(g_1g_2) +\act{g_1g_2}n_{\lambda}^{(a)}(g_3)-(-)^{a+\rho(g_1g_2g_3)}\sigma(g_1g_2,g_3)\\
    ={}&(-)^{\rho(g_2g_3)}n_{\lambda}^{(a)}(g_1) + (-)^{\rho(g_3)}\cdot\act{g_1}{n}_{\lambda}^{(a)}(g_2)+\act{g_1g_2}{n}_{\lambda}^{(a)}(g_3)-(-)^{a+\rho(g_1g_2g_3)}\cdot(\sigma(g_1,g_2)+ \sigma(g_1g_2,g_3))\notag
\end{align}
We also have
\begin{align}
    n_{\lambda}^{(a)}(g_1g_2g_3)
    ={}& (-)^{\rho(g_2g_3)}n_{\lambda}^{(a)}(g_1) +\act{g_1}n_{\lambda}^{(a)}(g_2g_3)-(-)^{a+\rho(g_1g_2g_3)}\sigma(g_1,g_2g_3)\\
    ={}&(-)^{\rho(g_2g_3)}n_{\lambda}^{(a)}(g_1) + (-)^{\rho(g_3)}\cdot\act{g_1}{n}_{\lambda}^{(a)}(g_2)  +\act{g_1g_2}{n}_{\lambda}^{(a)}(g_3)-(-)^{a+\rho(g_1g_2g_3)}((-)^{\rho(g_1)}\sigma(g_2,g_3)+ \sigma(g_1,g_2g_3))\notag
\end{align}
Thus, by comparing the above two equations, we recover Eq.~\eqref{eq:u1f_sigma_2-cocycle}:
\begin{align}
    \sigma(g_1,g_2)+\sigma(g_1g_2,g_3)=(-)^{\rho(g_1)}\sigma(g_2,g_3)+\sigma(g_1,g_2g_3)
    \label{eq:u1f_sigma_2-cocycle_tensor_eq}
\end{align}

\subsubsection{Fusion of symmetry flux in $G_b$}\label{subsubsec:tensor_eq_gb_fusion}
We now consider representation of Eq.~\eqref{eq:group_g12} on internal legs.
According to Eq.~(\ref{eq:sym_on_local_tensors}), when acting $(g_1,0)\cdot (g_2,0)$ on local site tensor $\hat{T}_s$ and bond tensor $\hat{B}_{ss'}$, we have
\begin{align}
   \hat{T}_{s}
   ={}&\left[ U_{s}(g_1)\otimes\left( \bigotimes_{\alpha}\subf W_{(s\alpha)}(g_1) \right)K^{s(g_1)} \right] \cdot \left[ U_{s}(g_2)\otimes\left( \bigotimes_{\alpha}\subf W_{(s\alpha)}(g_2) \right)K^{s(g_2)} \right]\cdot \hat{T}_{s}~,\notag\\
   ={}&\exp[\ii\alpha(g_1,g_2)n_{f;s}]\cdot U_{s}(g_1g_2)\otimes\left( \bigotimes_{\alpha}\subf \exp[\ii\alpha(g_1,g_2)n_{f;(s\alpha)}]\cdot W_{(s\alpha)}(g_1g_2) \right)K^{s(g_1g_2)}\cdot \hat{T}^s\\
   \label{eq:u1f_fusion_mul_rule}
   \hat{B}_{ss^{\prime}}={}&K^{s(g_1g_2)}\cdot \hat{B}_{ss{\prime}}\cdot K^{s(g_2)}\cdot \left( W^{-1}_{s\alpha}(g_2)\otimes_f W^{-1}_{s^{\prime}\alpha^{\prime}}(g_2) \right) K^{s(g_1)} \left( W_{s\alpha}^{-1}(g_1)\otimes_f W^{-1}_{s^{\prime}\alpha^{\prime}}(g_1) \right)\nonumber\\
   ={}&K^{s(g_1g_2)}\cdot\hat{B}_{ss^{\prime}}\cdot K^{s(g_1g_2)}\cdot \left( W^{-1}_{s\alpha}(g_1g_2)\otimes_f W^{-1}_{s^{\prime}\alpha^{\prime}}(g_1g_2) \right) \exp[-\ii\alpha(g_1,g_2)(n_{f;(s\alpha)}+n_{f;(s^{\prime}\alpha^{\prime})})]\,.
\end{align}
Naively, one might expect that $W_{(s\alpha)}(g_1)K^{s(g_1)}\cdot W_{(s\alpha)}(g_2)K^{s(g_2)}$ differs from $\exp[\ii\alpha(g_1,g_2)n_{f;(s\alpha)}]\cdot W_{(s\alpha)}(g_1g_2)K^{s(g_1g_2)}$ by an IGG element $D_{(s\alpha)}(g_1,g_2)$.
However, this relation is not true.
According to Eq.~(\ref{eq:u1f_g_act_nf}), the $W(g)$'s may not be parity-even, and thus, in general, one should add fermion swapping gate when move the actions of $g_1$ and $g_2$ on the same internal legs together, and thus in general,
\begin{align}
    \left( \bigotimes_{\alpha}\subf W_{(s\alpha)}(g_1) \right) K^{s(g_1)}\cdot \left( \bigotimes_{\alpha}\subf W_{(s\alpha)}(g_2) \right)K^{s(g_2)}
    \neq\left( \bigotimes_{\alpha}\subf W_{(s\alpha)}(g_1)\cdot \act{K^{s(g_1)}}{W_{(s\alpha)}(g_2)} \right) K^{s(g_1g_2)}
    \label{}
\end{align}
Therefore, IGG elements may not be able to capture the difference between these two ways of $g_1g_2$ action.

In the following, we propose a method to circumvent fermion swapping gates when exchanging $W(g_a)$'s acting on different internal legs.
To achieve this, we decompose $W_{(s\alpha)}(g)$ based on its fermion number:
\begin{align}
    W_{(s\alpha)}(g) = \sum_m W_{m;(s\alpha)}(g)\,,\quad
    \text{where }\ \left[ W_{m;(s\alpha)}(g),\,n_{f,j} \right]_{\rho(g)}=  m \cdot W_{m;(s\alpha)}(g)\,.
\end{align}
Here, $[A,B]_{\rho}\equiv AB-(-)^\rho BA$.
Fermion parity of $W_{m;(s\alpha)}(g)$ follows the even-oddness of $m$.
With such decomposition, the symmetry condition on local tensors $\hat{T}_s$ and $\hat{B}_{ss'}$ reads
\begin{align}
    \hat{T}_s=U_{s}(g)\otimes \sum_{m_0+m_1+\dots=0}\left( \bigotimes_{\alpha}\subf W_{m_\alpha;(s\alpha)}(g) \right)K^{s(g)}\cdot \hat{T}_s
    \label{eq:local_tensor_sym_beta}
\end{align}
\begin{align}
    \hat{B}_{ss^{\prime}}=K^{s(g)}\hat{B}_{ss^{\prime}}K^{s(g)}\cdot\sum_{m+m^{\prime}=0}\left( W_{m;s\alpha}(g)\otimes_f W_{m^{\prime};s^{\prime}\alpha^{\prime}}(g) \right)^{-1}
    \label{eq:bond_tensor_sym_beta}
\end{align}
\end{widetext}
Note that $\hat{T}_s$ and $\hat{B}_{ss'}$ are charge neutral, which implies only terms with zero total fermion numbers remain.

We then introduce traceless matrix $\beta$ with $\beta^2=\hat{1}$, and define
\begin{equation}
    W_{\beta}(g) \equiv \sum_{m} \beta^{m} W_{m}(g)\,,
    \label{eq:u1f_wbeta}
\end{equation}
\begin{align*}
    W_\beta(g)=\adjincludegraphics[valign=c]{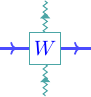}\,,
\end{align*}
where zigzag legs are space for $\beta$'s.
Similarly $W^{-1}_{\beta}(g)$ is defined as
\begin{align}
    W^{-1}_{\beta}(g) = \sum_m \beta^m W^{-1}_m(g)\,.
\end{align}
Thus, Eq.~\eqref{eq:local_tensor_sym_beta} can be expressed as
\begin{equation}
    \hat{T}^s=d_{\beta}^{-1}\, U(g)\otimes_f \Tr_\beta \left[ \bigotimes_{\alpha}\subf W_{\beta;(s\alpha)}(g)\right] K^{s(g)}\cdot \hat{T}^s\,,
\end{equation}
where $d_\beta$ is dimension for $\beta$. 
The above equation can be drawn graphically as
\begin{equation*}
    \adjincludegraphics[valign=c]{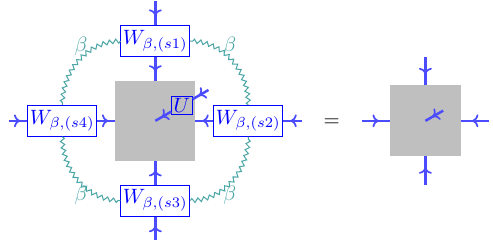}\,,
\end{equation*}
while for bond tensors
\begin{align*}
    \hat{B}_{ss^{\prime}} = d_{\beta}^{-1} K^{s(g)}\hat{B}_{ss^{\prime}}K^{s(g)}\cdot \Tr_\beta \left[  W^{-1}_{\beta,(s\alpha)}(g)\otimes_f  W^{-1}_{\beta,(s^{\prime}\alpha^{\prime})}(g)\right]\,,
\end{align*}
which is then illustrated as:
\begin{align*}
    \adjincludegraphics[valign=c]{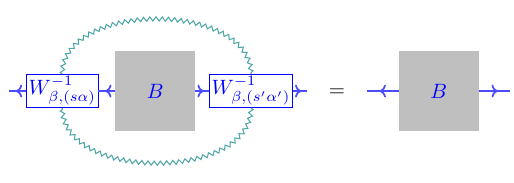}\,.
\end{align*}

To describe the action of $g_1\cdot g_2$ on internal legs, we introduce two traceless matrices, $\beta_1$ and $\beta_2$, satisfying the anti-commutation relation ${\beta_1,\beta_2}=0$.
This anti-commutation relation effectively cancels out the fermion sign arising from exchanging $W(g_{1/2})$'s, allowing us to move $W(g_{1/2})$ together on the same internal legs:
\begin{widetext}
\begin{align}
    \hat{T}^s&=d_{\beta}^{-1}\Tr_\beta\left[ U_s(g_1) \left( \bigotimes_{\alpha}\subf W_{\beta_1;(s\alpha)}(g_1)\right) K^{s(g_1)}\cdot U_s(g_2) \left( \bigotimes_{\alpha}\subf W_{\beta_2;(s\alpha)}(g_2)\right) K^{s(g_2)}\right]\cdot \hat{T}^s\notag\\
    &=d^{-1}_{\beta}\Tr_\beta \left[ U_s(g_1)\cdot \act{K^{s(g_1)}}{U_s(g_2)} \left( \bigotimes_{\alpha}\subf W_{\beta_1;(s\alpha)}(g_1)\cdot\act{K^{s(g_1)}}{W}_{\beta_2;(s\alpha)}(g_2) \right) K^{s(g_1g_2)} \right]\cdot\hat{T}^s \label{eq:u1f_fusion_beta}\\
    \hat{B}_{ss^{\prime}}&=K^{s(g_1g_2)}\cdot \hat{B}_{ss{\prime}}\cdot K^{s(g_2)}\cdot d_\beta^{-1}\Tr_\beta \left[ \left( W^{-1}_{\beta_2,(s\alpha)}(g_2)\otimes_f W^{-1}_{\beta_2,(s^{\prime}\alpha^{\prime})}(g_2) \right) K^{s(g_1)} \left( W^{-1}_{\beta_1,(s\alpha)}(g_1)\otimes_f W^{-1}_{\beta_1,(s^{\prime}\alpha^{\prime})}(g_1) \right)\right]\nonumber \\
    &=K^{s(g_1g_2)}\cdot \hat{B}_{ss{\prime}}\cdot K^{s(g_1g_2)}\cdot d_\beta^{-1}\Tr_\beta \left[ \act{K^{s(g_1)}}{W}^{-1}_{\beta_2,(s\alpha)}(g_2)\cdot W^{-1}_{\beta_1,(s\alpha)}(g_1) \otimes_f \act{K^{s(g_1)}}{W}^{-1}_{\beta_2,(s^{\prime}\alpha^{\prime})}(g_2)\cdot W^{-1}_{\beta_1,(s^{\prime}\alpha^{\prime})}(g_1)\right]\label{eq:u1f_fusion_beta_bond}
\end{align}
The above equation can easily be generalized to action of $g_1\cdot g_2\cdots g_n$: we simply consider $W_{\beta_1}(g_1)\cdot W_{\beta_2}(g_2)\cdots W_{\beta_n}(g_n)$, where $\left\{ \beta_i,\beta_j \right\}=2\delta_{ij}$.

Comparing Eq.~\eqref{eq:u1f_fusion_beta} and the second line in \eqref{eq:u1f_fusion_mul_rule}, we conclude that for $g_1\cdot g_2$ on a single internal leg $(s\alpha)$
\begin{align}
    &W_{\beta_1;(s\alpha)}(g_1)K^{s(g_1)}\cdot W_{\beta_2;(s\alpha)}(g_2)K^{s(g_2)}=D_{(\beta_1,\beta_2);(s\alpha)}(g_1,g_2)\cdot\exp\{ \ii\alpha(g_1,g_2)\cdot n_{f;(s\alpha)} \}\cdot W_{(s\alpha)}(g_1g_2)K^{s(g_1g_2)}    \label{eq:u1f_wg_fusion}\,,\\
    &~~\,\adjincludegraphics[valign=c]{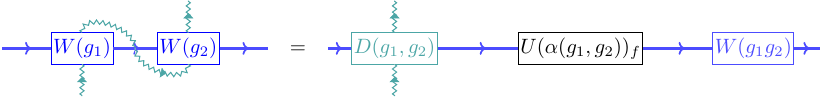}\,,\nonumber
\end{align}
where $D_{(\beta_1,\beta_2)}(g_1,g_2)$'s are charge neutral, which satisfy
\begin{align}
    d_\beta^{-1}\cdot \Tr_\beta\left[ \bigotimes_{\alpha} D_{(\beta_1,\beta_2);(s\alpha)}(g_1,g_2) \right]\cdot \hat{T}_s={}&\hat{T}_s\\
    \hat{B}_{ss^{\prime}}\cdot d_\beta^{-1}\cdot \Tr_\beta\left[D_{(\beta_1,\beta_2);(s\alpha)}(g_1,g_2)\otimes D_{(\beta_1,\beta_2);(s^{\prime}\alpha^{\prime})}(g_1,g_2) \right] ={}& \hat{B}_{ss^{\prime}}\,.
\end{align}
Namely, $D_{(\beta_1,\beta_2);(s\alpha)}$ behaves like an IGG element, but with additional $\beta$ legs.

\subsubsection{Decomposing $D_{(\beta_1,\beta_2)}(g_1,g_2)$ to plaquette IGG}
To avoid long-range entanglement, we assume that $D_{(\beta_1,\beta_2)}(g_1,g_2)$'s can be further decomposed to ``plaquette IGG elements'' with additional $\beta$ legs:
\begin{align}
    &D_{(\beta_1,\beta_2);(s\alpha)}(g_1,g_2)
    =\Lambda_{(\beta_1,\beta_2);(s\alpha)}^{(2)}(g_1,g_2)\cdot \Lambda_{(\beta_1,\beta_2);(s\alpha)}^{(1)}(g_1,g_2)
    \label{eq:u1f_db1b2_decomp}\\
    &\hspace{-1.0cm}\adjincludegraphics[valign=c,scale=1]{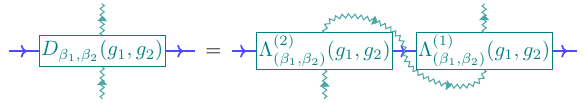}\nonumber
\end{align}
where $\Lambda$'s satisfy 
\begin{align}
    \Lambda_{(\beta_1,\beta_2);(s\alpha)}^{(1)}(g_1,g_2)\otimes_f \Lambda_{(\beta_1,\beta_2);(s\alpha+1)}^{(2)}(g_1,g_2)\cdot \hat{T}_s&=\hat{T}_s\otimes \I_\beta,,\notag\\
    \hat{B}_{ss'}\cdot \Lambda_{(\beta_1,\beta_2),(s\alpha)}^{(a)}(g_1,g_2)\otimes_f \Lambda_{(\beta_1,\beta_2);(s'\alpha')}^{(\bar{a})}(g_1,g_2)&=\hat{B}_{ss'}\otimes \I_\beta\,,\notag\\
    \label{eq:u1f_lambda_plq_igg}
\end{align}
\begin{align*}
    \hspace{-1cm}\adjincludegraphics[valign=c]{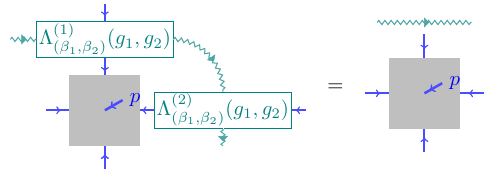};
    \adjincludegraphics[valign=c]{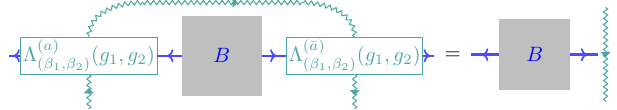}\,,
\end{align*}
where the dangling teal zigzag line represents the remaining identity matrix $\I_\beta$ in the $\beta$ space.
Similar as Eq.~\eqref{eq:global_igg_decomp_ambiguity}, decomposition in Eq.~\eqref{eq:u1f_db1b2_decomp} has a $U(1)$ phase ambiguity:
\begin{align}
    &D_{(\beta_1,\beta_2);(s\alpha)}(g_1,g_2)\notag\\
    ={}&\Lambda_{(\beta_1,\beta_2);(s\alpha)}^{(2)}(g_1,g_2)\cdot \Lambda_{(\beta_1,\beta_2);(s\alpha)}^{(1)}(g_1,g_2) \label{eq:u1f_db1b2_decomp_phase_ambiguity} \\
    ={}&\left( \ee^{\ii\theta}\cdot\Lambda_{(\beta_1,\beta_2);(s\alpha)}^{(2)}(g_1,g_2) \right) \cdot \left( \ee^{-\ii\theta}\cdot \Lambda_{(\beta_1,\beta_2);(s\alpha)}^{(1)}(g_1,g_2) \right)\notag
\end{align}

Here in the main text, we show the expression for $\Lambda^{(a)}$'s without derivation, and we leave details in Sec.III of SM\footnotemark[10]:
\begin{align}
    &\Lambda^{(1)}_{(\beta_1,\beta_2)}(g_1,g_2)=\iota^{(1)}_{\beta_1}(g_1) \cdot \act{g_1}{\iota}{^{(1)}_{\beta_2}}(g_2) \cdot \lambda^{(1)}(g_1,g_2)\notag\\
    &\Lambda^{(2)}_{(\beta_1,\beta_2)}(g_1,g_2)= \lambda^{(2)}(g_1,g_2)\cdot \act{g_1}{\iota}^{(2)}_{\beta_2}(g_2)\cdot \iota^{(2)}_{\beta_1}(g_1)\label{eq:lambda_iota}\\
    &\adjincludegraphics[valign=c]{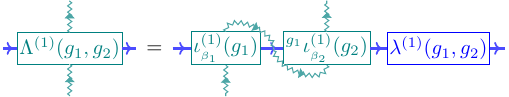}\nonumber\\
    &\adjincludegraphics[valign=c]{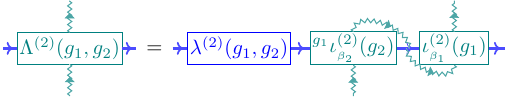}\,,
\end{align}
where $\lambda^{(a)}(g_1,g_2)$ belongs to plaquette IGG, namely
\begin{align}
    \lambda_{(s\alpha)}^{(1)}(g_1,g_2)\otimes_f \lambda_{(s\alpha)}^{(2)}(g_1,g_2)\cdot \hat{T}_s&=\hat{T}_s\notag\\
    \hat{B}_{ss'}\cdot \lambda_{(s\alpha)}^{(a)}(g_1,g_2)\otimes_f \lambda_{(s'\alpha')}^{(\bar{a})}(g_1,g_2)&=\hat{B}_{ss'}
    \label{}
\end{align}
$\iota^{(a)}$'s in Eq.~\eqref{eq:lambda_iota} are defined as
\begin{align}
    \iota_{\beta}^{(a)}(g)\equiv \beta^{n^{(a)}_{\lambda}(g)}\,,
    \label{}
\end{align}
which constitute a plaquette IGG, since
\begin{align*}
    \iota_{\beta;(s\alpha)}^{(1)}\otimes \iota_{\beta;(s\alpha+1)}^{(2)}\cdot \hat{T}^s = \beta^{n_{\lambda;(s\alpha)}^{(1)}+n_{\lambda;(s\alpha)}^{(2)}}\cdot \hat{T}^s =\hat{T}^s \otimes \I_\beta\,,
\end{align*}
and similar IGG holds for bond tensor:
\begin{align*}
    \iota_{\beta;(s\alpha)}^{(a)}\otimes \iota_{\beta;(s^{\prime}\alpha^{\prime})}^{(\bar{a})}\cdot \hat{B}_{ss^\prime} = \beta^{n_{\lambda;(s\alpha)}^{(a)}+n_{\lambda;(s^{\prime}\alpha^{\prime})}^{(\bar{a})}}\cdot \hat{B}_{ss^{\prime}} =\hat{B}_{ss^\prime} \otimes \I_\beta\,.
\end{align*}
We also introduce $I_{\beta}(g)$ for later use
\begin{align}
    &I_{\beta}(g)=\iota_{\beta}^{(2)}(g)\cdot \iota_{\beta}^{(1)}(g)
    =(\beta)^{n_{D;(s\alpha)}(g)}\,,\label{eq:u1f_ibeta_iota}\\
    &\hspace*{-0.7cm}\adjincludegraphics[valign=c]{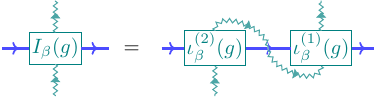}\,,\nonumber
\end{align}
To proceed, we also assume the following commutation relation:
\begin{align}
    [\lambda^{(a)}(g_1,g_2),\,\act{g_3}{n_\lambda^{(a)}(g_4)}]=0\,,\quad
    \forall g_j\in G_b
    \label{}
\end{align}

\subsubsection{Associativity of symmetry fusions}
Equipped with expressions for $\Lambda_{(\beta_1,\beta_2)}(g_1,g_2)$'s and $D_{(\beta_1,\beta_2)}(g_1,g_2)$'s, we are now ready to derive consistent conditions from the associativity of $g_1\cdot g_2\cdot g_3$ action on local tensors, $\forall g_{1,2,3}\in G_b$.
There are two different ways to fuse them on a single leg $(s\alpha)$:
\begin{align}
    &W_{\beta_1}(g_1)K^{s(g_1)}\cdot W_{\beta_2}(g_2)K^{s(g_2)}\cdot W_{\beta_3}(g_3)K^{s(g_3)}\nonumber\\
    =&D_{(\beta_1,\beta_2)}(g_1,g_2)\cdot \ee^{\ii\alpha(g_1,g_2)n_{f}}\cdot W(g_1g_2)K^{s(g_1g_2)}\cdot W_{\beta_3}(g_3)K^{s(g_3)}\nonumber\\
    =&D_{(\beta_1,\beta_2)}(g_1,g_2)\cdot I_{\beta_2}(g_1g_2)\cdot D_{(\beta_2,\beta_3)}(g_1g_2,g_3)\cdot \ee^{\ii(\alpha(g_1g_2,g_3)+\alpha(g_1,g_2))n_{f}}\cdot W(g_1g_2g_3)K^{s(g_1g_2g_3)}\,,\label{eq:u1f_3_sym_fusion_1}\\
    =& \adjincludegraphics[valign=c]{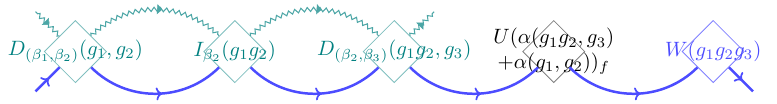}\,,\nonumber
\end{align}
where we use the following equation in the third line:
\begin{align}
    W_{\beta;(s\alpha)}(g) = I_{\beta;(s\alpha)}(g)\cdot W_{(s\alpha)}(g)\,,\quad
    \adjincludegraphics[valign=c]{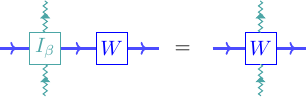}\,.
\end{align}
Another fusion way gives
\begin{align}
    &W_{\beta_1}(g_1)K^{s(g_1)}\cdot W_{\beta_2}(g_2)K^{s(g_2)}\cdot W_{\beta_3}(g_3)K^{s(g_3)}\nonumber\\
    =&W_{\beta_1}(g_1)K^{s(g_1)}\cdot D_{(\beta_2,\beta_3)}(g_2,g_3)\cdot \ee^{\ii\alpha(g_2,g_3)n_{f}}\cdot W(g_2g_3)K^{s(g_2g_3)}\nonumber\\
    =&\act{W_{\beta_1}(g_1)K^{s(g_1)}}{[D_{(\beta_2,\beta_3)}(g_2,g_3)\cdot \ee^{\ii\alpha(g_2,g_3)n_{f}}\cdot I_{\beta_3}(g_3)]}\cdot D_{(\beta_1,\beta_3)}(g_1,g_2g_3)\cdot \ee^{\ii\alpha(g_1,g_2g_3) n_{f}}\cdot W(g_1g_2g_3)K^{s(g_1g_2g_3)}\,,\label{eq:u1f_3_sym_fusion_2}
\end{align}
where $\beta_{1,2,3}$ anticommute with each other.

Then, by decomposing $D$'s in Eq.~\eqref{eq:u1f_3_sym_fusion_1} and \eqref{eq:u1f_3_sym_fusion_2} to $\Lambda$'s by Eq.~(\ref{eq:u1f_db1b2_decomp}), and further to $\lambda$'s by Eq.~\eqref{eq:lambda_iota}, we are able to derive consistent conditions for $\lambda^{(1/2)}$'s:
\begin{equation}
    \lambda^{(2)}(g_1,g_2)\lambda^{(2)}(g_1g_2,g_3) = \frac{(-)^{{n_{\lambda}^{(2)}(g_1)}\sigma(g_2,g_3)}}{\omega^{\prime}(g_1,g_2,g_3)}\left(\ee^{\ii (-)^{s(g_1)}\alpha(g_2,g_3)n^{(2)}_{\lambda}(g_1)}\right)\act{g_1}{\lambda^{(2)}(g_2,g_3)}\lambda^{(2)}(g_1,g_2g_3)\,,
    \label{eq:u1f_lam_omega_2}
\end{equation}
as well as
\begin{equation}
    \lambda^{(1)}(g_1,g_2)\lambda^{(1)}(g_1g_2,g_3) = \omega^{\prime}(g_1,g_2,g_3)(-)^{{n_{\lambda}^{(1)}(g_1)}\sigma(g_2,g_3)}\left(\ee^{\ii (-)^{s(g_1)}\alpha(g_2,g_3)n^{(1)}_{\lambda}(g_1)}\right)\act{g_1}{\lambda^{(1)}(g_2,g_3)}\lambda^{(1)}(g_1,g_2g_3)\,.
    \label{eq:u1f_lam_omega_1}
\end{equation}
The derivations are quite lengthy, and we list details in Sec.IV of SM\footnotemark[10].

We mention that that the use of $\omega^{\prime}(g_1,g_2,g_3)$ is intentionally, as it satisfies 3-cocycle condition similar~(but not the same) as Eq.~(\ref{eq:u1f_fermionic_3-cocycle}), see details in Sec.IV of SM\footnotemark[10]:
\begin{align}
    \frac{\omega^{\prime}(g_1,g_2,g_3)\omega^{\prime}(g_1,g_2g_3,g_4)[\omega^{\prime}(g_2,g_3,g_4)]^{1-2s(g_1)}}{\omega^{\prime}(g_1g_2,g_3,g_4)\omega^{\prime}(g_1,g_2,g_3g_4)}=(-)^{\sigma(g_1,g_2)\sigma(g_3,g_4)}~\ee^{\ii (-)^{\tilde\rho(g_1g_2)} \alpha(g_3,g_4)\sigma(g_1,g_2) }
    \label{eq:omegap_3-cocycle}
\end{align}
The relation between $\omega^{\prime}$ and the $\omega$ in Eq.~\eqref{eq:u1f_fermionic_3-cocycle} is derived in Sec.VI of SM\footnotemark[10], where we list the result here:
\begin{align}
    \omega(g_1,g_2,g_3) = \omega^{\prime}(g_1,g_2,g_3)\cdot \exp\big[ -\ii(-)^{\tilde\rho(g_1)}\alpha(g_2,g_3)\sigma(g_1,g_2g_3)+\ii\alpha(g_1,g_2)\sigma(g_1g_2,g_3) \big]\,.
    \label{eq:omega_omegaprime}
\end{align}
As we will see in Section~\ref{sec:edge_theory}, physical meaning of $\omega$'s are indeed related to fusion of symmetry defects, which characterize quantum anomaly at edge.
\end{widetext}

\subsection{Solutions to tensor equations of interacting TI}
Let us summarize results for tensor equations describing symmetry actions on internal legs of interacting TI. 
Given a fermionic 3-cocycle data $(\sigma,\omega')$ with consistent conditions in Eq.~\eqref{eq:u1f_sigma_2-cocycle_tensor_eq} and \eqref{eq:omegap_3-cocycle}, we are able to list a set of tensor equations. 
Firstly, due to the presence of IGG, $G_f$ are represented projectively on internal legs as
\begin{widetext}
\begin{align}
    W(g)K^{s(g)}\circ n_f&=(-)^{\rho(g)}n_f+n_D(g)\,,\notag\\
    W_{\beta_1}(g_1)K^{s(g_1)}\cdot W_{\beta_2}(g_2)K^{s(g_2)}&=D_{(\beta_1,\beta_2)}(g_1,g_2)\cdot\exp\{ \ii\alpha(g_1,g_2)\cdot n_{f} \}\cdot W(g_1g_2)K^{s(g_1g_2)}\,,\quad
    \forall g_j\in G_b
    \label{eq:group_relation_internal_legs}
\end{align}
\end{widetext}
where $W_{\beta_j}(g_j)$ is defined in Eq.~\eqref{eq:u1f_wbeta} with $\{ \beta_i,\,\beta_j\}=2\delta_{ij}$. 

Secondly, as TI phases are short-range entangled, $n_D(g)$ and $D_{(\beta_1,\beta_2)}(g_1,g_2)$, which may lead to long-range entanglement, should be decomposed as plaquette IGA/IGG elements:
\begin{align}
    n_D(g)&=n_\lambda^{(2)}(g)+n_\lambda^{(1)}(g)\,,\notag\\
    D_{(\beta_1,\beta_2)}(g_1,g_2)&=\Lambda^{(2)}_{(\beta_1,\beta_2)}(g_1,g_2)\cdot\Lambda^{(1)}_{(\beta_1,\beta_2)}(g_1,g_2)
    \label{eq:nd_d_decomp_to_nlambda_lambda}
\end{align}
Note that such decomposition is unique up to $U(1)$ phase, as seen from Eq.~\eqref{eq:global_igg_decomp_ambiguity} and \eqref{eq:global_iga_decomp_ambiguity}.
Here, $n_\lambda^{(a)}(g)$ is an plaquette IGA element.
And 
\begin{align}
    \Lambda^{(a)}_{(\beta_1,\beta_2)}(g_1,g_2)=\iota_{\beta_1}^{(a)}(g_1)\cdot \act{g_1}\iota_{\beta_2}^{(a)}(g_2)\cdot \lambda^{(a)}(g_1,g_2)
    \label{eq:lambdab1b2_decomp}
\end{align}
where $\lambda^{(a)}$ belongs to plaquette IGG, and $\iota_{\beta}\equiv\beta^{n_\lambda^{(a)}(g)}$.
We further assume the following simple commutation relations:
\begin{align}
    [\act{g_1}n_\lambda^{(a)}(g_2),\,\act{g_3}n_\lambda^{(b)}(g_4)]
    =[\lambda^{(a)}(g_1,g_2),\,\act{g_3}{n_\lambda^{(b)}(g_4)}]
    =0
    \label{}
\end{align}

Lastly, from fusion rules of symmetry defects~(or associativity of group multiplication), $n_\lambda^{(a)}(g)$'s and $\lambda^{(a)}(g_1,g_2)$ satisfy the following relations:
\begin{widetext}
\begin{align}
    &(-)^{\rho(g_2)}\cdot n_{\lambda;(s\alpha)}^{(a)}(g_1)+\act{g_1}{n}_{\lambda;(s\alpha)}^{(a)}(g_2)
    =n_{\lambda;(s\alpha)}^{(a)}(g_1g_2)+(-)^{a+\rho(g_1g_2)}\sigma(g_1,g_2)\notag\\
    &\lambda^{(1)}(g_1,g_2)\lambda^{(1)}(g_1g_2,g_3) 
    = \omega^{\prime}(g_1,g_2,g_3)(-)^{{n_{\lambda}^{(1)}(g_1)}\sigma(g_2,g_3)}\left(\ee^{\ii (-)^{s(g_1)}\alpha(g_2,g_3)n^{(1)}_{\lambda}(g_1)}\right)\act{g_1}{\lambda^{(1)}(g_2,g_3)}\lambda^{(1)}(g_1,g_2g_3)\notag\\
    &\lambda^{(2)}(g_1,g_2)\lambda^{(2)}(g_1g_2,g_3) 
    = \frac{(-)^{{n_{\lambda}^{(2)}(g_1)}\sigma(g_2,g_3)}}{\omega^{\prime}(g_1,g_2,g_3)}\left(\ee^{\ii (-)^{s(g_1)}\alpha(g_2,g_3)n^{(2)}_{\lambda}(g_1)}\right)\act{g_1}{\lambda^{(2)}(g_2,g_3)}\lambda^{(2)}(g_1,g_2g_3)
    \label{eq:nlambda_lambda_consistent_eq}
\end{align}
\end{widetext}

In the following, we focus on the case where $G_b$ is a finite group with order $\abs{G_b}$.
Then, for a given three cocycle data $(\sigma,\omega')$ for $G_f$, we list a solution for above tensor equations, where details are presented in Sec.V of SM\footnotemark[10].

We consider systems on bipartite lattices with sublattice indices $u,v$.
For an internal leg $(s\alpha)$, its dimension equals $\abs{G_b}^2$, and its basis states are labeled as $\ket{l,r}_{(s\alpha)}$, where $l,r\in G_b$.
As before, internal legs point from bond tensors to site tensors.

Fermion charge number operators act as
\begin{align}
    n_{f;(u\alpha)}\ket{l,r}&=-\sigma(r,r^{-1}l)\ket{l,r}\notag\\
    n_{f;(v\alpha)}\ket{l,r}&=\sigma(l,l^{-1}r)\ket{l,r}
\end{align}
where $\sigma$ is the given twisted 2-cocycle satisfying Eq.~\eqref{eq:u1f_sigma_2-cocycle_tensor_eq}.
For $g\in G_b$, its action on internal legs reads
\begin{align}
    W_{(u\alpha)}(g)\ket{l,r}&=(-)^{\sigma(g,r)\sigma(r,r^{-1}l)}\cdot\omega^{-1}(g,r,r^{-1}l)\ket{gl,gr}\notag\\
    W_{(v\alpha)}(g)\ket{l,r}&=(-)^{\sigma(g,r)\sigma(l,l^{-1}r)}\cdot \omega(g,l,l^{-1}r)\ket{gl,gr}
    \label{}
\end{align}
where $\omega$ are related to $\omega'$ by Eq.~\eqref{eq:omega_omegaprime}.
By acting $W(g)$ on $n_f$, one obtains $n_D(g)$, which can further be decomposed to $n_\lambda^{(a)}(g)$.
And by studying $g_1\cdot g_2$ on internal legs, one is able to obtain $D_{(\beta_1,\beta_2)(g_1,g_2)}$ and further $\lambda^{(a)}(g_1,g_2)$.
In Sec.V of SM\footnotemark[10], we carefully verify that $n_\lambda^{(a)}(g)$'s and $\lambda^{(a)}(g_1,g_2)$'s are indeed solutions for Eq.~\eqref{eq:nlambda_lambda_consistent_eq}.

\section{Edge theories from tensor equation}\label{sec:edge_theory}
In this section, we delve into the physical interpretation of the fermionic three-cocycle derived from the tensor equations.
By constructing edge theories based on FPEPS, we uncover an intriguing connection between the fusion rules of symmetry defects and the fermionic three-cocycle data.
These fusion rules play a crucial role in characterizing the quantum anomaly of edge theories for topological insulator phases.

\subsection{Edge Hilbert space}\label{subsec:edge_hilbert_space}
Let us first identify the edge Hilbert space from an infinite FPEPS.
Given an infinite FPEPS, we cut a region $A$ from it, whose boundary legs are labeled as $\partial A = \{1\cdots L\}$:
\begin{equation*}
    \adjincludegraphics[valign=c]{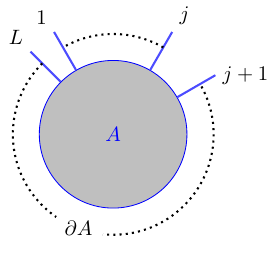}\,,
\end{equation*}
Here, we assume that all boundary legs are bra states pointing outwards.

By contracting all internal legs within $A$, we get a linear map $\hat{T}_A$ from boundary legs~($\HH_{\partial A}$) to physical legs within $A$~($\HH_A$), which reads
\begin{align}
    \hat{T}_A=\sum (T_A)_{i_b i_e}\ket{i_b}\bra{i_e}
    \label{}
\end{align}
For large enough $A$, $\dim\HH_A\gg \dim\HH_{\partial A}$, so such map can never be surjective.

One can construct a parent Hamiltonian $H$\cite{schuch2010peps}, where the infinite FPEPS belongs to ground state manifold.
By restrict $H$ to region $A$ with some additional boundary terms, one obtains $H_A$. 
It is then reasonable to assume that $\imag{\hat{T}_A}$ to be the low-energy space of $H_A$.
For gapped phases, $\imag{\hat{T}_A}$ is identified as edge Hilbert space of $H_A$, labeled as $\HH_{edge}$.
From the isomorphism theorem, we have
\begin{equation}
    \HH_{edge} \simeq \HH_{\partial A}/{\rm{ker}}\,\hat{T}_A\,.
\end{equation}

According to definition of IGG, for $J\in\IGG$, $\hat{T}_A\cdot J_{\partial A} = \hat{T}_A$, where $J_{\partial A}=\otimes_{j\in \partial A} J_j$.
Namely, 
\begin{align}
    \imag{(\hat{1}_{\partial A}-J_{\partial A})}\simeq \HH_{\partial A}/\ker{(\hat{1}_{\partial A}-J_{\partial A})}\subseteq\ker\hat{T}_A
    \label{}
\end{align}
For large enough region $A$, we further assume that
\begin{align}
    \ker{\hat{T}_A}=\bigcup_{J\in IGG}\imag{(\hat{1}_{\partial A}-J_{\partial A})}\,,
    \label{}
\end{align}
and thus edge Hilbert space are formed by those boundary states that are invariant under action of all IGG elements:
\begin{align}
    &\HH_{edge} \simeq\ker{(\hat{1}_{\partial A}-J_{\partial A})}\notag\\
    ={}&\{\ket{\psi_{\partial A}}~|~ J_{\partial A}\ket{\psi_{\partial A}} = \ket{\psi_{\partial A}},~\forall J\in {\rm{IGG}} \}\,.
    \label{eq:h_edge}
\end{align}

For later convenience, we construct projector to $\HH_{edge}$ as 
\begin{equation}
    P_{edge} = T_A^{-1}\cdot T_A\,,
    \label{eq:pedge_ta}
\end{equation}
where $T_A^{-1}$: $\HH_A\to \HH_{\partial A}$ is a pseudo-inverse of $T_A$.
According to Eq.~\eqref{eq:h_edge}, $\pe$ are defined by the following relation:
\begin{equation}
    \pe = J_{\partial_A}\cdot\pe = \pe \cdot J_{\partial A}\,,~\forall J\in\IGG
    \label{eq:edge_proj_igg_comm}
\end{equation}

For FSPT phases, all IGG elements can be decomposed to product of plaquette IGG elements, and thus
\begin{equation}
    \pe = \prod_{j=1}^{L} P_{j+\frac{1}{2}}\,,
    \label{eq:p_edge_dec}
\end{equation}
where $j+\frac{1}{2}$ is as boundary plaquette index.
As IGG belongs to different plaquettes commute, we have
\begin{align}
    P_{j_1+\frac{1}{2}}P_{j_2+\frac{1}{2}}=P_{j_2+\frac{1}{2}}P_{j_1+\frac{1}{2}}
    \label{}
\end{align}

With Eq.~\eqref{eq:edge_proj_igg_comm}, Eq.~\eqref{eq:nd_decomp}, and Eq.~\eqref{eq:lambda_iota}, we have
\begin{align}
   &P_{j+\frac{1}{2}}\cdot \left( n^{(1)}_{\lambda,j+\frac{1}{2}}(g)+n^{(2)}_{\lambda,j+\frac{1}{2}}(g) \right) =0,~~\forall g\in G_b \label{eq:local_edge_space_nlambda} \\
   &P_{j+\frac{1}{2}}\cdot \left(\lambda^{(1)}_{j+\frac{1}{2}}(g_1,g_2)\otimes\lambda^{(2)}_{j+\frac{1}{2}}(g_1,g_2)\right)=P_{j+\frac{1}{2}},~~\forall g_{1,2}\in G_b\notag
\end{align}
Here, $n_{\lambda,j+\frac{1}{2}}^{(a)}$'s and $\lambda^{(a)}_{j+\frac{1}{2}}$'s act on boundary leg $(j+a-1)$.

\subsection{Symmetry defects on the edge}\label{subsec:sym_defect_edge}
For system defined on $A$, symmetry $g$ acts on $\HH_A$ as $U_A(g)K^{s(g)}$, where $U_A(g)\equiv\bigotimes_{i\in A}U_i(A)$.
As $\hat{T}_A$ is cut from symmetric infinite FPEPS, we have
\begin{align}
    U_A(g)K^{s(g)}\cdot \hat{T}_A=\hat{T}_{A}\cdot W_{\partial A}(g)K^{s(g)}
    \label{eq:sym_on_ta}
\end{align}
Then, symmetry operation on $\HH_{edge}$ is defined as 
\begin{align}
    &U_{edge}(g)K^{s(g)}\equiv P_{edge}\cdot W_{\partial A}(g)K^{s(g)}\cdot P_{edge}\notag\\
    ={}&\hat{T}_A^{-1}\cdot U_A(g)K^{s(g)}\cdot \hat{T}_A\cdot P_{edge}
    =\hat{T}_A^{-1}\cdot U_A(g)K^{s(g)}\cdot \hat{T}_A\notag\\
    ={}&P_{edge}\cdot W_{\partial A}(g)K^{s(g)}
    \label{eq:pedge_w}
\end{align}
where we use Eq.~\eqref{eq:pedge_ta} and \eqref{eq:sym_on_ta} in the second line.
Here,
\begin{align}
    W_{\partial A}=\bigotimes_{j\in \partial A}\subf \overline{W}_j(g)
    \label{}
\end{align}
where $\overline{W}_j$ is either $W_j$ or $\act{F}{W}_j$ according to the rule of Kasteleyn orientation, see details in Section~\ref{sec:kasteleyn}.
We will drop the ``bar'' notation over $W$, as it does not affect the algebra later on.

In this work, $P_{edge}$ can be decomposed as in Eq.~\eqref{eq:p_edge_dec}, and we have additional constraint comparing to Eq.~\eqref{eq:pedge_w}: 
\begin{align}
    &P_{j+\frac{1}{2}}\cdot W_{j}(g)\otimes_f W_{j+1}(g) K^{s(g)}\cdot P_{j+\frac{1}{2}}\notag\\
    ={}& P_{j+\frac{1}{2}}\cdot W_{j}(g)\otimes_f W_{j+1}(g)K^{s(g)}
    \label{eq:plaquette_proj_w}
\end{align}
To see this, we consider $\ket{\psi}\in\HH_{\partial A}$, but $P_{j+\frac{1}{2}}\ket{\psi}=0$, and the above equation holds iff 
\begin{align}
    P_{j+\frac{1}{2}}\cdot W_{j}(g)\otimes_f W_{j+1}(g)K^{s(g)}\ket{\psi}=0
    \label{eq:wg_psi_ni}
\end{align}

By definition of $\ket{\psi}$, there exists a plaquette IGG element $\lambda_{j+\frac{1}{2}}$, such that
\begin{align}
    \left( \lambda_{j+\frac{1}{2}}^{(1)}\otimes\lambda_{j+\frac{1}{2}}^{(2)} \right) \cdot\ket{\psi}=\alpha\ket{\psi}\,,~\text{with }\alpha\neq 1
    \label{}
\end{align}
We define 
\begin{align}
    \tilde{\lambda}_{j+\frac{1}{2}}^{(a)}&\equiv W_{\partial A}(g)K^{s(g)}\circ \lambda_{j+\frac{1}{2}}^{(a)}\notag\\
    &=\left( W_{j}(g)\otimes_f W_{j+1}(g) K^{s(g)}\right)\circ \lambda_{j+\frac{1}{2}}^{(a)}
    \label{}
\end{align}
And thus
\begin{align}
    &\left( \tilde{\lambda}_{j+\frac{1}{2}}^{(1)}\otimes \tilde{\lambda}_{j+\frac{1}{2}}^{(2)} \right)\cdot W_{j}(g)\otimes W_{j+1}(g) K^{s(g)}\ket{\psi}\notag\\
    ={}&\alpha\cdot W_{j}(g)\otimes W_{j+1}(g) K^{s(g)}\ket{\psi}
    \label{}
\end{align}
which proves Eq.~\eqref{eq:wg_psi_ni}.

With the aforementioned definitions, we aim to extract the anomaly cocycle data of the edge theories following the procedure outlined in Section~\ref{subsec:anomaly_cocycle}.
As shown in Eq.~\eqref{eq:edge_hilbert_tensor_prod} and \eqref{eq:edge_nf_onsite}, edge Hilbert space has a tensor product structure, and $U_f(1)$ acts as onsite symmetry, 
These two properties in general are not satisfied for $\HH_{edge}$ obtained in Eq.~\eqref{eq:h_edge}.
To reconcile this, we consider FPEPS with each internal leg expressed as a tensor product of three legs: two side-legs supporting spins and one middle leg supporting a spinless fermion.
Then, operators $n_{\lambda;j-\frac{1}{2}}^{(2)}(g)$ and $n_{\lambda;j+\frac{2}{1}}^{(1)}(g)$ act on the left and right side-legs, respectively, while $n_{f;j}$ acts on the middle leg.
As a result, $P_{j+\frac{1}{2}}$ identifies the right side-leg of site $j$ with the left side-leg of site $j+1$.
In this way, the edge Hilbert space $\HH_{edge}$ becomes a tensor product of local spins on plaquettes and local fermions on sites.

With Eq.~\eqref{eq:p_edge_dec} and \eqref{eq:pedge_w}, it is straightforward to see that as in Section~\ref{subsec:anomaly_cocycle}, $U_{edge}$ is a finite depth local unitary circuits, which satisfies 
\begin{align}
    &U_{edge}(g_a)K^{s(g_a)}\circ n_{f,\pa}=(-1)^{\rho(g_a)}\cdot n_{f,\pa}\cdot \pe\,, \nonumber\\
    &U_{edge}(g_1)K^{s(g_1)}\cdot U_{edge}(g_2)K^{s(g_2)}\nonumber\\
    =&\exp\left[ \ii\alpha(g_1,g_2)\cdot n_{f;\partial A} \right]\cdot U_{edge}(g_1g_2)K^{s(g_1g_2)}\nonumber\,,
\end{align}
where $n_{f;\pa} \equiv \sum_{j}n_{f;j}$. 

In analogy to Section~\ref{subsec:anomaly_cocycle}, the extraction of anomaly data relies on the fusion rule of symmetry defects.
Therefore, to extract this data, we begin by constructing symmetry defects using the FPEPS formalism.
For a subregion $M={1,\dots,l}$ of $\partial A$, $g$-defects at the two ends can be created through the following operation:
\begin{align}
    U_M(g)=\pe \cdot w_{l+1}(g)w_L(g)\cdot W_{M}(g)\cdot \pe\,,
    \label{eq:um_def}
\end{align}
where $W_M(g) = \bigotimes_{f,j=1}^l W_j(g)$.
$w_L(g)$ and $w_{l+1}(g)$ are local operators at two ends of $M$, which are chosen to give the following equation
\begin{align*}
    U_M(g) = \pe\cdot w_{l+1}(g)w_L(g)\cdot W_M(g)
\end{align*}
With Eq.~\eqref{eq:plaquette_proj_w}, it requires
\begin{align}
    P_{\frac{1}{2}}\cdot (w_L(g)\cdot W_1(g)) &=  P_{\frac{1}{2}}\cdot (w_L(g)\cdot W_1(g)) \cdot P_{\frac{1}{2}}\label{eq:edge_proj_require}\\
    P_{l+\frac{1}{2}}\cdot (w_{l+1}(g)\cdot W_l(g)) &=  P_{l+\frac{1}{2}}\cdot (w_{l+1}(g)\cdot W_l(g)) \cdot P_{l+\frac{1}{2}}\,.\nonumber
\end{align}

Furthermore, $U_M(g)$ are defined to be charge neutral, namely,
\begin{align}
    &0=[U_M(g),n_{_{f,\pa}}]_{\rho(g)}\nonumber\\
    ={}&\pe\cdot \big([w_{l+1}(g),n_{_{f,l+1}}]_{\rho(g)} w_L(g)+ n_{_{D,M}}w_{l+1}(g)w_L(g)\nonumber\\
    &+w_{l+1}(g)[w_{L}(g),n_{_{f,L}}]_{\rho(g)}\big)\cdot W_M(g)\cdot \pe
\end{align}
which in turn put constraint on $w_{L/(l+1)}(g)$, where
\begin{align}
    [w_{L}(g),n_{_{f,L}}]_{\rho(g)}& = n^{(1)}_{\lambda,\frac{1}{2}}(g)\cdot w_L(g)\,\nonumber\\
    [w_{l+1}(g),n_{_{f,l+1}}]_{\rho(g)}&  = n^{(2)}_{\lambda,l+\frac{1}{2}}(g)\cdot w_{l+1}(g)\,,
    \label{eq:u1f_w_fermion_number}
\end{align}
We also provide fermion parity of $w_{L/(l+1)}(g)$ for later reference:
\begin{align}
    \act{F}{w_L(g)} &= (-)^{n^{(1)}_{\lambda,\frac{1}{2}}(g)}w_L(g)\,\nonumber\\
    \act{F}{w_{l+1}(g)}& = (-)^{n^{(2)}_{\lambda,l+\frac{1}{2}}(g)}w_{l+1}(g)\,.
\end{align}

\subsection{Fusion of symmetry defects}
In the following, we will study fusion rules of symmetry defects using FPEPS.
These fusion rules are directly determined by the tensor equations derived in Section~\ref{sec:tensor_equation}.
Interestingly, the fermionic three-cocycle data that characterizes the tensor equations also serves as the quantum anomaly data for the edge theories of TI phases.

Let us rewrite the symmetry fusion equation in Eq.~\eqref{eq:u1f_um_condition} here:
\begin{align}
    &U_M(g_1)K^{s(g_1)}\cdot U_M(g_2)K^{s(g_2)}\label{eq:umg1_umg2_fusion}\\
    ={}&\Omega_a(g_1,g_2)\Omega_b(g_1,g_2)\cdot \exp\left[ \ii\alpha(g_1,g_2)\cdot n_{_{f,\overline{M}}} \right]\cdot\nonumber\\
     {}&U_M(g_1g_2)K^{s(g_1g_2)}\,,\notag
\end{align}
where $ n_{_{f,\overline{M}}} = \sum_{j=L}^{l+1} n_{_{f,j}}$.
Namely, fusion of $g_1$- and $g_2$-defect differ from $g_1g_2$-defect up to local excitations $\Omega_{a/b}$ at ends of $M$.

We try to calculate such fusion process based on the definition of $U_M(g)$ in Eq.~\eqref{eq:um_def}, which gives
\begin{align}
    &U_M(g_1)K^{s(g_1)}\cdot U_M(g_2)K^{s(g_2)}\notag\\
    ={}&\pe \cdot \left[ w_{l+1}(g_1)w_L(g_1)W_M(g_1) \right]\cdot\notag\\
    &\left[w_{l+1}(g_2)w_L(g_2)W_M(g_2)\right]\cdot \pe
    \label{eq:umg1_umg2_mult}
\end{align}
However, the transformation of the above equation into the form of Eq.~\eqref{eq:umg1_umg2_fusion} is not straightforward.
This is because $W_j(g)$ does not possess a determined fermion parity, and thus rearranging $W_j(g)$ tensors generally results in fermion swapping gates.

To avoid these swapping gates, we follow the strategy presented in Section~\ref{subsubsec:tensor_eq_gb_fusion}, where we introduce additional anti-commuting matrices $\beta$'s to the definition of $U_M(g)$:
\begin{align}
    U_M(g) ={}& d_{\beta}^{-1}\pe\cdot \Tr_\beta \Big[w_{_{\beta,l+1}}(g)
    \otimes_f w_{_{\beta,L}}(g)\notag\\
    &\BOF{j=1}{l}W_{\beta,j}(g) \Big] K^{s(g)}\cdot\pe\,,
\end{align}
in which
\begin{align}
    &W_{\beta,j}(g) = I_{\beta,j}(g)\cdot W_j(g)\,,\notag\\
    &w_{_{\beta,L}}(g) = \iota_{\beta,\frac{1}{2}}^{(1)}(g)\cdot w_L(g)\,,\notag\\
    &w_{_{\beta,l+1}}(g) = \iota_{\beta,l+\frac{1}{2}}^{(2)}(g)\cdot w_{l+1}(g)\,.
    \label{}
\end{align}
\begin{align*}
    \adjincludegraphics[valign=c]{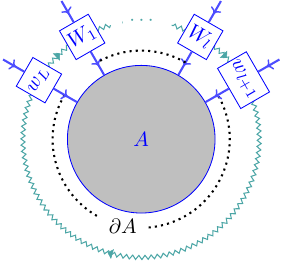}\,.
\end{align*}
With this formalism, fusion of two symmetry defects reads
\begin{align}
    &U_M(g_1)K^{s(g_1)}\cdot U_M(g_2)K^{s(g_2)}\notag\\
    ={}&d_{\beta}^{-1}\pe\cdot \Tr_{\beta}\Big[ \left( w_{_{\beta_1,l+1}}(g_1) \act{K^{s(g_1)}}{w}_{_{\beta_2,l+1}}(g_2) \right)\otimes_f\notag\\
    &\left( w_{_{\beta_1,L}}(g_1) \act{K^{s(g_1)}}{w}_{_{\beta_2,L}}(g_2) \right)\BOF{j=1}{l} \Big( W_{\beta_1,j}(g_1)\cdot\notag\\
&\act{K^{s(g_1)}}{W}_{\beta_2,j}(g_2) \Big) \Big]K^{s(g_1g_2)} \cdot\pe\,.
    \label{eq:u1f_UM_fusion_def}
\end{align}
By Eq.~\eqref{eq:u1f_wg_fusion}, fusion of $W_j(g_1)$ and $W_j(g_2)$ gives additional $D_{(\beta_1,\beta_2)}(g_1,g_2)$, which can be further decomposed to plaquette IGG element $\Lambda$'s, as shown in Eq.~\eqref{eq:u1f_db1b2_decomp}.
With Eq.~\eqref{eq:local_edge_space_nlambda}, such plaquette IGG element can be absorbed by $\pe$, except for terms at two ends.
The derivation to obtain $\Omega_{a/b}$ is quite lengthy, so we provide brief summary here, with detailed calculation presented in Sec.VI of SM\footnotemark[10], 

For simplicity, let us ignore all the $\alpha(g_1,g_2)$'s, $K^{s(g)}$'s, and $\lambda(g_1,g_2)$'s, then

\begin{widetext}
\begin{align}
    w_{_{\beta_1,L}}(g_1) {w}_{_{\beta_2,L}}(g_2)\sim{}& \Lambda^{(1)}_{(\beta_1,\beta_2),\frac{1}{2}}(g_1,g_2) w_L(g_1)w_L(g_2)\notag\\
    w_{_{\beta_1,l+1}}(g_1) {w}_{_{\beta_2,l+1}}(g_2)\sim{}& \Lambda^{(2)}_{(\beta_1,\beta_2),l+\frac{1}{2}}(g_1,g_2)(-)^{\act{g_1}{n}_{\lambda,l+\frac{1}{2}}^{(2)}(g_2)\,{n}_{\lambda,l+\frac{1}{2}}^{(2)}(g_1)} w_{l+1}(g_1)w_{l+1}(g_2)\notag\\
    W_{\beta_1,j}(g_1){W}_{\beta_2,j}(g_2) \sim{}&  D_{(\beta_1,\beta_2),j}(g_1,g_2) W(g_1g_2)=\Lambda^{(2)}_{(\beta_1,\beta_2),j-\frac{1}{2}}(g_1,g_2)\Lambda^{(1)}_{(\beta_1,\beta_2),j+\frac{1}{2}}(g_1,g_2)W_j(g_1g_2)\,.
\end{align}
Eq.~\eqref{eq:u1f_UM_fusion_def} is thus evaluated as
\begin{align}
    U_M(g_1)\cdot U_M(g_2)\sim \pe \cdot & \Tr_{\beta}\Big[\prod_{j=1}^{l+1} \Lambda^{(1)}_{(\beta_1,\beta_2),j-\frac{1}{2}}(g_1,g_2)\Lambda^{(2)}_{(\beta_1,\beta_2),j-\frac{1}{2}}(g_1,g_2)\otimes_f \nonumber\\
    &(w_{l+1}(g_1)w_{l+1}(g_2))\otimes_f (w_L(g_1)w_L(g_2))\BOF{j=1}{l} W_j(g_1g_2) \Big]\cdot \pe\,.
\end{align}
$\Lambda^{(1/2)}$'s form pairs of plaquette IGG, so are absorbed to $\pe$. 
Therefore,
\begin{align}
     U_M(g_1)\cdot U_M(g_2)\sim \pe\cdot (w_{l+1}(g_1)w_{l+1}(g_2)) \cdot (w_L(g_1)w_L(g_2))\cdot(w^{-1}_L(g_1g_2)w^{-1}_{l+1}(g_1g_2))\cdot U_M(g_1g_2)\cdot\pe\,,
\end{align}
which roughly gives
\begin{align}
    \Omega_a(g_1,g_2)\sim&{} \pe\cdot w_{l+1}(g_1)w_{l+1}(g_2)w_{l+1}^{-1}(g_1g_2) \cdot\pe\\
    \Omega_b(g_1,g_2)\sim&{} \pe\cdot w_{L}(g_1)w_{L}(g_2)w_{L}^{-1}(g_1g_2) \cdot\pe
\end{align}
A much more detailed and careful calculation is presented in Sec.VI of SM\footnotemark[10], from which we obtain
\begin{align}
    \Omega_a(g_1,g_2) ={}& \pe \cdot (-)^{\act{g_1}{n}_{\lambda,l+\frac{1}{2}}^{(2)}(g_2)\cdot {n}_{\lambda,l+\frac{1}{2}}^{(2)}(g_1)}\cdot w_{l+1}(g_1) \cdot \act{K^{s(g_1)}}{w}_{l+1}(g_2) \cdot w^{-1}_{l+1}(g_1g_2)\cdot\notag\\
    &\ee^{-\ii\alpha(g_1,g_2)\,n_{_{f,l+1}}}\cdot\lambda^{(1)}_{l+\frac{1}{2}}(g_1,g_2) \cdot \pe\label{eq:def_omega_a}\\
    \Omega_b(g_1,g_2) ={}& \pe \cdot w_L(g_1)\cdot \act{K^{s(g_1)}}{w_L(g_2)}\cdot w^{-1}_L(g_1g_2)\cdot
    \ee^{-\ii\alpha(g_1,g_2)\,n_{_{f,L}}}\,\lambda^{(2)}_{\frac{1}{2}}(g_1,g_2)  \cdot \pe\,,\label{eq:def_omega_b}
\end{align}
\end{widetext}
As shown in Section~\ref{sec:classification}, fermionic three-cocycle data are extracted by calculating fermion number of $\Omega_{a/b}$ and by fusing three symmetry defects. 
We now perform similar calculation with expressions in Eq.~\eqref{eq:def_omega_a} and \eqref{eq:def_omega_b}.

Fermion number calculations are presented in Sec.VI of SM\footnotemark[10], which gives
\begin{align}
    [n_{f;\partial A},\Omega_a(g_1,g_2)] &= -\sigma(g_1,g_2)\Omega_a(g_1,g_2)\,,\notag\\
    [n_{f;\partial A},\Omega_b(g_1,g_2)] &= \sigma(g_1,g_2)\Omega_b(g_1,g_2)\,,
    \label{}
\end{align}
where $\sigma$ is defined in Eq.~\eqref{eq:u1f_n_lam_fusion}.
Thus, we identify $\sigma$ in tensor equations with the fermion number of $\Omega_{a/b}$, which resembles Eq.~\eqref{eq:Omega_ab_fermion_number}.

Fusion of three symmetry defects can also be directly calculated using Eq.~\eqref{eq:def_omega_a} and \eqref{eq:def_omega_b}, where details are presented in Sec.VI of SM\footnotemark[10].
There, we are able to relate $\omega'$ defined in Eq.~\eqref{eq:u1f_lam_omega_2} and \eqref{eq:u1f_lam_omega_1} to $\omega$ appeared in Eq.~\eqref{eq:omegaab_twist_two-cocycle}, which characterizes fusion anomaly of symmetry defects.

As we will show in Section~\ref{subsec:z2_edge}, for the case where $G_f=\ZZ_2\times U_f(1)$ and $G_b=\ZZ_2=\{e,g\}$, fusion two $g-$symmetry flux gives additional single fermions at two ends.

\section{Gauge transformation and Kasteleyn orientation}\label{sec:kasteleyn}
In accordance with our approach outlined in Section~\ref{sec:tensor_equation} and detailed in Sec.I of SM\footnotemark[10], we incorporate symmetries on the physical wavefunction by imposing constraints on the local tensors.
Specifically, the symmetry on the inward/outward leg $(s\alpha)$ is implemented through the action of $W_{(s\alpha)}(g)$/$W^{-1}_{(s\alpha)}(g)$, which results in the identity when contracting the leg $(s\alpha)$.
In our investigation, we encountered a significant aspect worth noting: when the transformation $W(g)$ is not fermion parity even, meaning $\act{F}{W(g)}\neq W(g)$, moving $W_{(s\alpha)}(g)$ and $W^{-1}_{(s\alpha)}$ together generally leads to fermion swapping gates, which disrupts the cancellation between $W$ and $W^{-1}$ when contracting internal legs.

To overcome this challenge, we devised a solution by extracting an oriented graph based on the order of the action of $W(g)/W^{-1}(g)$.
We demonstrated that when this graph results in a \emph{Kasteleyn orientation} \cite{cimasoni2007dimers,WareSonChengMishmashAliceaBauer2016,TarantinoFidkowski2016,cimasoni2007dimers,EllisonFidkowski2019}, the FPEPS wavefunction exhibits symmetry.
In this section, we first illustrate the concept of Kasteleyn orientation using a 1D example, and then extend it to higher-dimensional cases.

\subsection{Warm up: a 1D example}\label{subsec:kasteleyn_1d}
We start from an FMPS wavefunction
\begin{equation}
    \ket{\psi}=\fTr \left[ \mathbb{B}\otimes_f\mathbb{T} \right]\,,
    \label{eq:fmps_wf}
\end{equation}
where site tensors $\mathbb{T}\equiv\otimes_{f,j} \hat{T}_j$, and bond tensors $\mathbb{B}\equiv\otimes_{f,j}\hat{B}_{j,j+1}$. 
Each site tensors have one physical leg $j$ and two incoming internal legs $(2j-1)$ and $(2j)$.
We consider systems with size $L$ on closed boundary, where leg $(1)$ and $(2L+1)$ are identified. 
Note that all local tensors are parity even:
\begin{align}
    \hat{T}_j&=F_j\otimes_f F_{(2j-1)}\otimes_f F_{(2j)}\cdot \hat{T}_j\,,\notag\\
    \hat{B}_{j,j+1}&=\hat{B}_{j,j+1}\cdot F_{(2j+1)}\otimes_f F_{(2j+1)}\,.
    \label{eq:fmps_parity}
\end{align}
We assume that $\ket{\psi}$ hosts onsite unitary symmetry $g$: 
\begin{align}
    \ket{\psi}=g\circ\ket{\psi}\equiv\bigotimes_j U_j(g)\cdot \ket{\psi}
    \label{eq:sym_wf}
\end{align}
which acts as gauge transformation $W(g)$ on internal legs.
\begin{align}
    \hat{T}_j&=U_j(g)\otimes_f W_{(2j-1)}(g)\otimes_f W_{(2j)}(g)\cdot \hat{T}_{j}\,,\notag\\
    \hat{B}_{j,j+1}&=\hat{B}_{j,j+1}\cdot W^{-1}_{(2j+1)}(g)\otimes_f W^{-1}_{(2j)}(g)\,,\label{eq:fmps_sym_cond}
\end{align}
While $U(g)$'s are parity even, $W(g)$ may be parity odd, which happens for certain FSPT phases:
\begin{align}
    \act{F}{W(g)}=-W(g)
    \label{eq:wg_parity_odd}
\end{align}
Note that by comparing Eq.~\eqref{eq:fmps_parity} and Eq.~\eqref{eq:fmps_sym_cond}, $W(g)$'s on different legs must share the same parity quantum number.

Naively, Eq.~\eqref{eq:fmps_sym_cond} implies symmetric physical wavefunction, as $W(g)$ and $W^{-1}(g)$ cancels when contracting internal legs.
However, due to the odd parity of $W(g)$, moving $W(g)$ and $W^{-1}(g)$ together may lead to extra $-$ sign from swapping fermions, and thus Eq.~\eqref{eq:sym_wf} holds up to $\pm$ sign.
Note that in such case, the following different order of $W(g)$'s action on local tensors leads to different symmetric condition:
\begin{align}
    \hat{T}_j&=U_j(g)\otimes_f W_{(2j)}(g)\otimes_f W_{(2j-1)}(g)\cdot \hat{T}_{j}\notag\\
    &=U_j(g)\otimes_f \act{F}{W_{(2j-1)}(g)}\otimes_f W_{(2j)}(g)\cdot \hat{T}_{j}\,,\notag\\
    \hat{B}_{j,j+1}&=\hat{B}_{j,j+1}\cdot W^{-1}_{(2j)}(g)\otimes_f W^{-1}_{(2j+1)}(g)\notag\\
    &=\hat{B}_{j,j+1}\cdot W^{-1}_{(2j+1)}(g)\otimes_f \act{F}{W^{-1}_{(2j)}(g)}\,,
    \label{eq:fmps_sym_cond_II}
\end{align}

\begin{figure}[h!]
    \centering
    \includegraphics[width=0.48\textwidth]{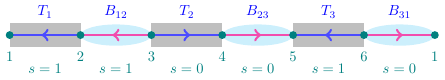}
    \caption{An example of 1D Kasteleyn orientation with $L=3$.
        Site tensors $\hat{T}_j$ are shaded grey, and physical indices are ignored.
        Bond tensors are drawn as cyan eclipses. 
        Node numbers $k$ from 1 to 6 are labelling the indices of $W_{(k)}(g)$'s.
        Symmetry constraints on local tensors can be read from direction of arrows in the figure.
    }
    \label{fig:1d_example}
\end{figure}

In the following, given an arbitrary choice of symmetry conditions on local tensors, we work out the $\pm$ sign.
A representative choice is shown in Fig.~(\ref{fig:1d_example}).
Symmetry constraints are encoded in arrow orientations in the figure. 
For example, if the blue arrow on site tensor $\hat{T}_j$ points from leg $(2j-1)$ to $(2j)$, $\hat{T}_j$ satisfies Eq.~\eqref{eq:fmps_sym_cond}. 
As for bond tensor $B_{j,j+1}$, if the magenta arrow points from leg $(2j+1)$ to $(2j)$, it satisfies bond tensor constraints in Eq.~\eqref{eq:fmps_sym_cond_II}.

Given a specific orientation for a 1D system with size $L$ and closed boundary, as shown in Fig.~(\ref{fig:1d_example}), such orientation can be represented by function $s$, where $s(k,k+1)=0/1$ if the arrow points from $(k)/(k+1)$ to $(k+1)/(k)$.
Symmetry constraint on local tensors then reads 
\begin{widetext}
\begin{align}
    U_j(g)\cdot \hat{T}_j  = (-1)^{s(2j-1,2j)}\cdot \left( W_{(2j-1)}(g)\otimes_f W_{(2j)}(g) \right)^{-1}\cdot \hat{T}_j\,,\quad
    \hat{B}_{j,j+1} = (-1)^{s(2j,2j+1)}\cdot \hat{B}_{j,j+1}\cdot W_{(2j)}\otimes_f W_{(2j+1)}
    \label{eq:T_B_trans_1d}
\end{align}
We now calculate $g\circ\ket{\psi}$ with Eq.~\eqref{eq:T_B_trans_1d}:
    \begin{align}
        g\circ\ket{\psi}&= (-1)^{\sum_k s(k,k+1)}\cdot \fTr\left[ \mathbb{B}\cdot \left( \BOF{j=1}{L}W_{(2j)}(g)\otimes_f W_{(2j+1)}(g) \right)\otimes_f \left( \BOF{j=1}{L}W_{(2j-1)}(g)\otimes_f W_{(2j)}(g)\right)^{-1} \cdot \mathbb{T} \right]\notag\\
        &=(-1)^{\left[\sum_k s(k,k+1)\right]+1}\cdot \fTr[\mathbb{B}\otimes_f\mathbb{T}]
        =(-1)^{\left[\sum_k s(k,k+1)\right]+1}\cdot\ket{\psi}
    \end{align}
\end{widetext}
Based on the above equation, we now introduce \emph{Kasteleyn orientation}.
In any dimension, an oriented graph satisfies Kasteleyn orientation if, for any enclosed loop within the graph, the count of counter-clockwise oriented edges encircling it is \emph{odd}.
For the 1D scenario, with just a single loop in consideration, the Kasteleyn orientation condition is equivalent to $\sum_k s(k,k+1)=1 \pmod 2$.
It has been formally established that Kasteleyn orientation exists for any planar graph possessing an even number of vertices\cite{cimasoni2007dimers}.
However, the selection of Kasteleyn orientation is not unique: by reversing the arrows on edges connected to a specific vertex $v$, one can derive an alternative orientation.
This choice extends to every vertex, allowing the option to flip or not flip arrows on edges connecting to that vertex, generating a total of $2^{N_v}$ potential Kasteleyn orientations, with $N_v$ representing the total vertex count.

In the context of our 1D example, if the order of $W(g)$'s action conforms to the Kasteleyn orientation, the condition expressed in Eq.~\eqref{eq:sym_wf} is fulfilled.
However, if this order fails to align with Kasteleyn orientation, an additional $-$ sign is introduced.

Before discussing higher dimension,  let us first turn our attention to a FMPS with open boundary conditions.
To facilitate our subsequent analysis, we introduce an additional bond tensor $\hat{B}_{0,1}$ with outgoing internal legs $(0)$ and $(1)$.
Upon contracting all internal legs within the bulk, we arrive at a large tensor $\hat{T}_A$, which resides on the bulk physical legs ${1,\dots,L}$ and the boundary legs $(0)$ and $(2L+1)$.
Symmetry transformations applied to $\hat{T}_A$ induce gauge transformations on the boundary legs, yielding:
\begin{widetext}
    \begin{align}
        \BOF{j=1}{L} U_j(g)\cdot \hat{T}_A&= (-1)^{\sum_{k=0}^{2L} s(k,k+1)}\cdot \fTr\left[ \mathbb{B}\cdot \left( \BOF{j=0}{L}W_{(2j)}(g)\otimes_f W_{(2j+1)}(g) \right)\otimes_f \left( \BOF{j=1}{L}W_{(2j-1)}(g)\otimes_f W_{(2j)}(g)\right)^{-1} \cdot \mathbb{T} \right]\notag\\
        &=(-1)^{\sum_k s(k,k+1)}\cdot \hat{T}_A\cdot W_{(0)}(g)\otimes_f W_{(2L+1)}(g)
        \label{eq:kasteleyn_g_on_open_chain}
    \end{align}
\end{widetext}
The sequence of $g$-action on the boundary legs is determined using the following method.
An additional virtual oriented line is introduced, connecting leg $(2L+1)$ to $(0)$, with its orientation chosen in such a way that this closed chain adheres to Kasteleyn orientation.
The order of $W_{(0)}(g)$ and $W_{(2L+1)}(g)$ is then aligned with the direction of this newly introduced virtual line.
This rule is visually depicted in the figure below, where the violet dashed line indicates the order of the symmetry action on the boundary.
\begin{align}
    \adjincludegraphics[width=8.5cm,valign=c]{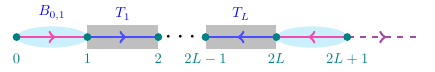},
    \label{eq:boundary_action_1d}
\end{align}

\subsection{Kasteleyn orientations on higher dimensions}\label{subsec:kasteleyn_high_dim}
In this subsection, we move forward to Kasteleyn orientations in higher dimensions.
For a directed graph with Kasteleyn orientations, any loop in such graph has odd number of counter-clockwise edges. 
For topological insulators, symmetry transformation on internal legs may not have fixed parity, leading to fermion swapping gates when internal legs are contracted.
In the following, we define rules to extract an orientation from orders of symmetry actions on internal legs.
We claim that if such orientation is Kasteleyn, swapping gates are avoided when contracting tensors, leading to a symmetric wavefunction.
Details of the prove is left to Sec.VII of SM\footnotemark[10].

For simplicity, we consider FPEPS defined on a honeycomb lattice.
For $g\in G_b$, it acts as $W_{(s\alpha)}(g)$ on internal leg $(s\alpha)$. 
We focus on the case where $W_{(s\alpha)}(g)$ do not have fixed parity, namely, $n_D(g)$ in Eq.~\eqref{eq:u1f_g_act_nf} is not a $c$-number.
We now define rules to extract orientations from action of $W(g)$'s on local tensors.
We first expand all sites of the honeycomb lattice to triangles as
\begin{widetext}
\begin{align}
    \adjincludegraphics[valign=c]{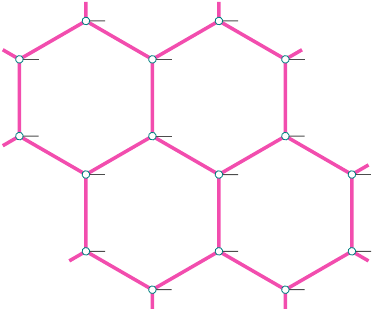}\rightarrow\adjincludegraphics[valign=c]{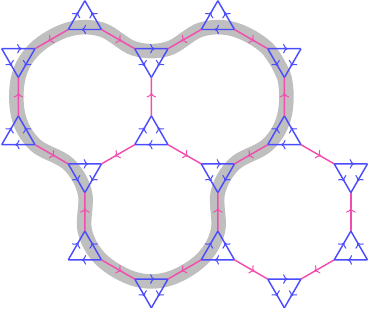}
    \label{eq:honeycomb_to_fisher_kasteleyn}
\end{align}
\end{widetext}
Vertices in the new lattice live on internal legs of the honeycomb FPEPS, and are labeled as $(s\alpha)$.
Arrows on blue triangles in Eq.~(\ref{eq:honeycomb_to_fisher_kasteleyn}) are extracted from $g$-action on site tensors, while arrows on magenta lines of the original hexagons from $g$-action on bond tensors.
More specifically, from $g$-symmetric condition on tensor at site $s$
\begin{align}
    &U_s(g)K^{s(g)}\cdot\hat{T}_s \notag\\
    ={}& \left[ W_{(s\alpha)}(g)\otimes_f W_{(s\beta)}(g) \otimes_f W_{(s\gamma)}(g) \right]^{-1} \cdot \hat{T}_s  \,,
    \label{eq:site_g_action}
\end{align}
the arrows on the corresponding blue triangle is $(s\alpha)\rightarrow(s\beta)$, $(s\beta)\rightarrow(s\gamma)$, and $(s\alpha)\rightarrow(s\gamma)$.
For example,
    \begin{align}
    U_s(g)\cdot \hat{T}^*_s &= \left[ W_{(sx)}(g)\otimes_f W_{(sy)}(g) \otimes_f W_{(sz)}(g) \right]^{-1} \cdot \hat{T}_s \notag\\ 
    &\Rightarrow \adjincludegraphics[valign=c,scale=2]{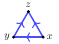}\,.
    \label{eq:site_arrow_rule}
    \end{align}
Similarly, if $g$-action on bond tensor $\hat{B}_{ss'}$ reads
\begin{align}
    \hat{B}_{ss'}\cdot K^{s(g)} 
    =  \hat{B}_{ss'} \cdot  W_{(s\alpha)}(g) \otimes_f  W_{(s'\alpha)}(g)\,,
    \label{eq:bond_g_action}
\end{align}
the arrow points from $(s\alpha)$ to $(s'\alpha)$.
Following this rule, $g$-symmetry constraint on all local tensors leads to the oriented graph as in Eq.~(\ref{eq:honeycomb_to_fisher_kasteleyn}).
In Sec.VII of SM\footnotemark[10], we show that the physical wavefunction is $g$-symmetric only when the right hand side of Eq.~(\ref{eq:honeycomb_to_fisher_kasteleyn}) is a Kasteleyn orientation.

\subsection{Symmetry on the open boundary from Kasteleyn orientation}

\begin{figure}
    \centering
    \includegraphics[width=0.5\textwidth]{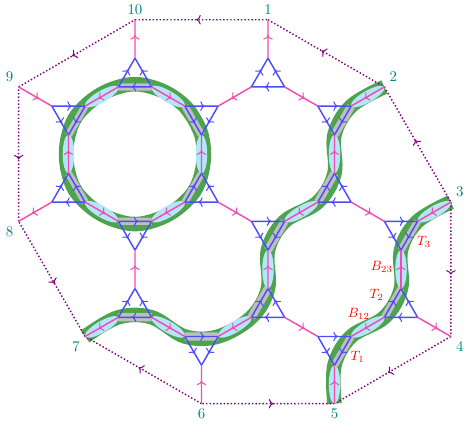}
    \caption{A Kasteleyn oriented honeycomb lattice with open boundary.
        Dark green shading indicates locations of $P_o(g)$'s, and gray rectangles and cyan eclipses denote action of $W_o(g)$'s, which are parity odd.
        Parity even operators $W_e(g)$'s act at nodes without shading. 
        Arrows on violet dashed lines around edge follow rules of Kasteleyn orientations: every loop, including these violet dashed lines, contains odd number of counter-clockwise orientations.
    }
    \label{fig:kasteleyn_domain_wall}
\end{figure}

We now consider a finite 2D FPEPS on region $A$, which is cut from an infinite FPEPS, as shown in Fig.~\ref{fig:kasteleyn_domain_wall}.
By contracting all internal legs within $A$,  we obtain a large tensor $\hat{T}_A$, which is a linear map from boundary virtual legs to bulk physical legs:
\begin{align}
    \hat{T}_A = \fTr_{A} [\mathbb{B}\otimes_f \mathbb{T}]\,,
\end{align}
Here, boundary legs, labeled as $j\in\left\{ 1,2,\dots,L \right\}$, are ordered clockwise, as shown in Fig.~(\ref{fig:kasteleyn_domain_wall}).

As shown in Eq.~\eqref{eq:sym_on_ta}, one expects that physical symmetry action $U_A(g)K^{s(g)}$ on $\hat{T}_A$ can be pushed to $W_{\partial A}(g)$ action on boundary legs.
To determine $W_{\partial A}(g)$ for the case where $W(g)$'s do not have fixed parity, we first connect neighbouring boundary legs with dashed violet lines in Fig.~\ref{fig:kasteleyn_domain_wall}, whose direction are chosen such that the whole graph gives Kasteleyn orientation.  
$W_{\partial A}$ then can be expressed as
\begin{align}
    W_{\partial A}(g) &= \BOF{j=1}{L}\overline{W}_j(g)\,,
    \label{eq:u1f_sym_edge_w} \\
    \text{with }\overline{W}_j(g) &\equiv F_j^{\xi(j)}\circ {W_j(g)} = [D_j(g)]^{\xi(j)}\cdot{W}_j(g)\,.\notag
\end{align}
where $\xi(j)$ counts the number of dashed violet lines with counter-clockwise arrows from leg $1$ to $j$.
For example, for the system in Fig.~\ref{fig:kasteleyn_domain_wall}, $\xi(j)$ are listed as Table~\ref{table:coeff}.
\begin{table}[!ht]
\centering
 \begin{tabular}{| c | c | c | c | c | } 
 \hline
    $\xi(1)=0$ & $\xi(2)=1$ & $\xi(3)=2$ & $\xi(4)=2$ & $\xi(5)=2$ \\
 \hline
    $\xi(6)=3$ & $\xi(7)=3$ & $\xi(8)=4$ & $\xi(9)=5$ & $\xi(10)=6$\\
\hline
\end{tabular}\,.
\caption{The value of $\xi(j)$ for the system in Fig.~\ref{fig:kasteleyn_domain_wall}}
\label{table:coeff}
\end{table}
Details of proof for the above statement are left in Sec.VII of SM\footnotemark[10].

\section{Examples for $\ZZ_2^{Ising}\times U_f(1)$ FSPT case}\label{sec:solution}
In this section, we reinvestigate the TI phase presented in Section~\ref{sec:example}, where $G_f=\ZZ_2^g\times U_f(1)$, and $\ZZ_2^g=\{e,g\}$.
We will show that this TI phase are characterized by non-trivial fermionic three cocycle data, and thus indeed fit in our general scheme of interacting topological insulators.
By solving symmetry constraints on local tensors, we obtain variational tensor wavefunctions for this phase for spin-1/2 fermionic systems on a square lattice.
Furthermore, based on symmetry transformation of internal legs~(Eq.~\eqref{eq:ti_u1f_charge_internal} and \eqref{eq:ti_z2_internal}) and methods developed in Section~\ref{sec:edge_theory}, we construct edge theories for this phase, which shows quantum anomaly behaviour.

\subsection{Local Tensor Solution}
Here, we consider spin-$\frac{1}{2}$ fermions on square lattice, with local basis $\{\ket{0},~c_{\uparrow}^\dg\ket{0},~c_{\downarrow}^\dg\ket{0},~c_\uparrow^\dg c_\downarrow^\dg\ket{0}\}$.
We use the bipartite feature of square lattice, where one unit cell contains two sites with sublattice indices $u$ and $v$.
As we consider charge neutral system in this work, we put holes on $u$ and electrons on $v$.
Namely,
\begin{align}
    [n_{f,u},~c_{u\sigma}^\dg]=c_{u\sigma}^\dg~,\quad
    [n_{f,v},~c_{v\sigma}^\dg]=-c_{v\sigma}^\dg~.
    \label{}
\end{align}
$g$ acts as $\sigma^x$ on spins:
\begin{align}
    U(g)\cdot c_\uparrow\cdot U(g)=c_\downarrow
\end{align}

With translational symmetry in this bipartite lattice, there are two kinds of site tensors $\hat{T}^u$ and $\hat{T}^v$, and four types of bond tensors $\hat{B}^{a}$ with $a\in\{1,2,3,4\}$.
Graphically, a site tensor reads
\begin{align*}
    \hspace{-0.1cm}\adjincludegraphics[valign=c]{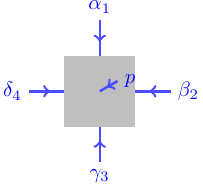}=T^s_{\alpha\beta\gamma\delta,p} \vket{\alpha}_1|\beta)_2|\gamma)_3|\delta)_4\ket{p}=\hat{T}^s\,,
\end{align*}
where we put physical leg at the last position.
Bond tensors are represented as
\begin{align*}
    \hat{B}^a=B_{\alpha\beta}^a\vbra{\alpha}_a \vbra{\beta}_{a+2} ={}& \adjincludegraphics[valign=c]{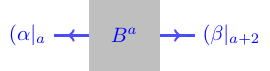}\,,
\end{align*}
where in $B_{\alpha\beta}$ the first index and second index are the legs that connect two adjacent $u-$ and $v-$site respectively. $a+2$ takes value module 4.
An internal leg is set to be a four dimensional Hilbert space, where basis states are labeled as $\{|ee),~|eg),~|ge),~|gg)\}$.
An internal leg can then be written as tensor product of two two-dimensional Hilbert space, and thus site tensors and bond tensors can be expressed as
\begin{align*}
    \adjincludegraphics[valign=c]{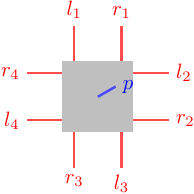}~;\quad
    \hspace{0.48cm}\adjincludegraphics[valign=c]{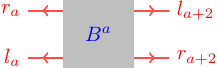}\,.
\end{align*}

We now give a nontrivial fermionic three cocycle $(\sigma,\omega)$, where $\sigma(g,g)=1$ and $\omega(g,g,g)=\ii$.
Other $\sigma$'s are set to be $0$, and $\omega$'s are set to be $1$.
Equipped with such cocycle, solutions for $n_f$, $W(g)$, $n_\lambda$ can be read from Sec.VI of SM\footnotemark[10].
From Eq.~(S64) and Eq.~(S80) in the SM\footnotemark[10], fermion charge number $n_f$'s are
\begin{align}
    n_{f,(ua)}=-\vket{eg}\vbra{eg}~,\quad
    n_{f,(va)}=\vket{ge}\vbra{ge}~.
    \label{}
\end{align}
According to Eq.~(S65) and Eq.~(S81) in the SM\footnotemark[10], $W(g)$'s are expressed as
\begin{align}
    W_{(ua)}(g)&=\vket{gg}\vbra{ee}+\ii\vket{ge}\vbra{eg}+\vket{eg}\vbra{ge}+\vket{ee}\vbra{gg}\notag\\
    W_{(va)}(g)&=\vket{gg}\vbra{ee}+\vket{ge}\vbra{eg}+\ii\vket{eg}\vbra{ge}+\vket{ee}\vbra{gg}
\end{align}
The plaquette IGG's (IGA's) are also given by two-cocycle data as
From Eq.~(S67) and Eq.~(S83) in the SM\footnotemark[10], we obtain
\begin{align}
    n_{\lambda,(sa)}^{(1)}(g)&=-\vket{ee}\vbra{ee}-\vket{ge}\vbra{ge}~,\nonumber\\
    n_{\lambda,(sa)}^{(2)}(g)&=\vket{ee}\vbra{ee}+\vket{eg}\vbra{eg}~\,,
\end{align}
which is independent to $(sa)$.
The relation between $n_{\lambda}^{(a)}(g)$'s in Eq.~\eqref{eq:u1f_n_lam_fusion} can be easily verified, where
\begin{align}
    \act{g}{n}_{\lambda}^{(1/2)}(g) = - {n}_{\lambda}^{(1/2)}(g) \mp 1\,.
\end{align}
$n_{_D}(g)=n_{\lambda}^{(1)}(g)+n_{\lambda}^{(2)}(g)$, so
\begin{align}
    (-)^{n_{_D}(g)} = \vket{ee}\vbra{ee}-\vket{eg}\vbra{eg}-\vket{ge}\vbra{ge}+\vket{gg}\vbra{gg}\,.
\end{align}
With this expression, we can decompose $W(g)$ to its fermion parity even ($E(g)$) and odd ($O(g)$) part:
\begin{align}
    E_{(sa)}(g) &= \frac{1+(-)^{n_{_D}}}{2} W_{(sa)}(g)\,;\nonumber\\
    O_{(sa)}(g) &= \frac{1-(-)^{n_{_D}}}{2} W_{(sa)}(g)\nonumber\,.
\end{align}

We now impose symmetry and IGG constraint on site and bond tensors.
The charge neutral condition for site tensors is
\begin{align}
    \left( n_{f,s}+\sum_{a=1}^4 n_{f,(sa)} \right)\cdot \hat{T}^s=0
    \label{}
\end{align}
For $\ZZ_2^{Ising}$ symmetry, we have
\begin{align}
    U_s(g)\cdot \bigotimes_{a=1}^4\subf W_{(sa)}(g) \cdot \hat{T}^s = \hat{T}^s.
\end{align}
Note that the above solution cannot be directly implemented, as entries of tensors contain fermion mode.
To take care of the fermion swapping signs, Jordan-Wigner transformation is implemented, where the Jordan-Wigner string $J$ is 
\begin{align}
    J_{(sa)} = \exp \left\{\ii \pi n_{f,(sa)}  \right\}\,,
\end{align}
The ``bosonized'' $W(g)$'s are
\begin{align}
    \act{b}{W}_{(sa)}(g) = E_{(sa)}(g) +  \bigotimes_{j=1}^{a-1}J_{(sj)}\otimes O_{(sa)}(g)\,.
\end{align}
Now we can define the symmetry operations
\begin{align}
    \mathbb{G}_{Ising}^s ={}& U(g)\cdot \prod_{a=1}^4 \act{b}{W}_{(sa)}(g) \\
    \mathbb{N}^s ={}&n_{f,s}+\sum_{a=1}^{4}n_{f,(sa)}\,,
\end{align}
and $\hat{T}$ should satisfies $\ZZ_2$ and $U_f(1)$ symmetry conditions:
\begin{align}
    \mathbb{G}_{Ising}^s \cdot \hat{T}^s = \hat{T}^s;~\mathbb{N}^s \cdot \hat{T}^s &= 0\,.
    \label{eq:teq_gf}
\end{align}
Furthermore, plaquette $\IGA$ impose more constraints on $\hat{T}$ 
\begin{equation}
    \left(n^{(1)}_{\lambda,(sa)}(g) + n^{(2)}_{\lambda,(sa+1)}(g) \right)\cdot \hat{T}^s = 0\,.
    \label{eq:teq_iga}
\end{equation}
Here, plaquette $\IGG$ simply identifies Ising spins within one plaquette, which is represented graphically as
\begin{equation*}
    \adjincludegraphics[valign=c]{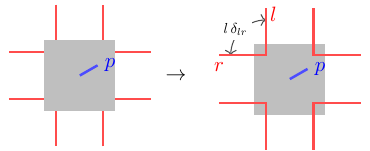}\,.
\end{equation*}

The bond tensor can be treated in the similar way, where the plaquette IGG gives 
\begin{align}
    \hspace{-0.2cm}\left(n_{\lambda,(ua)}^{(1)}+n_{\lambda,(va+2)}^{(2)}\right)\cdot \hat{B}^a = \left(n_{\lambda,(ua)}^{(2)}+n_{\lambda,(va+2)}^{(1)}\right)\cdot \hat{B}^a=0\label{eq:bond_condition_igg}\,,
\end{align}
which identifies the legs belong to the same plaquette:
\begin{align*}
    \adjincludegraphics[valign=c]{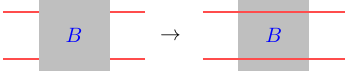}\,.
\end{align*}
Symmetry constraints on bond tensors give
\begin{align}
    \hat{B}^a\cdot W_{(ua)}(g)\otimes_f W_{(va+2)}(g) ={}& \hat{B}^{a}\label{eq:bond_condition_sym}\\
    \hat{B}^a \cdot (n_{f,(ua)}+n_{f,(va+2)}) ={}& 0\label{eq:bond_condition_charge}\,.
\end{align}

The dimension of site tensor Hilbert space is 1024, and we have 14 linear independent solutions that satisfy Eq.~\eqref{eq:teq_gf} and \eqref{eq:teq_iga} for $\hat{T}^u$ (magenta color indicates the presence of fermion on the internal leg):
\begin{equation*}
    \adjincludegraphics[valign=c]{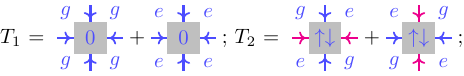}
\end{equation*}
\begin{equation*}
    \adjincludegraphics[valign=c]{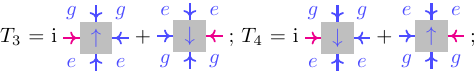}
\end{equation*}
\begin{equation*}
    \adjincludegraphics[valign=c]{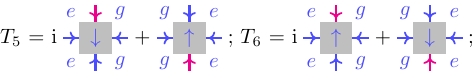}
\end{equation*}
\begin{equation*}
    \adjincludegraphics[valign=c]{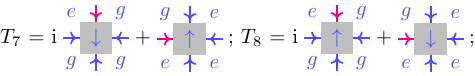}
\end{equation*}
\begin{equation*}
    \adjincludegraphics[valign=c]{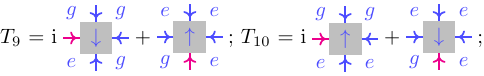}
\end{equation*}
\begin{equation*}
    \adjincludegraphics[valign=c]{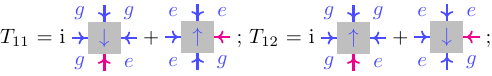}
\end{equation*}
\begin{equation*}
    \adjincludegraphics[valign=c]{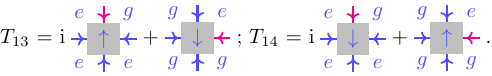}
\end{equation*}
A valid local tensor can then be expressed as
\begin{equation}
    \hat{T}^u = \sum_{j=1}^{14} c_{j}^u T_j\,,
\end{equation}
with $c_j$ being an arbitrary complex number. 
The basis for site tensor $\hat{T}^v$ is obtained by exchanging $e \leftrightarrow g$ in the basis above.
The solution for bond tensor $B_{uv}$ that satisfies Eq.~\eqref{eq:bond_condition_igg}, \eqref{eq:bond_condition_sym}, and \eqref{eq:bond_condition_charge} is given as
\begin{align}
    \hat{B}^a ={}& \vbra{gg}_{a}\vbra{gg}_{a+2}+ \vbra{ee}_a\vbra{ee}_{a+2} \notag\\
    &+ \vbra{eg}_a \vbra{ge}_{a+2} + \vbra{ge}_a\vbra{eg}_{a+2}
    \label{}
\end{align}
where we use maximal entangled bond states here.

In the context of other target symmetry groups, the methodology for solving the tensor equation remains consistent with the approach described above.
This procedure involves feeding the machinery with a particular set of representations of the (twisted) group cohomology data under consideration.
Subsequently, symmetry and IGG on internal legs, such as $W(g)$, $n_{\lambda}(g)$, and $n_f$ possess feasible solutions for any given cohomology data.
More comprehensive details regarding these solutions are elaborated in Sec.VI of SM\footnotemark[10].

\subsection{Edge theory}\label{subsec:z2_edge}
In this subsection, we delve into the derivation of the edge theory for the considered topological phase, following the approach outlined in Section~\ref{sec:edge_theory}.
To initiate this analysis, we isolate a region $A$ from an infinite FPEPS structure.
Subsequently, we perform the contraction of all internal legs confined within $A$, resulting in the formation of a large tensor $\hat{T}_A$.
This tensor serves as a mapping from the boundary's virtual legs $\HH_{\partial A}$ to the bulk's physical legs $\HH_A$, where the legs of $\partial A$ are labeled by ${1,\dots,L}$.
Specifically, we designate a subregion $M=\{1,\dots,l \}$ of $\partial A$ that encompasses the first $l$ sites. The associated Hilbert space is graphically represented as follows:
\begin{align*}
    \adjincludegraphics[scale=1,valign=c]{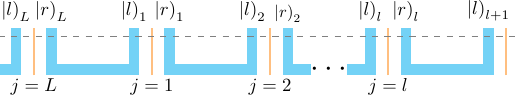}\,,
\end{align*}
Here, spins within the same plaquette are identified by $\pe$: $|r)_n = |l)_{n+1}$

We now create symmetry defects at two ends of $M$ by
\begin{align*}
    U_M(g) = \pe \cdot w_{l+1}(g)w_L(g) \BOF{j=1}{l}W_j(g)\cdot \pe\,.
\end{align*}
$U_M(g)$ are required to be charge neutral, which impose constraints on $w_L(g)$ and $w_{l+1}(g)$ as
\begin{align}
    [w_L,n_f] ={}& n_{\lambda,L}^{(1)}\cdot w_L\,;\notag\\
    [w_{l+1},n_f] ={}& n_{\lambda,l+1}^{(2)}\cdot w_{l+1}\,.
\end{align}
(see also Eq.~\eqref{eq:u1f_w_fermion_number}).
Following general procedure presented in Section~\ref{subsec:sym_defect_edge}, we can set
\begin{align}
w_L(g) ={}&\sum_{l} c_L |le)_L(lg|_L + |lg)_L(le|_L\,;\\
w_{l+1}(g) ={}& \sum_{r} c^{\dagger}_{l+1} |er)_{l+1}(gr|_{l+1} + |gr)_{l+1}(er|_{l+1}\,,\notag
\end{align}
where $L$ and $l+1$ are chosen as $v-$sites.

From group relation $g^2=e$, one expects that by fusing two $g$-defects, one at most gets local excitations $\Omega_{a/b}$, where $a,b$ labels ends of $M$:
\begin{equation}
    U_M(g)\cdot U_M(g) = \pe\cdot\Omega_a(g,g)\cdot\Omega_b(g,g)\cdot \pe\,.
\end{equation}
From Eq.~\eqref{eq:def_omega_b}, $\Omega_b$ simplifies to:
\begin{align}
    \Omega_b(g,g) ={}& \pe \cdot w_L(g)\cdot w_L(g)\lambda^{(2)}_{\frac{1}{2}}(g,g)  \cdot\pe\,.\nonumber\\
    ={}&c_L ~\lambda^{(2)}_{\frac{1}{2}}(g,g)\,, 
\end{align}
which contains fermion annihilation operator, and thus gives $\sigma(g,g)=1$.
From Eq.~\eqref{eq:omegaab_twist_two-cocycle}, the three-cocycle data $\omega(g,g,g)$  can be extracted by 
\begin{align}
    \omega(g,g,g)\cdot \Omega_b(g,g)  = \act{U_M(g)}{\Omega}_b(g,g)\,.
    \label{eq:omegab_omegaggg}
\end{align}

According to Eq.~(S68) in the SM\footnotemark[10], a consistent choice of $\lambda^{(2)}_{\frac{1}{2}}(g,g)$ reads $\lambda^{(2)}_{\frac{1}{2}}(g,g) = \exp \left[ \ii\frac{\pi}{2}n^{(2)}_{\lambda,1}(g) \right]$.
Inserting this expression to Eq.~\eqref{eq:omegab_omegaggg}, one obtains
\begin{align}
    &\act{U_M(g)}{\Omega}_b(g)\nonumber\\
     ={}& c_L\left((-)^{n_{\lambda,1}^{(2)}(g)}\right)\left( \act{g}{\lambda}^{(2)}_{\frac{1}{2}}(g,g)\right)=\ii\Omega_b(g)
\end{align}
where the sign factor $(-)^{n_{\lambda,1}^{(2)}(g)}$ comes from the swapping between $W_M$ and $c_L$.
So, one conclude $\omega(g,g,g)=\ii$.

\section{Discussion}\label{sec:discussion}
In conclusion, we have presented a comprehensive classification scheme for topological insulators based on tensor equations that capture symmetry transformation rules on internal legs.
By extracting quantum anomaly data from these equations, we have characterized the edge Hilbert spaces and symmetry actions associated with each topological phase.
Furthermore, by imposing symmetry constraints on local tensors, we have obtained variational wavefunctions that go beyond fixed-point solutions, providing a broader range of possibilities for simulating and understanding interacting topological phases.

Our work opens up several promising directions in the field of topological phases and fermionic tensor network states.
One natural extension is to incorporate space group symmetries to study higher-order topological insulator (HOTI) phases\cite{schindler2018higher,RasmussenLu2018classification,SongFangQi2018,ElseThorngren2018crystalline,park2019higher,chen2020higher}.
The lattice-based nature of FPEPS makes them well-suited for exploring the interplay between space and onsite symmetries.
Investigating whether Kasteleyn orientations impose additional constraints on the classification and characterization of HOTI phases would be particularly intriguing.
Additionally, the development of variational tensor wavefunctions for two-dimensional topological superconductors presents another avenue for future research\cite{TarantinoFidkowski2016,WareSonChengMishmashAliceaBauer2016,WangGu2018,WangGu2020construction}.
Exploring the possibility of decorating domain walls with Kitaev chains introduces new types of tensor equations and edge theories, offering novel insights into the properties of these systems.

An important and challenging direction is the realization of anomalous edge theories for topological insulator phases in lower-dimensional systems.
While bosonic SPT phases can be realized in systems with extended onsite symmetry groups, known as intrinsic gapless SPT (igSPT) phases\cite{thorngren2021intrinsically,wen2023bulk}, achieving the realization of edge theories for fermionic symmetry-protected topological (FSPT) phases with onsite symmetries and tensor products of local Hilbert spaces remains a significant challenge\cite{JonesMetlitski2021one,Metlitski20191d}.
The tensor equations derived in our work provide valuable insights for realizing such edge theories for two-dimensional topological insulator phases.

Expanding our framework to higher dimensions represents a captivating direction for future research.
Investigating tensor equations for three-dimensional FSPT phases could unveil new phenomena and enrich our understanding of higher-dimensional topological phases.
Of particular interest is the investigation of edge theories for three-dimensional topological insulator phases.
The recent connection between the half-filled Landau level problem\cite{willett1990anomalous,halperin1993theory}, a renowned platform for studying non-Fermi liquid behavior, and the boundary of a three-dimensional time-reversal symmetric topological insulator with time-reversal acting as a particle-hole symmetry has opened up exciting possibilities\cite{lee1998neutral,son2015composite,barkeshli2015particle,wang2015dual,metlitskivishwanath2016particle,wang2016half}.
However, it remains an open question whether such systems can be effectively realized using onsite symmetries and tensor product structures of local Hilbert spaces in two dimensions.
Deriving tensor equations for these three-dimensional topological insulator phases holds great promise, as it could provide a framework for simulating their boundary theory.
This, in turn, opens up the possibility of simulating intriguing phenomena, such as the half-filled Landau level problem, on novel two-dimensional systems.

Furthermore, to enhance the practical applicability of our method and facilitate the exploration of topological insulator phases in correlated models, it is crucial to develop efficient numerical methods and algorithms\cite{orus2009simulation,CorbozEvenblyVerstraeteVidal2010,wu2020tensor,mortier2022tensor} specifically tailored to fermionic tensor network states.
Our framework inherently reduces the number of variational parameters needed in numerical simulations, leading to improved computational efficiency.
By combining the power of our scheme with advanced numerical techniques, we may overcome the challenges posed by interaction effects, leading to the discovery of TI phases and a deeper understanding of their properties in strongly correlated systems.

We would like to thank Zheng-Cheng Gu, Chenjie Wang, and Chong Wang for helpful discussions. The work is supported by MOST grant No.~2022YFA1403900, NSFC No.~12104451, NSFC No.~11920101005, and funds from Strategic Priority Research Program of CAS No.~XDB28000000.

\bibliography{bibfpeps}{}

\clearpage

\setcounter{secnumdepth}{3}
\setcounter{equation}{0}
\setcounter{figure}{0}
\setcounter{section}{0}
\renewcommand{\thesection}{\Roman{section}}
\renewcommand{\theequation}{S\arabic{equation}}
\renewcommand{\thefigure}{S\arabic{figure}}
\newcommand\Scite[1]{[\citealp{#1}]}
\makeatletter \renewcommand\@biblabel[1]{[#1]} \makeatother
\onecolumngrid
\begin{center}
    \Large \textbf{Supplementary Material}
\end{center}

\section{Brief introduction to symmetric fermionic tensor network states}\label{app:ftn}
In this appendix, we review the fundamentals of fermionic tensor network states\cite{kraus2010fermionic,BultinckWilliamsonHaegemanVerstraete2017fermionicmps,Bultinck2017fermionic}, and fix our notations used in the main text.
\subsection{Fermionic tensors and tensor contraction} 
Building blocks of fermionic tensor networks are fermionic tensors, which live in fermionic tensor product~(labeled as $\otimes_f$) of legs.
Legs with inward/outward arrows are local fermion Hilbert spaces for ket/bra states, where the parity of state $\ket{i}/\bra{i}$ is $(-1)^{\abs{i}}$ with $\abs{i}\in\{0,1\}$. 
Exchanging states of two legs gives $-1$ if these two states are both parity odd:
\begin{align}
\ket{i}_a\otimes_f\ket{j}_b=(-1)^{\abs{i}\abs{j}}\ket{j}_b\otimes_f\ket{i}_a
\end{align}

As an example of fermionic tensors, let us consider tensor $\hat{T}$ with three legs, say $a,b,c$:
\begin{align}
    \hat{T}=(T_{abc})_{ijk} \ket{i}_a\otimes_f\ket{j}_b\otimes_f\ket{k}_c
\end{align}
Leg indices $abc$ are sometimes ignored when there is no confusion.
We may also omit $\otimes_f$'s and use a more compact form $\hat{T}=T_{ijk}\ket{ijk}$.

\begin{figure}[ht]
    \centering
    \includegraphics[scale=1]{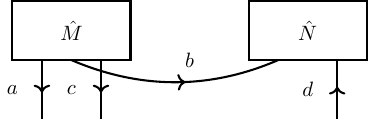}
    \caption{Graphic representation of contraction between fermionic tensors $\hat{M}$ and $\hat{N}$. 
        The outward legs are bra spaces and inward legs are ket spaces.
    The intersection of leg $b$ and $c$ indicates a possible extra minus sign due to exchanging fermions.}
    \label{fig:app_contraction}
\end{figure}

Ket and bra states can be contracted using $\fTr$, defined as
\begin{align}
    \fTr[\bra{i}\otimes_f\ket{j}]=(-)^{\abs{i}\abs{j}}\fTr[\ket{j}\otimes_f\bra{i}]=\delta_{ij}
    \label{eq:ftr}
\end{align}
It is noteworthy that the order of contracted states matters as extra $-1$ may be produced.
Generalization to tensor contractions is straightforward.
As shown in Fig.~\ref{fig:app_contraction}, for two fermionic tensors $\hat{M}=M^{ijk}\bra{i}_a\bra{j}_b\bra{k}_c$, $\hat{N}=N_{lm}\ket{l}_b \ket{m}_{d}$, 
\begin{align}
    \fTr_{b}[\hat{M}\otimes_f \hat{N}]\equiv(-1)^{\abs{j}\abs{k}} M^{ijk}N_{lm}\delta_{jl}\bra{i}\bra{k}\ket{m}
\end{align}
We may omit cumbersome $\fTr$'s and $\otimes_{f}$'s and use $\hat{M}\cdot \hat{N}$ to represent tensor contraction. 

For local tensors in tensor network states, there are two types of legs: internal ones and physical ones.
States in physical legs is denoted by $\ket{\bullet}/\bra{\bullet}$, while states in internal legs by $\vket{\bullet}/\vbra{\bullet}$.
To get a physical wavefunction, all internal legs are contracted.
By fixing parity of all local tensors, physical wavefunctions have fixed parity
In this work, we focus on the case where all local tensors are \emph{parity even}.

\subsection{Gauge transformation and symmetries of FPEPS}\label{subapp:sym_ftn_gauge_trans}
\begin{figure}
    \centering
    \includegraphics[width=0.8\textwidth]{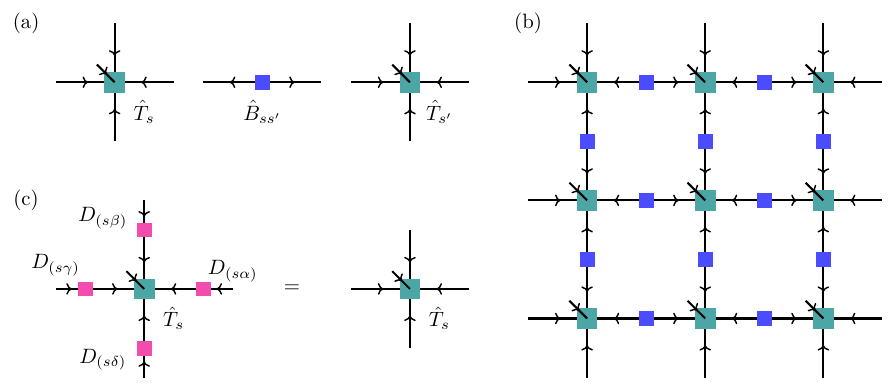}
    \caption{ (a) A bond tensor $\hat{B}_{ss'}$ and its neighbouring site tensors. (b) A $3\times 3$ FPEPS on square lattice with boundary legs. (c) $\IGG$ invariance of a single site tensor. 
    }
    \label{fig:fPEPS}
\end{figure}

We consider a particular type of fermionic tensor network -- fermionic projected entangled-pair state(s)~(FPEPS)\cite{SchuchGarciaCirac2011}.
As shown in Fig.~\ref{fig:fPEPS}(a), we focus on FPEPS with both site tensors $\hat{T}_s$ and bond tensors $\hat{B}_{ss'}$, where $s,s'$ label neighbouring site coordinates.
Without loss of generality, we assume that internal legs of site tensors all point inward~(ket spaces), while those of bond tensors point outward~(bra spaces).
Physical wavefunction then reads $\ket{\Psi}=\fTr\left[ \mathbb{B}\otimes_f \mathbb{T}  \right]$, where $\mathbb{T}=\bigotimes_{s}\subf \hat{T}_s$ and $\mathbb{B}=\bigotimes_{\lrangle{ss'}}\subf\hat{B}_{ss'}$
Note that as all tensors are parity even, different orders of tensor contraction give the same state.

Representation of physical wavefunction $\ket{\Psi}$ by FPEPS is far from unique.
In particular, two FPEPS represent the same wavefunction if they are related by some gauge transformation:
\begin{align}
    \ket{\psi}=\fTr\left[ \mathbb{B}\otimes_f\mathbb{T} \right]
    =\fTr\left[ \mathbb{B}\otimes_f\mathbb{W}^{-1}\otimes_f\mathbb{W}\otimes_f \mathbb{T} \right]
\end{align}
Here, $\mathbb{W}$ and $\mathbb{W}^{-1}$ are tensor products of gauge transformation $W$'s on internal legs:
\begin{align}
    \mathbb{W}&=W_{(s_{1}\alpha_{1})}\otimes_f W_{(s_{1}\alpha_{2})}\otimes_f\cdots\otimes_f  W_{(s_{1}\alpha_{m})} \otimes_f W_{(s_{2}\alpha_{1})}\otimes_f\cdots \otimes_f W_{(s_{n}\alpha_{m})}~,\notag\\
    \mathbb{W}^{-1}&=W^{-1}_{(s_{n}\alpha_{m})}\otimes_f \cdots\otimes_f W^{-1}_{(s_{n}\alpha_{1})}\otimes_f\cdots W^{-1}_{(s_{n-1}\alpha_{m})}\otimes_f\cdots \otimes_f W^{-1}_{(s_{1}\alpha_{1})}
    \label{}
\end{align}
where $(s\alpha)$ labels internal leg, and 
\begin{align}
    W_{(s\alpha)} \vket{i}_{(s\alpha)}=\sum_{b}[W_{(s\alpha)}]_{ji}\vket{j}_{(s\alpha)}~,\quad
    \vbra{i}_{(s\alpha)} W^{-1}_{(s\alpha)}=\sum_{j}\vbra{j}_{(s\alpha)} \left[ W_{(s\alpha)}^{-1} \right]_{ij}
\end{align}
$W$'s in general do not have fixed parity, and thus permuting $W$'s and $W^{-1}$'s may lead to fermion swapping gate.

For the case where $\ket{\Psi}$ is invariant under symmetry $g$, we assume that $g$-action on physical legs can be pushed to gauge transformation on internal legs of local tensors:
\begin{align}
    U_{s}(g)\otimes\left( \bigotimes_{\alpha}\subf W_{(s\alpha)}(g) \right)\cdot \hat{T}_{s}=\hat{T}_{s}~,\quad
    \hat{B}_{ss'}\cdot W^{-1}_{(s\alpha)}(g)\otimes_f W^{-1}_{(s'\alpha')}(g)=\hat{B}_{ss'}
\end{align}
where $A\cdot B$ means $\fTr[A\otimes_f B]$.
The above equations give symmetry constraints for local tensors.
We remark that to get a symmetric wavefunction, orders of $W(g)$'s in the above equation are essential.
In particular, as we will show in Appendix~\ref{app:kasteleyn}, a valid order of $W(g)$'s gives a Kasteleyn orientation on the lattice. 

As shown in Fig.~\ref{fig:fPEPS}(c), there exists a special kind of gauge transformation $\mathbb{D}$, which leaves every single tensor invariant: 
\begin{align}
    \left(\bigotimes_{\alpha}\subf D_{(s\alpha)}\right) \cdot \hat{T}_{s}=\hat{T}_s~,\quad
    \hat{B}_{ss'}\cdot D^{-1}_{(s\alpha)}\otimes_{f} D^{-1}_{(s'\alpha')} =\hat{B}_{ss'}~.
\end{align}
Such gauge transformation form invariant gauge group~($\IGG$).
In this work, we focus on the case where $D$'s are parity even.

Note that the group always have a trivial center $H$ formed by phase factors $\chi_{(s\alpha)}$ that satisfy $\prod_{(s\alpha)}\chi_{(s\alpha)}=1$. 
In addition, if the $\IGG$ is a $U(1)$ group generated by $n_{D;(s\alpha)}$, namely, $D_{(s\alpha)}(\theta)=\exp[\ii\theta n_{D;(s\alpha)}]$.
We then have $\left(\sum_{\alpha}n_{D;(s\alpha)}\right)\cdot \hat{T}^{s}=0$

\subsection{Plaquette $\IGG$ and vanishing long-range entanglement}\label{subapp:sym_ftn_igg}
Beside global $\IGG$ which is a set of gauge transformations acting on all internal legs, there exists another type of $\IGG$ which we call plaquette $\IGG$\cite{JiangRan2017anyon}. A plaquette $\IGG$ is from the decomposition of a global $\IGG$

\begin{align}
    D_{(s\alpha)}= \lambda^{(a)}_{(s\alpha)}\cdot \lambda^{(\bar{a})}_{(s\alpha)}
\end{align} 
Here, $a=1/2$ while $\bar{a}=2/1$. We assume that $\IGG$'s acting on different plaquettes are commutative. What's more, we have the plaquette $\IGG$ equation on a site tensor as
\begin{align}
    \adjincludegraphics[scale=0.8,valign=c]{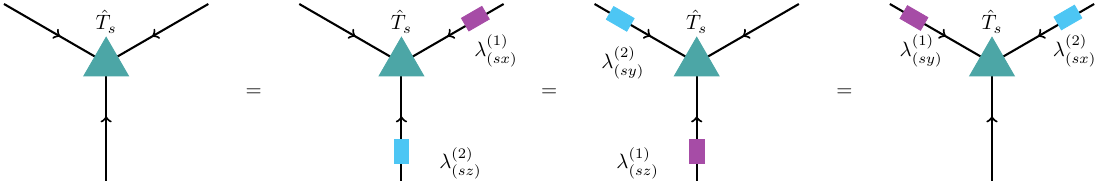}
    \label{eq:plaquette_equation}
\end{align}
For a bond tensor, the plaquette $\IGG$ equation is
\begin{align}
\adjincludegraphics[scale=0.8,valign=c]{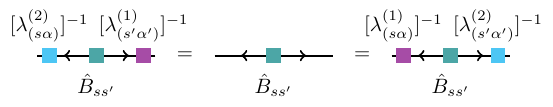}
    \label{eq:plaquette_equation_bond}
\end{align}
Thus the action of plaquette $\IGG$ is local but still keeps local tensors invariant. In this work, we assume all global $\IGG$'s have such plaquette decomposition.

\begin{figure}[h]
    \centering
    \includegraphics[width=0.6\textwidth]{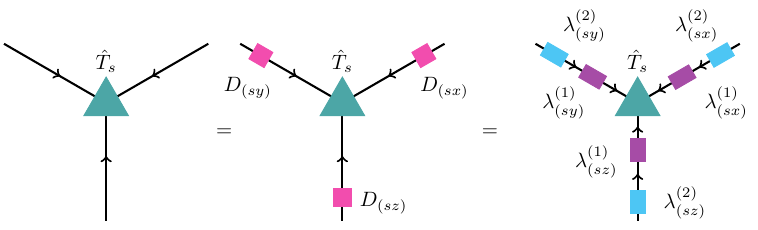}
    \caption{Any $\IGG$ element can be decomposed to plaquette IGG.}
    \label{fig:Plaquette_decomposition}
\end{figure}

Nontrivial global $\IGG$ element often leads to topological ground state degeneracy.
To see this, let us consider a FPEPS with periodic boundary condition on square lattice:
\begin{align*}
    \ket{\Psi}=\adjincludegraphics[scale=0.5,valign=c]{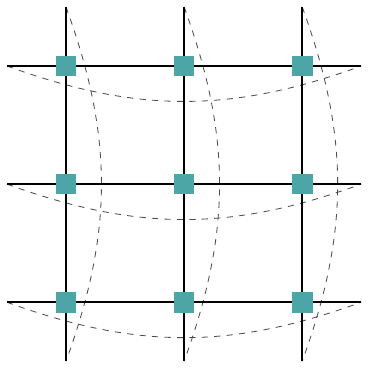}~,
\end{align*}
where the bond tensors, physical legs and leg orientations are omitted for brevity.
If such tensor network has a nontrivial $\IGG$ element, we can insert ``$\IGG$ loops'' in internal legs, which leaves the wavefunction invariant:
\begin{align*}
    \ket{\Psi}=\adjincludegraphics[scale=0.5,valign=c]{fPEPS_periodic.pdf}~= \adjincludegraphics[scale=0.5,valign=c]{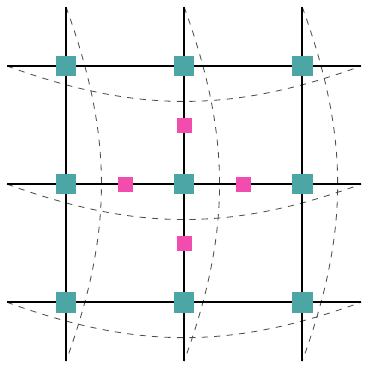}~.
\end{align*}
However, by inserting non-contractible loops of $\IGG$ action, we in general obtain a different state as
\begin{align}
    \ket{\Psi_{v}}=\adjincludegraphics[scale=0.5,valign=c]{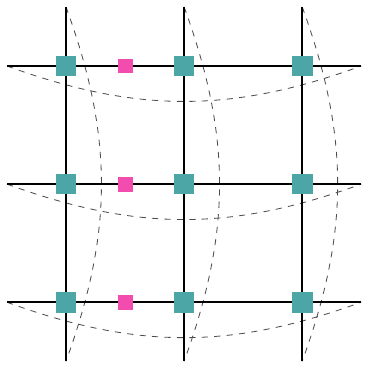}.
    \label{eq:psiv_igg}
\end{align}
$\ket{\Psi_{v}}$ can't be distinguished with $\ket{\Psi}$ using local operators, as the noncontratible loop can move freely.
Thus, if $\ket{\psi}$ is a ground state of a local Hamiltonian, a nontrivial $\IGG$ indicates that the Hamiltonian has topological ground state degeneracy.
In other words, $\ket{\psi}$ is a long-range entangled state.

However, the plaquette decomposition of the $\IGG$ element kills the long-range entanglement. 
To see this, we apply Eq.~(\ref{eq:plaquette_equation}) for $\ket{\Psi_v}$ in Eq.~(\ref{eq:psiv_igg}), and obtain 
\begin{align*}
    \ket{\Psi_v}=\adjincludegraphics[scale=0.5,valign=c]{fPEPS_loop_nonc.pdf} ~
    =\adjincludegraphics[scale=0.5,valign=c]{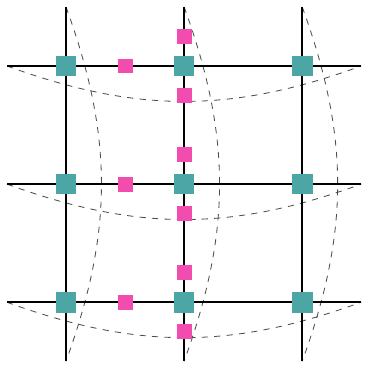}
    =\adjincludegraphics[scale=0.5,valign=c]{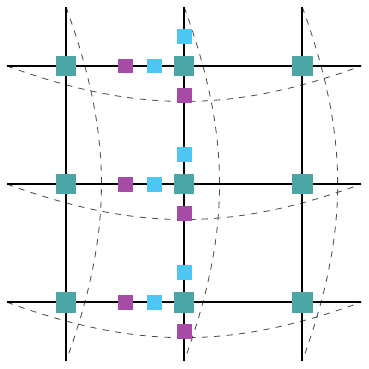}~=\adjincludegraphics[scale=0.5,valign=c]{fPEPS_periodic.pdf}~
    =\ket{\Psi}
\end{align*}
In other words, the non-contractible loop the $\IGG$ element acts trivially on the tensor network state if the $\IGG$ element has plaquette decomposition.
Thus, we obtain an SRE state.

\section{Cocycle and coboundary conditions of $\omega(g_1,g_2,g_3)$}
\label{app:omega_cocycle}
\subsection{3-cocycle condition}\label{subapp:3-cocycle}
In the following, by calculating $\Omega_a(g_1,g_2)\Omega_a(g_1g_2,g_3)\Omega_a(g_1g_2g_3,g_4)$ in two different ways, we will derive the consistent conditions for $\omega$ as in Eq.~(\ref{eq:u1f_fermionic_3-cocycle}).

Firstly, we have
\begin{align}
    &\Omega_a(g_1,g_2)\Omega_a(g_1g_2,g_3)\Omega_a(g_1g_2g_3,g_4)\notag\\
    ={}&\omega(g_1,g_2,g_3)\cdot{}^{U_M(g_1)K^{s(g_1)}}\Omega_a(g_2,g_3)\Omega_a(g_1,g_2g_3)\Omega_a(g_1g_2g_3,g_4)\notag\\
    ={}&\omega(g_1,g_2,g_3)\omega(g_1,g_2g_3,g_4)\cdot {}^{U_M(g_1)K^{s(g_1)}}\left[ \Omega_a(g_2,g_3)\Omega_a(g_2g_3,g_4) \right] \Omega_a(g_1,g_2g_3g_4)\notag\\
    ={}&\omega(g_1,g_2,g_3)\omega(g_1,g_2g_3,g_4)\omega(g_2,g_3,g_4)\cdot {}^{U_M(g_1)K^{s(g_1)}}\left[ {}^{U_M(g_2)K^{s(g_2)}}\Omega_a(g_3,g_4)\Omega_a(g_2,g_3g_4) \right] \Omega_a(g_1,g_2g_3g_4)\notag\\
    ={}&\omega(g_1,g_2,g_3)\omega(g_1,g_2g_3,g_4)[\omega(g_2,g_3,g_4)]^{1-2s(g_1)}\cdot {}^{\Omega(g_1,g_2)\exp[\ii\alpha(g_1,g_2)\cdot n_f]U_M(g_1g_2)K^{s(g_1g_2)}}\Omega_a(g_3,g_4) \times\notag\\
    {}&{}^{U_M(g_1)K^{s(g_1)}}\Omega_a(g_2,g_3g_4) \Omega_a(g_1,g_2g_3g_4)\notag\\
    ={}&\exp\left[ \ii\cdot(-1)^{\rho(g_1g_2)}\alpha(g_1,g_2)\sigma(g_3,g_4) \right]\cdot(-1)^{\sigma(g_1,g_2)\sigma(g_3g_4)}\cdot \omega(g_1,g_2,g_3)\omega(g_1,g_2g_3,g_4) [\omega(g_2,g_3,g_4)]^{1-2s(g_1)}\cdot\notag\\
    {}&{}^{\Omega_a(g_1,g_2)U_M(g_1g_2)K^{s(g_1g_2)}}\Omega_a(g_3,g_4)\cdot {}^{U_M(g_1)K^{s(g_1)}}\Omega_a(g_2,g_3g_4) \Omega_a(g_1,g_2g_3g_4)
    \label{eq:Omega_pentagon_1}
\end{align}
Then above equation can also be calculated in another way:
\begin{align}
    & \Omega_a(g_1,g_2)\Omega_a(g_1g_2,g_3)\Omega_a(g_1g_2g_3,g_4)\notag\\
    ={}&\omega(g_1g_2,g_3,g_4)\cdot \Omega_a(g_1,g_2)\cdot {}^{U_M(g_1g_2)K^{s(g_1g_2)}}\Omega_a(g_3,g_4)\Omega_a(g_1g_2,g_3g_4)\notag\\
    ={}&\omega(g_1g_2,g_3,g_4)\omega(g_1,g_2,g_3g_4)\cdot{}^{\Omega_a(g_1,g_2)U_M(g_1g_2)K^{s(g_1g_2)}}\Omega_a(g_3,g_4)\cdot {}^{U_M(g_1)K^{s(g_1)}}\Omega_a(g_2,g_3g_4) \Omega_a(g_1,g_2g_3g_4)\,,
    \label{eq:Omega_pentagon_2}
\end{align}
Thus, by equating Eq.~\eqref{eq:Omega_pentagon_1} and \eqref{eq:Omega_pentagon_2}, we obtain Eq.~(\ref{eq:u1f_fermionic_3-cocycle}).

\subsection{coboundary condition}\label{subapp:3-coboundary}
The left side of Eq.~\eqref{eq:tildew} is straightforward:
\begin{equation}
    \beta(g_1,g_2)\beta(g_1g_2,g_3)\cdot\Sigma_a(g_1)\cdot{}^{U_M(g_1)K^{s(g_1)}}\Sigma_a(g_2)\cdot\Omega_a(g_1,g_2)\cdot{}^{U_M(g_1g_2)K^{s(g_1g_2)}}\Sigma_a(g_3)\cdot\Omega_a(g_1g_2,g_3)\cdot\Sigma_a^{-1}(g_1g_2g_3)\,.
\end{equation}
The right side is 
\begin{align}
    &\tilde{\omega}(g_1,g_2,g_3)\Sigma_a(g_1)\Sigma_b(g_1){}^{U_M(g_1)K^{s(g_1)}}(\beta(g_2,g_3)\cdot \Sigma_a(g_2)\cdot{}^{U_M(g_2)K^{s(g_2)}}\Sigma_a(g_3)\cdot\Omega_a(g_2,g_3)\cdot\Sigma_a^{-1}(g_2g_3))\Sigma_b^{-1}(g_1)\Sigma_a^{-1}(g_1)\nonumber\\
    &(\beta(g_1,g_2g_3)\cdot \Sigma_a(g_1)\cdot{}^{U_M(g_1)K^{s(g_1)}}\Sigma_a(g_2g_3)\cdot\Omega_a(g_1,g_2g_3)\cdot\Sigma_a^{-1}(g_1g_2g_3))\nonumber\\
    ={}&\tilde{\omega}(g_1,g_2,g_3) (-)^{\mu(g_1)(\mu(g_2)+\mu(g_3)+\sigma(g_2g_3)-\mu(g_2g_3))}(\beta(g_2,g_3))^{1-2s(g_1)}\cdot\Sigma_a(g_1)\cdot \nonumber\\
    &{}^{U_M(g_1)K^{s(g_1)}}\left(\Sigma_a(g_2)\cdot{}^{U_M(g_2)K^{s(g_2)}}\Sigma_a(g_3)\Omega_a(g_2,g_3)\right)\beta(g_1,g_2g_3)\Omega_a(g_1,g_2g_3)\Sigma^{-1}_{a}(g_1g_2g_3)\nonumber\\
    ={}&\tilde{\omega}(g_1,g_2,g_3)(-)^{\mu(g_1)\tilde{\sigma}(g_2,g_3)}(\beta(g_2,g_3))^{1-2s(g_1)}\beta(g_1,g_2g_3)\Sigma_a(g_1)\cdot\left({}^{U_M(g_1)K^{s(g_1)}}\Sigma_a(g_2)\right)\cdot\nonumber\\
    &\left({}^{\Omega(g_1,g_2)\exp\{i\alpha(g_1,g_2)n_f\}U_M(g_1g_2)K^{s(g_1g_2)}}\Sigma_a(g_3)\right)\cdot\left({}^{U_M(g_1)K^{s(g_1)}}\Omega_a(g_2,g_3)\right)\cdot \Omega_a(g_1g_2,g_3)\Sigma_a^{-1}(g_1g_2g_3)\nonumber\\
    ={}&\tilde{\omega}(g_1,g_2,g_3)(-)^{\mu(g_1)\tilde{\sigma}(g_2,g_3)}(\beta(g_2,g_3))^{1-2s(g_1)}\beta(g_1,g_2g_3)\Sigma_a(g_1)\cdot\left({}^{U_M(g_1)K^{s(g_1)}}\Sigma_a(g_2)\right)\nonumber\\
    &\exp \{i(-)^{\rho(g_1g_2)}\alpha(g_1,g_2)\mu(g_3)\} (-)^{\sigma(g_1,g_2)\mu(g_3)}\Omega_a(g_1,g_2)\cdot\left({}^{U_M{g_1g_2}K^{s(g_1g_2)}}\Sigma_a(g_3)\right)\cdot\Omega^{-1}_a(g_1,g_2)\omega^{-1}(g_1,g_2,g_3)\nonumber\\
    &\Omega_a(g_1,g_2)\cdot\Omega_a(g_1g_2,g_3)\cdot\Sigma_a^{-1}(g_1g_2g_3)\nonumber\\
    ={}&\tilde{\omega}(g_1,g_2,g_3)\omega^{-1}(g_1,g_2,g_3)(-)^{\mu(g_1)\tilde{\sigma}(g_2,g_3) + \sigma(g_1,g_2)\mu(g_3)}(\beta(g_2,g_3))^{1-2s(g_1)}\beta(g_1,g_2g_3)\exp \{i(-)^{\rho(g_1g_2)}\alpha(g_1,g_2)\mu(g_3)\}\nonumber\\
    &\Sigma_a(g_1)\cdot{}^{U_M(g_1)K^{s(g_1)}}\Sigma_a(g_2)\cdot\Omega_a(g_1,g_2)\cdot{}^{U_M(g_1g_2)K^{s(g_1g_2)}}\Sigma_a(g_3)\cdot\Omega_a(g_1g_2,g_3)\cdot\Sigma_a^{-1}(g_1g_2g_3)\,.
\end{align}
By comparing the above two equations, we obtain the coboundary condition Eq.~\eqref{eq:u1f_fermionic_coboundary}:
\begin{equation}
    \tilde{\omega}(g_1,g_2,g_3) = \omega(g_1,g_2,g_3)\frac{\beta(g_1,g_2)\beta(g_1g_2,g_3)}{\beta(g_1g_2,g_3)(\beta(g_2,g_3))^{1-2s(g_1)}}(-)^{\mu(g_1)\tilde{\sigma}(g_2,g_3) + \sigma(g_1,g_2)\mu(g_3)}\exp \{-i(-)^{\rho(g_1g_2)}\alpha(g_1,g_2)\mu(g_3)\}\,.
\end{equation}

\section{Additional details and analysis on tensor equations}
\subsection{Action of $g_1g_2$ on $n_f$}\label{subapp:g1g2_on_nf}
In this part, we will prove Eq.~(\ref{eq:g1g2_on_nf}) and Eq.~(\ref{eq:u1f_nD_fusion}).
Similar as in Section~\ref{subsec:edge_hilbert_space}, we start from a region $A$, and its boundary Hilbert space defined by projector $P_{edge}$.
Let us consider action of $n_f$ on a single leg $j$ together with action of $g_1g_2$ on $\partial A$:
\begin{align}
    &W_{\partial A}(g_1)K^{s(g_1)}\cdot W_{\partial A}(g_2)K^{s(g_2)}\cdot n_{f;j}\cdot P_{edge}\notag\\
    ={}&\left[ (-)^{\rho(g_1g_2)}n_{f;j}+(-)^{\rho(g_2)}n_{D;j}(g_1)+\act{g_1}{n_{D;j}(g_2)} \right]\cdot W_{\partial A}(g_1)K^{s(g_1)}\cdot W_{\partial A}(g_2)K^{s(g_2)}\cdot P_{edge}
    \label{eq:g1g2_on_nf1}
\end{align}
where $W_{\partial A}(g)\equiv\bigotimes\subf_{,j\in\partial A} W_{j}(g)$, and $\act{g}n\equiv W_{\partial A}(g)K^{s(g)}\circ n_{f;j}$.
Similarly,
\begin{align}
    W_{\partial A}(g_1g_2)K^{s(g_1g_2)}\cdot n_{f;j}\cdot P_{edge}
    =\left[ (-)^{\rho(g_1g_2)}n_{f;j}+n_{D;j}(g_1g_2) \right]\cdot W_{\partial A}(g_1g_2)K^{s(g_1g_2)}\cdot P_{edge}
    \label{eq:g1g2_on_nf2}
\end{align}
Note that $n_{f;j}$ commute with $n_{\lambda}$'s, and $P_{edge}$ is defined by $n_\lambda$'s as in Eq.~(\ref{eq:p_edge_dec}) and Eq.~(\ref{eq:local_edge_space_nlambda}).
Thus, we have $n_{f;j}\cdot P_{edge}=P_{edge}\cdot n_{f;j}$.
Together with the following equation indicated by Eq.(\ref{eq:pedge_w}) where
\begin{align}
    U_{edge}(g_1g_2)K^{s(g_1g_2)}\equiv P_{edge} W(g_1g_2)K^{s(g_1g_2)}\cdot 
    = U_{edge}(g_1)K^{s(g_1)}\cdot U_{edge}(g_2)K^{s(g_2)}= P_{edge} W_{\partial A}(g_1)K^{s(g_1)}\cdot W_{\partial A}(g_2)K^{s(g_2)}\cdot \,,
    \label{}
\end{align}
we conclude that Eq.~\eqref{eq:g1g2_on_nf1} and Eq.~\eqref{eq:g1g2_on_nf2} are equal to each other as in Eq.~\eqref{eq:g1g2_on_nf}.
And by equating these two equations, we have
\begin{align}
    \left[ (-)^{\rho(g_1g_2)}n_{f;j}+(-)^{\rho(g_2)}n_{D;j}(g_1)+\act{g_1}{n_{D;j}(g_2)} \right]\cdot P_{edge}
    =\left[ (-)^{\rho(g_1g_2)}n_{f;j}+n_{D;j}(g_1g_2) \right] \cdot P_{edge}
    \label{}
\end{align}
In other words, $\left[ (-)^{\rho(g_1g_2)}n_{f;j}+(-)^{\rho(g_2)}n_{D;j}(g_1)+\act{g_1}{n_{D;j}(g_2)} \right]$ and $\left[ (-)^{\rho(g_1g_2)}n_{f;j}+n_{D;j}(g_1g_2) \right]$ differ up to an IGG element.
However, these two terms only act on a single leg, thus they can only differ by a single-leg IGG, which we always set to be trivial.
Thus, we conclude
\begin{align}
    (-)^{\rho(g_2)}n_{D;j}(g_1)+\act{g_1}{n_{D;j}(g_2)}=n_{D;j}(g_1g_2)
    \label{}
\end{align}
as in Eq.~(\ref{eq:u1f_nD_fusion}).

\subsection{Commutation relation between $n_D$'s}\label{subapp:nd_comm}
As assumed in the main text, $[n_{D,j}(g_1), \act{g_1}{n}_{D,j}(g_2)]=0$.
From Eq.~\eqref{eq:u1f_nD_fusion}, we conclude that $[n_{_{D,j}}(g_1),n_{_{D,j}}(g_1g_2)]=0$, so 
\begin{equation}
    [n_{_{D,j}}(g_1),n_{_{D,j}}(g_2)] = 0, ~~\forall g_{1,2}\in G_b\,.
\end{equation}
In addition, $\forall g_{1,2,3}\in G_b$,
\begin{align}
    [\act{g_1}{n}_{_{D,j}}(g_2),\act{g_1g_2}{n}_{_{D,j}}(g_3)] 
    =[\act{g_1}{n}_{_{D,j}}(g_2),n_{_{D,j}}(g_1g_2g_3)]=0
\end{align}
Therefore 
\begin{equation}
    [\act{g_1}{n}_{_{D,j}}(g_2),n_{_{D,j}}(g_3)] = 0, ~~\forall g_{1,2,3}\in G_b\,.
\end{equation}
Acting $g_4\in G_b$ to the equation above,
\begin{align*}
    &[\act{g_4}{n}_{_{D,j}}(g_3),\act{g_4g_1}{n}_{_{D,j}}(g_2)] = 0\,,
\end{align*}
so, we reach the final conclusion
\begin{equation}
    [\act{g_1}{n}_{_{D,j}}(g_2),\act{g_3}{n}_{_{D,j}}(g_4)]= 0,~~\forall g_{1,2,3,4}\in G_b\,.
\end{equation}

\subsection{Details about $D_{(\beta_1,\beta_2)}(g_1,g_2)$ and $\Lambda^{(a)}_{(\beta_1,\beta_2)}(g_1,g_2)$}\label{subapp:db1b2_lambdab1b2}
In this part, we try to derive expressions for $\Lambda^{(a)}$'s presented in Eq.~\eqref{eq:lambda_iota}.
We first express $D_{(\beta_1,\beta_2)}(g_1,g_2)$ as
\begin{align}
    D_{(\beta_1,\beta_2)}(g_1,g_2)&=\sum_{m_1,m_2}\beta_1^{m_1}\beta_2^{m_2}\cdot d_{m_1m_2}(g_1,g_2)
    \label{eq:d_decomp}
\end{align}
Inserting the above decomposition into Eq.~\eqref{eq:u1f_wg_fusion}, we obtain
\begin{align}
    W_{m_1}(g_1)K^{s(g_1)}\cdot W_{m_2}(g_2)K^{s(g_2)}
    =d_{m_1m_2}(g_1,g_2) \cdot \exp\left[ \ii\alpha(g_1,g_2)\cdot n_{f} \right]\cdot W_{m^{\prime}}(g_1g_2)K^{s(g_1g_2)}
    \label{eq:u1f_wgm1m2_fusion}
\end{align}
with $m^{\prime}=(-)^{\rho(g_2)}\cdot m_1+m_2$.

We now try to identify domain and image of $d_{m_1m_2}(g_1,g_2)$.
According to Eq.~(\ref{eq:u1f_g_act_nf}), we have
\begin{align}
    n_{D}(g)\cdot W_{m}(g)&=m\cdot W_{m}(g)\,.
    \label{}
\end{align}
Namely, image of $W_{m}(g)$ is identified as subspace with $n_{D;s\alpha}(g)=m$~(labeled as $\HH_{D;m}(g)$). 
Defining $p_{D;m}(g)$ as the projector to $\HH_{D;m}(g)$, we have 
\begin{align}
    n_{D}(g)=\sum_{m}m\cdot p_{D;m}(g)
    \label{}
\end{align}
Thus, $\HH_{D;m_1}(g)\cap\HH_{D;m_2}(g)=\emptyset$ for $m_1\neq m_2$.

From Eq.~\eqref{eq:u1f_wgm1m2_fusion}, $d_{m_1m_2}(g_1,g_2)$ is a mapping from $\HH_{D;m'}(g_1g_2)$ to $\HH_{D;m_1}(g_1)\cap\act{g_1}{\HH_{D;m_2}(g_2)}$, where $\act{g_1}{\HH}_{D;m_2}(g_2)$ labels space with $\act{g_1}{n}_{D}(g_2)=m_2$.
From Eq.~\eqref{eq:u1f_nD_fusion} as well as Eq.~\eqref{eq:nd_comm}, we obtain
\begin{align}
    p_{D;m}(g_1g_2) =\sum_{\substack{(-)^{\rho(g_2)} m_1\\+m_2=m}}p_{D;m_1}(g_1)\cdot \act{g_1}{p}_{D;m_2}(g_2)
\end{align}
or equivalently
\begin{align}
    \HH_{D;m}(g_1g_2)=\bigcup_{\substack{(-)^{\rho(g_2)} m_1\\+m_2=m}} \left( \HH_{D;m_1}(g_1)\cap \act{g_1}{\HH}{_{D;m_2}}(g_2) \right)\,.
\end{align}
Here, we simply assume $d_{m_1m_2}(g_1,g_2)$ to be an automorphism in $\HH_{D;m_1}(g_1)\cap \act{g_1}{\HH}{_{D;m_2}}(g_2)$, namely,
\begin{align}
    d_{m_1m_2}(g_1,g_2)=\big[ p_{D;m_1}(g_1)\,\act{g_1}{p}_{D;m_2}(g_2) \big] \cdot d_{m_1m_2}(g_1,g_2)\cdot \big[ p_{D;m_1}(g_1)\,\act{g_1}{p}_{D;m_2}(g_2) \big]
    \label{eq:dm1m2_domain_image}
\end{align}

Similar as Eq.~\eqref{eq:d_decomp}, we express $\Lambda_{(\beta_1,\beta_2)}(g_1,g_2)$ as
\begin{align}
    \Lambda^{(a)}_{(\beta_1,\beta_2)}(g_1,g_2)&=\sum_{l_1,l_2}\beta_1^{l_1}\beta_2^{l_2}\cdot \lambda^{(a)}_{l_1l_2}(g_1,g_2)
    \label{eq:lambda_decomp}
\end{align}
Defining $p^{(a)}_{\lambda;l}(g)$ as projector to space with $n_{\lambda}^{(a)}(g)=l$~(labeled as $\HH^{(a)}_{\lambda;l}(g)$), we have
\begin{align}
    n_{\lambda}^{(a)}(g)=\sum_m m\cdot p^{(a)}_{\lambda;m}(g)
\end{align}
From relation in Eq.~\eqref{eq:nd_decomp} where $n_{D}(g)=n_{\lambda}^{(2)}(g)+n_{\lambda}^{(1)}(g)$, we have
\begin{align}
    p_{D;m}(g)=\sum_{l}p_{\lambda;l}^{(2)}(g)\cdot p_{\lambda;m-l}^{(1)}(g),
    \label{eq:u1f_pd_plambda}
\end{align}
We then have
\begin{align}
    \beta_1^{m_1}\beta_2^{m_2}\cdot p_{D;m_1}(g_1)\,\act{g_1}{p}_{D;m_2}(g_2)
    =\sum_{l_1,l_2}(-1)^{l_2(m_1-l_1)}\Big[ \beta_1^{l_1}\beta_2^{l_2}\cdot p_{\lambda;l_1}^{(2)}(g_1)\,\act{g_1}{p_{\lambda;l_2}^{(2)}(g_2)} \Big] \cdot\Big[ \beta_1^{m_1-l_1}\beta_2^{m_2-l_2}\cdot p_{\lambda;m_1-l_1}^{(1)}(g_1)\,\act{g_1}{ p_{\lambda;m-l_2}^{(1)}(g_2) } \Big]
    \label{}
\end{align}
Comparing this equation with Eq.~\eqref{eq:u1f_db1b2_decomp}, \eqref{eq:d_decomp}, \eqref{eq:lambda_decomp}, and \eqref{eq:dm1m2_domain_image}, it is natural to set $\lambda$ as an automorphism in $\HH^{(a)}_{\lambda;l_1}(g_1)\cap \act{g_1}{\HH}^{(a)}{_{\lambda;l_2}}(g_2)$, namely,
\begin{align}
    \lambda^{(a)}_{l_1l_2}(g_1,g_2)=\big[ p^{(a)}_{\lambda;l_1}(g_1)\,\act{g_1}{p}^{(a)}_{\lambda;l_2}(g_2) \big] \cdot \lambda^{(a)}_{l_1l_2}(g_1,g_2) \cdot \big[ p^{(a)}_{\lambda;l_1}(g_1)\,\act{g_1}{p}^{(a)}_{\lambda;l_2}(g_2) \big]
    \label{eq:u1f_lambdal1l2_automorphism}
\end{align}

To make it more concise, we introduce $I_\beta(g)$ and $\iota_\beta^{(a)}(g)$ as
\begin{align}
    I_{\beta;(s\alpha)}(g) &= \sum_{m} \beta^{m} p_{D;(s\alpha)}(g,m)=(\beta)^{n_{D;(s\alpha)}(g)}\notag\\
    \iota_{\beta}^{(a)}(g) &= \sum_l \beta^l\cdot p^{(a)}_{\lambda;l}(g) = \beta^{n^{(a)}_{\lambda}(g)}\,.
    \label{eq:iota_def}
\end{align}

According to Eq.~\eqref{eq:u1f_pd_plambda}, we get
\begin{align}
    &I_{\beta}(g)=\iota_{\beta}^{(2)}(g)\cdot \iota_{\beta}^{(1)}(g)\,,\label{eq:app_u1f_ibeta_iota}\\
    &\hspace*{-0.7cm}\adjincludegraphics[valign=c]{I_decomp.pdf}\,,\nonumber
\end{align}
According to Eq.~\eqref{eq:lambda_decomp} as well as \eqref{eq:u1f_lambdal1l2_automorphism}, we reach the following expression for $\Lambda^{(2)}$ and $\Lambda^{(1)}$, where
\begin{align}
    &\Lambda^{(1)}_{(\beta_1,\beta_2)}(g_1,g_2)=\iota^{(1)}_{\beta_1}(g_1) \cdot \act{g_1}{\iota}{^{(1)}_{\beta_2}}(g_2) \cdot \lambda^{(1)}(g_1,g_2)\notag\\
    &\Lambda^{(2)}_{(\beta_1,\beta_2)}(g_1,g_2)= \lambda^{(2)}(g_1,g_2)\cdot \act{g_1}{\iota}^{(2)}_{\beta_2}(g_2)\cdot \iota^{(2)}_{\beta_1}(g_1)\label{}\\
    &\adjincludegraphics[valign=c]{Lambda_1.pdf}\,,\nonumber\\
    &\adjincludegraphics[valign=c]{Lambda.pdf}\,,\nonumber
\end{align}

where 
\begin{align}
    \lambda^{(a)}(g_1,g_2)\equiv\bigoplus_{l_1l_2}\lambda^{(a)}_{l_{1},l_{2}}(g_1,g_2)
    \label{}
\end{align}
Or equivalently,
\begin{align}
    [\lambda^{(a)}(g_1,g_2),\,n_\lambda^{(a)}(g_1)]
    =[\lambda^{(a)}_{l_{1},l_{2}}(g_1,g_2),\,\act{g_1}{n_\lambda^{(a)}(g_2)}]
    =0
    \label{}
\end{align}
Here, we will assume a more strict constraint on $\lambda^{(a)}$'s, where
\begin{align}
    [\lambda^{(a)}(g_1,g_2),\,\act{g_3}{n_\lambda^{(a)}(g_4)}]=0\,,\quad
    \forall g_j\in G_b
    \label{eq:u1f_lambdag1g2_nlambda_comm}
\end{align}

From the definition of $\iota$'s in Eq~\eqref{eq:iota_def}, we derive that
\begin{align}
    \iota_{\beta;(s\alpha)}^{(1)}(g)\cdot \iota_{\beta;(s\alpha+1)}^{(2)}(g)\cdot \hat{T}_s&=\hat{T}_s\notag\\
    \hat{B}_{ss'}\cdot \iota_{\beta;(s\alpha)}^{(a)}(g)\cdot \iota_{\beta;(s'\alpha')}^{(\bar{a})}(g)&=\hat{B}_{ss'}
\end{align}
Thus, by inserting the above equation to Eq.~\eqref{eq:u1f_lambda_plq_igg}, it is straightforward to see that $\lambda^{(a)}(g_1,g_2)$'s belong to plaquette IGG, namely,
\begin{align}
    \lambda_{(s\alpha)}^{(1)}(g_1,g_2)\otimes_f \lambda_{(s\alpha+1)}^{(2)}(g_1,g_2)\cdot \hat{T}_s&=\hat{T}_s\notag\\
    \hat{B}_{ss'}\cdot \lambda_{(s\alpha)}^{(a)}(g_1,g_2)\otimes_f \lambda_{(s'\alpha')}^{(\bar{a})}(g_1,g_2)&=\hat{B}_{ss'}
    \label{}
\end{align}

\section{$\lambda^{(a)}(g_1,g_2)$ and fermionic three cocycle data}\label{app:relation_lam_g1g2}
In this appendix, we first use the associativity relation in Eq.~\eqref{eq:u1f_3_sym_fusion_1} and \eqref{eq:u1f_3_sym_fusion_2} to derive the consistent conditions for $\lambda^{(a)}(g_1,g_2)$.
Furthermore, by considering process related to four symmetry actions, we obtain pentagon equations which give fermionic three-cocycle conditions.

\subsection{Consistent conditions for $\lambda(g_1,g_2)$'s}
Using Eq.~\eqref{eq:u1f_db1b2_decomp} and Eq.~\eqref{eq:lambda_iota}, Eq.~\eqref{eq:u1f_3_sym_fusion_1} can be expressed in terms of $\lambda$'s and $\iota$'s as 
\begin{align}
    &\nonumber\\
    &\lambda^{(2)}(g_1,g_2)\act{g_1}{\io{2}{2}{g_2}} {\io{2}{1}{g_1}}\tikzmark{here}{} \io{1}{1}{g_1}\act{g_1}{\io{1}{2}{g_2}}\lambda^{(1)}(g_1,g_2) \tikzmark{term1}{{\io{2}{2}{g_1g_2}}}\io{1}{2}{g_1g_2}\tikzmark{term2}{\boxed{\lambda^{(2)}(g_1g_2,g_3)\act{g_1g_2}{\io{2}{3}{g_3}}\io{2}{2}{g_1g_2}}}\nonumber\\
    &\io{1}{2}{g_1g_2}\act{g_1g_2}{\io{1}{3}{g_3}} \lambda^{(1)}(g_1g_2,g_3)\ee^{\ii (\alpha(g_1,g_2)+\alpha(g_1g_2,g_3))n_f}W(g_1g_2g_3)\,.
\begin{tikzpicture}[remember picture,overlay,>=stealth]
\draw[->,blue] (term1.north) -- ++(0,2pt)  -| (here.north);
\draw[->,blue] (term2.north) -- ++(0,2pt)  -| (here.north);
\end{tikzpicture}
\label{eq:u1f_3sym_fusion1_move}
\end{align}
We move terms according to blue arrows, such that terms with super-indices $(1)$ and $(2)$ are separated.
Note that as $\beta_1\beta_2=-\beta_2\beta_1$, we have
\begin{equation}
    \act{g_1}{\io{a}{1}{g_2}}\act{g_3}{\io{b}{2}{g_4}} = \left[(-)^{\act{g_1}{n_{\lambda}^{(a)}(g_2)}\act{g_3}{n_{\lambda}^{(b)}(g_4)}}\right] ~\act{g_3}{\io{b}{2}{g_4}}\act{g_1}{\io{a}{1}{g_2}}\,,
    \label{eq:iota_commute}
\end{equation}
As we move two $\io{2}{2}{g_1g_2}$ across ${\io{1}{1}{g_1}}$ in Eq.~\eqref{eq:u1f_3sym_fusion1_move}, no additional factor arise from $\io{2}{2}{g_1g_2}$.
Meanwhile, $\act{g_1g_2}{\io{2}{3}{g_3}}$ is moved across $\io{1}{1}{g_1}$, $\act{g_1}{\io{1}{2}{g_2}}$, and $\io{1}{2}{g_1g_2}$, which contributes a factor
\begin{equation*}
(-)^{\act{g_1g_2}{n_{\lambda}^{(2)}(g_3)}(n^{(1)}_\lambda(g_1) + \act{g_1}{n}^{(1)}_{\lambda}(g_2) + n^{(1)}_\lambda(g_1g_2))} = (-)^{\act{g_1g_2}{n_{\lambda}^{(2)}(g_3)}\sigma(g_1,g_2)}  
\end{equation*}
where we use Eq.~\eqref{eq:u1f_n_lam_fusion} to get right-hand side.
Thus, by rearranging terms according to blue arrows, Eq.~\eqref{eq:u1f_3sym_fusion1_move} becomes
\begin{align}
    \left[(-)^{\act{g_1g_2}{n_{\lambda}^{(2)}(g_3)}\sigma(g_1,g_2)} \right] &\left(\lambda^{(2)}(g_1,g_2)\act{g_1}{\io{2}{2}{g_2}} {\io{2}{1}{g_1}} \right)\io{2}{2}{g_1g_2}\left(\lambda^{(2)}(g_1g_2,g_3)\act{g_1g_2}{\io{2}{3}{g_3}}\io{2}{2}{g_1g_2} \right)\nonumber\\
    &\left(\io{1}{1}{g_1}\act{g_1}{\io{1}{2}{g_2}}\lambda^{(1)}(g_1,g_2) \right)\io{1}{2}{g_1g_2}\left( \io{1}{2}{g_1g_2}\act{g_1g_2}{\io{1}{3}{g_3}} \lambda^{(1)}(g_1g_2,g_3)\right)\nonumber\\
    &\ee^{\ii (\alpha(g_1,g_2)+\alpha(g_1g_2,g_3))n_f}W(g_1g_2g_3)\notag\\
    =\left[(-)^{\act{g_1g_2}{n_{\lambda}^{(2)}(g_3)}\sigma(g_1,g_2)} \right] &\Lambda^{(2)}_{(\beta_1,\beta_2)}(g_1,g_2)\io{2}{2}{g_1g_{2}}\Lambda^{(2)}_{(\beta_2,\beta_3)}(g_1g_2,g_3)\Lambda^{(1)}_{(\beta_1,\beta_2)}(g_1,g_2)\io{1}{2}{g_1g_2} \Lambda^{(1)}_{(\beta_2,\beta_3)}(g_1g_2,g_3)\nonumber\\
    &\ee^{\ii (\alpha(g_1,g_2)+\alpha(g_1g_2,g_3))n_f}W(g_1g_2g_3)\label{eq:D_fusion_separation_1}\,.
\end{align}

Similarly, Eq.~\eqref{eq:u1f_3_sym_fusion_2} is expanded as
\begin{align}
    &\io{2}{1}{g_1}\io{1}{1}{g_1}\tikzmark{h1}{}\act{g_1}{\lambda^{(2)}(g_2,g_3)}\act{g_1g_2}{\io{2}{3}{g_3}}\act{g_1}{\io{2}{2}{g_2}}\tikzmark{h2}{}\ee^{\ii(-)^{s(g_1)} \alpha(g_2,g_3)\act{g_1}{n}_f}\cdot \act{g_1}{\io{1}{2}{g_2}}\act{g_1g_2}{\io{1}{3}{g_3}}\act{g_1}{\lambda^{(1)}(g_2,g_3)} \tikzmark{t1}{\act{g_1}{\io{2}{3}{g_2g_3}}}\act{g_1}{\io{1}{3}{g_2g_3}}\nonumber\\
    &\nonumber\\
    &\tikzmark{t2}{\boxed{\io{2}{1}{g_1}\io{1}{1}{g_1}}}~\tikzmark{t3}{\boxed{\lambda^{(2)}(g_1,g_2g_3) \act{g_1}{\io{2}{3}{g_2g_3}} \io{2}{1}{g_1} }}\io{1}{1}{g_1} \act{g_1}{\io{1}{3}{g_2g_3}}\lambda^{(1)}(g_1,g_2g_3)\ee^{\ii\alpha(g_1,g_2g_3) n_{f}} W(g_1g_2g_3)\,.
\begin{tikzpicture}[remember picture,overlay,>=stealth]
\draw[->,blue] (t1.north) -- ++(0,5pt)  -| (h2.north);
\draw[->,red] (t2.north) -- ++(0,5pt)  -| (h1.north);
\draw[->,blue] (t3.north) -- ++(0,5pt)  -| (h2.north);
\end{tikzpicture}
\label{eq:u1f_3sym_fusion2_move}
\end{align}
The red move does not contribute any additional factor.
Two blue moves contain two $\act{g_1}{\io{2}{3}{g_2g_3}}$ terms, and both move across $\act{g_1}{\io{1}{2}{g_2}}$, thus no additional factor either.
The only additional factor comes from moving $\io{2}{1}{g_1}$ across $\act{g_1}{\io{1}{2}{g_2}}$, $\act{g_1g_2}{\io{1}{3}{g_3}}$, and$\act{g_1}{\io{1}{3}{g_2g_3}}$, which contributes
\begin{equation*}
(-)^{{n_{\lambda}^{(2)}(g_1)}(\act{g_1}{n}^{(1)}_\lambda(g_2) + \act{g_1g_2}{n}^{(1)}_{\lambda}(g_3) + \act{g_1}{n}^{(1)}_\lambda(g_2g_3))} = (-)^{{n_{\lambda}^{(2)}(g_1)}\sigma(g_2,g_3)}\,. 
\end{equation*}
Further, from Eq.~\eqref{eq:group_rho_alpha}, the exponential factor in Eq.~\eqref{eq:u1f_3sym_fusion2_move} gives
\begin{equation*}
    \ee^{\ii ((-)^{s(g_1)}\alpha(g_2,g_3)\act{g_1}{n}_f +\alpha(g_1,g_2g_3)n_f)}=\ee^{\ii ((-)^{\tilde{\rho}(g_1)}\alpha(g_2,g_3)+\alpha(g_1,g_2g_3) )n_f + \ii (-)^{s(g_1)}\alpha(g_2,g_3)n_{_D}(g_1)}\,,
\end{equation*}
After arranging terms accordingly, Eq.~\eqref{eq:u1f_3sym_fusion2_move} becomes
\begin{align}
    \left[(-)^{{n_{\lambda}^{(2)}(g_1)}\sigma(g_2,g_3)}\right]&\left(\act{g_1}{\lambda^{(2)}(g_2,g_3)}\act{g_1g_2}{\io{2}{3}{g_3}}\act{g_1}{\io{2}{2}{g_2}}\right)\act{g_1}{\io{2}{3}{g_2g_3}}\left(\lambda^{(2)}(g_1,g_2g_3) \act{g_1}{\io{2}{3}{g_2g_3}} \io{2}{1}{g_1} \right)\nonumber\\
    &\left(\act{g_1}{\io{1}{2}{g_2}}\act{g_1g_2}{\io{1}{3}{g_3}}\act{g_1}{\lambda^{(1)}(g_2,g_3)} \right)\act{g_1}{\io{1}{3}{g_2g_3}}\left(\io{1}{1}{g_1} \act{g_1}{\io{1}{3}{g_2g_3}}\lambda^{(1)}(g_1,g_2g_3) \right)\nonumber\\
    &\ee^{\ii ((-)^{\tilde{\rho}(g_1)}\alpha(g_2,g_3)+\alpha(g_1,g_2g_3) )n_f + \ii (-)^{s(g_1)}\alpha(g_2,g_3)n_{_D}(g_1)}\notag\\
    =\left[(-)^{{n_{\lambda}^{(2)}(g_1)}\sigma(g_2,g_3)}\right]&\act{g_1}{\Lambda^{(2)}_{(\beta_2,\beta_3)}(g_2,g_3)}\act{g_1}{\io{2}{3}{g_2g_3}}\Lambda^{(2)}_{(\beta_1,\beta_3)}(g_1,g_2g_3)\act{g_1}{\Lambda^{(1)}_{(\beta_2,\beta_3)}(g_2,g_3)}\act{g_1}{\io{1}{3}{g_2g_3}}\Lambda^{(1)}_{(\beta_1,\beta_3)}(g_1,g_2g_3)\nonumber\\
    &\ee^{\ii ((-)^{\tilde{\rho}(g_1)}\alpha(g_2,g_3)+\alpha(g_1,g_2g_3) )n_f + \ii (-)^{s(g_1)}\alpha(g_2,g_3)n_{_D}(g_1)}\,.\label{eq:D_fusion_separation_2}
\end{align}

Comparing Eq.~\eqref{eq:D_fusion_separation_1} and \eqref{eq:D_fusion_separation_2}, we conclude tensor equation for $\Lambda^{(2)}$ reads
\begin{align}
    &\left[(-)^{\act{g_1g_2}{n_{\lambda}^{(2)}(g_3)}\sigma(g_1,g_2)} \right]\Lambda^{(2)}_{(\beta_1,\beta_2)}(g_1,g_2)\io{2}{2}{g_1g_2} \Lambda^{(2)}_{(\beta_2,\beta_3)}(g_1g_2,g_3)\nonumber\\
    =&\frac{(-)^{{n_{\lambda}^{(2)}(g_1)}\sigma(g_2,g_3)}}{\omega^{\prime}(g_1,g_2,g_3)}\left(\ee^{\ii (-)^{s(g_1)}\alpha(g_2,g_3)n^{(2)}_{\lambda}(g_1)}\right) \act{g_1}{\Lambda^{(2)}_{(\beta_2,\beta_3)}(g_2,g_3)}\act{g_1}{\io{2}{3}{g_2g_3}}\Lambda^{(2)}_{(\beta_1,\beta_3)}(g_1,g_2g_3)\label{eq:teq_Lam_2}
\end{align}
where the presence of phase factor $\omega'$ is due to the ambiguity in Eq.~\eqref{eq:u1f_db1b2_decomp_phase_ambiguity}.
By rearranging $\iota^{(2)}$'s on both sides of Eq.~\eqref{eq:teq_Lam_2}, we obtain consistent condition for $\lambda^{(2)}$ as
\begin{equation}
    \lambda^{(2)}(g_1,g_2)\lambda^{(2)}(g_1g_2,g_3) = \frac{(-)^{{n_{\lambda}^{(2)}(g_1)}\sigma(g_2,g_3)}}{\omega^{\prime}(g_1,g_2,g_3)}\left(\ee^{\ii (-)^{s(g_1)}\alpha(g_2,g_3)n^{(2)}_{\lambda}(g_1)}\right)\act{g_1}{\lambda^{(2)}(g_2,g_3)}\lambda^{(2)}(g_1,g_2g_3)\,,
    \label{eq:u1f_lambda_omega_prime_relation}
\end{equation}

Similarly, tensor equation for $\Lambda^{(1)}$ reads
\begin{align}
    &\Lambda^{(1)}_{(\beta_1,\beta_2)}(g_1,g_2)\io{1}{2}{g_1g_2} \Lambda^{(1)}_{(\beta_2,\beta_3)}(g_1g_2,g_3)\nonumber\\
    =& \omega^{\prime}(g_1,g_2,g_3)\left(\ee^{\ii (-)^{s(g_1)}\alpha(g_2,g_3)n^{(1)}_{\lambda}(g_1)}\right)\act{g_1}{\Lambda^{(1)}_{(\beta_2,\beta_3)}(g_2,g_3)}\act{g_1}{\io{1}{3}{g_2g_3}}\Lambda^{(1)}_{(\beta_1,\beta_3)}(g_1,g_2g_3)\label{eq:teq_Lam_1}\,.
\end{align}
By rearranging $\iota^{(1)}$'s on both sides, we obtain
\begin{equation}
    \lambda^{(1)}(g_1,g_2)\lambda^{(1)}(g_1g_2,g_3) = \omega^{\prime}(g_1,g_2,g_3)(-)^{{n_{\lambda}^{(1)}(g_1)}\sigma(g_2,g_3)}\left(\ee^{\ii (-)^{s(g_1)}\alpha(g_2,g_3)n^{(1)}_{\lambda}(g_1)}\right)\act{g_1}{\lambda^{(1)}(g_2,g_3)}\lambda^{(1)}(g_1,g_2g_3)\,,
\end{equation}

\subsection{3-cocycle data from $\lambda(g_1,g_2)$'s}
In this part, we will derive the consistent condition for $\omega'(g_1,g_2,g_3)$ by considering $\lambda^{(2)}(g_1,g_2)\cdot \lambda^{(2)}(g_1g_2,g_3)\cdot \lambda^{(2)}(g_1g_2g_3,g_4)$.
There are two ways to calculate this expression, which form the pentagon equation (c.f. Appendix \ref{app:omega_cocycle}). 
Using Eq.~\eqref{eq:u1f_lambda_omega_prime_relation}, we have
\begin{align}
    &\lambda^{(2)}(g_1,g_2)\cdot \lambda^{(2)}(g_1g_2,g_3)\cdot \lambda^{(2)}(g_1g_2g_3,g_4)\nonumber\\
    ={}&\frac{(-)^{n_{\lambda}^{(2)}(g_1)\sigma(g_2,g_3)}}{\omega^{\prime}(g_1,g_2,g_3)}\cdot \left(\ee^{\ii (-)^{s(g_1)}\alpha(g_2,g_3)n_{\lambda}^{(2)}(g_1)}\right)\cdot \act{g_1}{\lambda}^{(2)}(g_2,g_3)\cdot \lambda^{(2)}(g_1,g_2g_3)\cdot \lambda^{(2)}(g_1g_2g_3,g_4)\nonumber\\
    ={}&\frac{(-)^{n_{\lambda}^{(2)}(g_1)\sigma(g_2,g_3)}}{\omega^{\prime}(g_1,g_2,g_3)}\cdot \left(\ee^{\ii (-)^{s(g_1)}\alpha(g_2,g_3)n_{\lambda}^{(2)}(g_1)}\right)\cdot \frac{(-)^{n_{\lambda}^{(2)}(g_1)\sigma(g_2g_3,g_4)}}{\omega^{\prime}(g_1,g_2g_3,g_4)}\cdot \left(\ee^{\ii (-)^{s(g_1)}\alpha(g_2g_3,g_4)n_{\lambda}^{(2)}(g_1)}\right)\cdot \nonumber\\
   &\act{g_1}{\lambda}^{(2)}(g_2,g_3)\cdot \act{g_1}{\lambda}^{(2)}(g_2g_3,g_4)\cdot {\lambda}^{(2)}(g_1,g_2g_3g_4)\nonumber\\
   ={}&\frac{(-)^{n_{\lambda}^{(2)}(g_1)\sigma(g_2,g_3)}}{\omega^{\prime}(g_1,g_2,g_3)}\cdot \left(\ee^{\ii (-)^{s(g_1)}\alpha(g_2,g_3)n_{\lambda}^{(2)}(g_1)}\right)\cdot \frac{(-)^{n_{\lambda}^{(2)}(g_1)\sigma(g_2g_3,g_4)}}{\omega^{\prime}(g_1,g_2g_3,g_4)}\cdot \left(\ee^{\ii (-)^{s(g_1)}\alpha(g_2g_3,g_4)n_{\lambda}^{(2)}(g_1)}\right)\cdot\nonumber\\
   &\act{g_1K^{s(g_1)}}{\Bigg(}\frac{(-)^{n_{\lambda}^{(2)}(g_2)\sigma(g_3,g_4)}}{\omega^{\prime}(g_2,g_3,g_4)}\left(\ee^{\ii (-)^{s(g_2)}\alpha(g_3,g_4)n_{\lambda}^{(2)}(g_2)}\right)\Bigg)\cdot \act{g_1g_2}{\lambda}^{(2)}(g_3,g_4)\cdot \act{g_1}{\lambda}^{(2)}(g_2,g_3g_4)\cdot \lambda^{(2)}(g_1,g_2g_3g_4)
   \label{eq:lam_pentagon_1}
\end{align}
Another fusion process gives
\begin{align}
    &\lambda^{(2)}(g_1,g_2)\cdot \lambda^{(2)}(g_1g_2,g_3)\cdot \lambda^{(2)}(g_1g_2g_3,g_4)\nonumber\\
    ={}&\frac{(-)^{n_{\lambda}^{(2)}(g_1g_2)\sigma(g_3,g_4)}}{\omega^{\prime}(g_1g_2,g_3,g_4)}\cdot \left(\ee^{\ii (-)^{s(g_1g_2)}\alpha(g_3,g_4)n_{\lambda}^{(2)}(g_1g_2)}\right)\cdot \lambda^{(2)}(g_1,g_2)\cdot \lambda^{(2)}(g_1g_2,g_3g_4)\cdot \act{g_1g_2}{\lambda}^{(2)}(g_3,g_4)\nonumber\\
    ={}&\frac{(-)^{n_{\lambda}^{(2)}(g_1g_2)\sigma(g_3,g_4)}}{\omega^{\prime}(g_1g_2,g_3,g_4)}\cdot \left(\ee^{\ii (-)^{s(g_1g_2)}\alpha(g_3,g_4)n_{\lambda}^{(2)}(g_1g_2)}\right)\cdot \frac{(-)^{n_{\lambda}^{(2)}(g_1)\sigma(g_2,g_3g_4)}}{\omega^{\prime}(g_1,g_2,g_3g_4)}\cdot \left(\ee^{\ii (-)^{s(g_1)}\alpha(g_2,g_3g_4)n_{\lambda}^{(2)}(g_1)}\right)\cdot \nonumber\\
    &\act{g_1g_2}{\lambda}^{(2)}(g_3,g_4)\cdot\act{g_1}{\lambda}^{(2)}(g_2,g_3g_4)\cdot \lambda^{(2)}(g_1,g_2g_3g_4)
    \label{eq:lam_pentagon_2}
\end{align}
Comparing Eq.~\eqref{eq:lam_pentagon_1} and \eqref{eq:lam_pentagon_2}, we obtain
\begin{align}
    &\frac{\omega^{\prime}(g_1,g_2,g_3)\omega^{\prime}(g_1,g_2g_3,g_4)[\omega^{\prime}(g_2,g_3,g_4)]^{1-2s(g_1)}}{\omega^{\prime}(g_1g_2,g_3,g_4)\omega^{\prime}(g_1,g_2,g_3g_4)}\nonumber\\
    =&(-)^{n^{(2)}_{\lambda}(g_1)(-\sigma(g_2,g_3g_4)+\sigma(g_2,g_3)+\sigma(g_2g_3,g_4))}(-)^{\sigma(g_3,g_4)(\act{g_1}{n}^{(2)}_{\lambda}(g_2)-n^{(2)}_{\lambda}(g_1g_2))}\nonumber\\
    &\ee^{\ii (-)^{s(g_1)}n^{(2)}_{\lambda}(g_1)(\alpha(g_2,g_3)+\alpha(g_2g_3,g_4)-\alpha(g_2,g_3g_4))}~\ee^{\ii(-)^{s(g_1g_2)}\alpha(g_3,g_4)(\act{g_1}{n}^{(2)}_{\lambda}(g_2)-n^{(2)}_{\lambda}(g_1g_2))}\nonumber\\
    =&(-)^{((-)^{{\rho}(g_2)}n^{(2)}_{\lambda}(g_1)+\act{g_1}{n}^{(2)}_{\lambda}(g_2)-n^{(2)}_{\lambda}(g_1g_2))\sigma(g_3,g_4)}~\ee^{\ii \alpha(g_3,g_4)((-)^{s(g_1)+\tilde{\rho}(g_2)}n^{(2)}_{\lambda}(g_1) + (-)^{s(g_1g_2)}(\act{g_1}{n}^{(2)}_{\lambda}(g_2)-n^{(2)}_{\lambda}(g_1g_2))) }\nonumber\\
    =&(-)^{\sigma(g_1,g_2)\sigma(g_3,g_4)}~\ee^{\ii (-)^{s(g_1g_2)} \alpha(g_3,g_4)(-)^{\rho(g_1g_2)}\sigma(g_1,g_2) }\nonumber\\
    =&(-)^{\sigma(g_1,g_2)\sigma(g_3,g_4)}~\ee^{\ii (-)^{\tilde\rho(g_1g_2)} \alpha(g_3,g_4)\sigma(g_1,g_2) }
    \label{eq:u1f_omega_prime_twist_3cocycle}
\end{align}
Thus, we recover Eq.~\eqref{eq:omegap_3-cocycle}.

\section{Details about solutions of tensor equations} \label{app:mpo_solution}
In this appendix, for a given symmetry group $G_f$ and fermionic three cocycle data $(\sigma,\omega)$, we construct $n_f$ and $W(g)$ based on this data, extract IGG/IGA $n_\lambda^{(a)}(g)$ and $\lambda^{(a)}(g_1,g_2)$ from them, and verify that they indeed satisfy tensor equations presented in Eq.~\eqref{eq:group_relation_internal_legs} and \eqref{eq:nlambda_lambda_consistent_eq}.

As in the main text, $G_b$ here is a finite group with order $\abs{G_b}$.
Dimension of an internal leg equals $\abs{G_b}^2$, whose basis are labeled as $\ket{l,r}$, where $l,r\in G_b$.
Internal legs points from bond tensors to site tensors.
We focus on FPEPS on a bipartite lattice, where sublattice indices are labeled as $u,v$.
In the following, we will consider these two sublattices separately.

\subsection{$u-$sublattice}
We first examine symmetry action on internal legs on $u-$sublattice. 
We will ignore internal leg index $(u\alpha)$ for simplicity.
Fermion charge number operator acts as
\begin{align}
    n_f\ket{l,r}=-\sigma(r,r^{-1}l)\ket{l,r}
    \label{eq:nf_u_action}
\end{align}
where $\sigma$ is the twisted 2-cocycle that satisfies Eq.~\eqref{eq:u1f_sigma_2-cocycle}.

$g\in G_b$ is represented as $W(g)$ on internal legs, which acts as
\begin{align}
    W(g)\ket{l,r}=(-)^{\sigma(g,r)\sigma(r,r^{-1}l)}\cdot \omega^{-1}(g,r,r^{-1}l) \ket{gl,gr}
    \label{eq:wg_from_mpo}
\end{align}
Here, $\omega$ is the given fermionic three cocycle that satisfies Eq.~\eqref{eq:u1f_fermionic_3-cocycle}

By acting $W(g)$ on $n_f$, we obtain
\begin{align}
    \act{g}{n}_f &= \sum_{l,r} -\sigma(r,r^{-1}l)\ket{gl,gr}\bra{gl,gr} = \sum_{l,r} -\sigma(g^{-1}r,r^{-1}l)\ket{l,r}\bra{l,r}\notag\\
    &= \sum_{l,r} (-)^{\rho(g)}\cdot \left(-\sigma(g,g^{-1}r) +\sigma(g,g^{-1}l) -\sigma(r,r^{-1}l)\right) \ket{l,r}\bra{l,r}\,.
\end{align}
where we use Eq.~\eqref{eq:u1f_sigma_2-cocycle} to get equation on the second line.
Since $\act{g}{n}_{f} = (-)^{\rho(g)}n_{f} + n^{(2)}_{\lambda}(g) +n^{(1)}_{\lambda}(g)$,
By comparing with the first line in Eq.~\eqref{eq:group_relation_internal_legs} and Eq.~\eqref{eq:nd_d_decomp_to_nlambda_lambda}, we identify
\begin{align}
    n^{(2)}_\lambda(g) = \sum_{l,r} (-)^{\rho(g)}\sigma(g,g^{-1}l)\ket{l,r}\bra{l,r}\,;\quad
    n^{(1)}_{\lambda}(g) = \sum_{l,r} (-)^{\rho(g)+1}\sigma(g,g^{-1}r)\ket{l,r}\bra{l,r}\,.
    \label{eq:nlambda_from_mpo_u}
\end{align}
It is now straightforward to check the first line in Eq.~\eqref{eq:nlambda_lambda_consistent_eq} by inserting the above definition of $n_\lambda^{(a)}(g)$ and using Eq.~\eqref{eq:u1f_sigma_2-cocycle}.

Instead of extracting $\lambda^{(a)}(g_1,g_2)$ from group relation in Eq.~\eqref{eq:group_relation_internal_legs}, \eqref{eq:nd_d_decomp_to_nlambda_lambda}, and Eq.~\eqref{eq:lambdab1b2_decomp}, here we directly construct $\lambda^{(a)}(g_1,g_2)$, and show they satisfy all tensor equations.
We start from expression for $\lambda^{(2)}$, which reads
\begin{align}
    \lambda^{(2)}(g_1,g_2) =& \sum_{l,r}\frac{(-)^{\sigma(g_2,g_2^{-1}g_1^{-1}l)\sigma(g_1,g_1^{-1}l)}}{\omega^{\prime}(g_1,g_2,g_2^{-1}g_1^{-1}l)}\exp\left[ \ii (-)^{\tilde\rho(g_1)}\alpha(g_2,g_2^{-1}g_1^{-1}l)\sigma(g_1,g_1^{-1}l) \right]\ket{l,r}\bra{l,r}\label{eq:u1f_lambda_2_mpo_form}\\
    =&\sum_{l,r}\frac{(-)^{\sigma(g_2,g_3l)\sigma(g_1,g_2g_3l)}}{\omega^{\prime}(g_1,g_2,g_3l)}\exp \left[\ii(-)^{\tilde\rho(g_1)}\alpha(g_2,g_3l)\sigma(g_1,g_2g_3l) \right]\ket{g_1g_2g_3l,g_1g_2g_3r}\bra{g_1g_2g_3l,g_1g_2g_3r}\nonumber\,.
\end{align}
where $\omega^{\prime}$ are related to $\omega$ by Eq.~\eqref{eq:omega_omegaprime}.

We are now ready to verify that such $\lambda^{(2)}$ satisfies Eq.~\eqref{eq:u1f_lam_omega_2}, where for later convenience, we rewrite it as
\begin{equation}
    \frac{\lambda^{(2)}(g_1,g_2)\lambda^{(2)}(g_1g_2,g_3)}{\act{g_1}{\lambda}^{(2)}(g_2,g_3)\lambda^{(2)}(g_1,g_2g_3)}\cdot \frac{\omega^{\prime}(g_1,g_2,g_3)}{(-)^{n^{(2)}_{\lambda}(g_1)\sigma(g_2,g_3)}}\cdot \ee^{-\ii(-)^{s(g_1)}\alpha(g_2,g_3)n_{\lambda}^{(2)}(g_1)}=1
    \label{eq:u1f_lam_omega_2_app}
\end{equation} 
We list all other terms in the above equations:
\begin{align}
    \lambda^{(2)}(g_1g_2,g_3) &= \sum_{l,r}\frac{(-)^{\sigma(g_3,l)\sigma(g_1g_2,g_3l)}}{\omega^{\prime}(g_1g_2,g_3,l)}\exp \left\{\ii(-)^{\tilde\rho(g_1g_2)}\alpha(g_3,l)\sigma(g_1g_2,g_3l) \right\} \ket{g_1g_2g_3l,g_1g_2g_3r}\bra{g_1g_2g_3l,g_1g_2g_3r}\nonumber\\
    \act{g_1}{\lambda}^{(2)}(g_2,g_3) &= \sum_{l,r}\frac{(-)^{\sigma(g_3,l)\sigma(g_2,g_3l)}}{[\omega^{\prime}(g_2,g_3,l)]^{(-)^{s(g_1)}}}\exp \left\{\ii(-)^{s(g_1)+\tilde\rho(g_2)}\alpha(g_3,l)\sigma(g_2,g_3l) \right\} \ket{g_1g_2g_3l,g_1g_2g_3r}\bra{g_1g_2g_3l,g_1g_2g_3r}\nonumber\\
    \lambda^{(2)}(g_1,g_2g_3) &= \sum_{l,r}\frac{(-)^{\sigma(g_2g_3,l)\sigma(g_1,g_2g_3l)}}{\omega^{\prime}(g_1,g_2g_3,l)}\exp \left\{\ii(-)^{\tilde\rho(g_1)}\alpha(g_2g_3,l)\sigma(g_1,g_2g_3l) \right\} \ket{g_1g_2g_3l,g_1g_2g_3r}\bra{g_1g_2g_3l,g_1g_2g_3r}\nonumber\\
    \ee^{\ii(-)^{s(g_1)}\alpha(g_2,g_3)n_{\lambda}^{(2)}(g_1)} &= \sum_{l,r}\exp \left\{\ii(-)^{\tilde\rho(g_1)}\alpha(g_2,g_3)\sigma(g_1,g_2g_3l) \right\}\ket{g_1g_2g_3l,g_1g_2g_3r}\bra{g_1g_2g_3l,g_1g_2g_3r}\nonumber\\
    (-)^{n^{(2)}_{\lambda}(g_1)\sigma(g_2,g_3)} &= \sum_{l,r}(-)^{\sigma(g_2,g_3)\sigma(g_1,g_2g_3l)}\ket{g_1g_2g_3l,g_1g_2g_3r}\bra{g_1g_2g_3l,g_1g_2g_3r}
\end{align}
Inserting above expressions into left-hand side of Eq.~\eqref{eq:u1f_lam_omega_2_app}, we obtain
\begin{align}
    &\sum_{l,r}\frac{\omega^{\prime}(g_1,g_2,g_3)\omega^{\prime}(g_1,g_2g_3,l)[\omega^{\prime}(g_2,g_3,l)]^{(-)^{s(g_1)}}}{\omega^{\prime}(g_1,g_2,g_3l)\omega^{\prime}(g_1g_2,g_3,l)}\cdot \frac{(-)^{\sigma(g_2,g_3l)\sigma(g_1,g_2g_3l)+\sigma(g_3,l)\sigma(g_1g_2,g_3l)}}{(-)^{\sigma(g_2,g_3)\sigma(g_1,g_2g_3l)+\sigma(g_3,l)\sigma(g_2,g_3l)+\sigma(g_2g_3,l)\sigma(g_1,g_2g_3l)}}\nonumber\\
    &\times \frac{\exp\left\{\ii (-)^{\tilde\rho(g_1)}\alpha(g_2,g_3l)\sigma(g_1,g_2g_3l)+\ii(-)^{\tilde\rho(g_1g_2)}\alpha(g_3,l)\sigma(g_1g_2,g_3l) \right\} }{\exp \left\{\ii(-)^{s(g_1)+\tilde\rho(g_2)}\alpha(g_3,l)\sigma(g_2,g_3l) + \ii(-)^{\tilde\rho(g_1)}\alpha(g_2g_3,l)\sigma(g_1,g_2g_3l)+\ii(-)^{\tilde\rho(g_1)}\alpha(g_2,g_3)\sigma(g_1,g_2g_3l) \right\}}\nonumber\\
    &\ket{g_1g_2g_3l,g_1g_2g_3r}\bra{g_1g_2g_3l,g_1g_2g_3r}\notag\\
    ={}&\sum_{l,r} \frac{\exp\left\{ {\color{red}{\ii(-)^{\tilde\rho(g_1)}\alpha(g_2,g_3l)\sigma(g_1,g_2g_3l)}}{\color{blue}{+\ii(-)^{\tilde\rho(g_1g_2)}\alpha(g_3,l)\sigma(g_1g_2,g_3l)+\ii(-)^{\tilde\rho(g_1g_2)}\alpha(g_3,l)\sigma(g_1,g_2)}} \right\} }{\exp \left\{{\color{blue}{\ii(-)^{s(g_1)+\tilde\rho(g_2)}\alpha(g_3,l)\sigma(g_2,g_3l)}} {\color{red}{+ \ii(-)^{\tilde\rho(g_1)}\alpha(g_2g_3,l)\sigma(g_1,g_2g_3l)+\ii(-)^{\tilde\rho(g_1)}\alpha(g_2,g_3)\sigma(g_1,g_2g_3l)}} \right\}}\nonumber\\
    \times&\frac{(-)^{{\color{violet}{\sigma(g_2,g_3l)\sigma(g_1,g_2g_3l)}}{\color{cyan}{+\sigma(g_3,l)\sigma(g_1g_2,g_3l)+\sigma(g_3,l)\sigma(g_1,g_2)}}}}{(-)^{{\color{cyan}{\sigma(g_3,l)\sigma(g_2,g_3l)}}{\color{violet}{+\sigma(g_2,g_3)\sigma(g_1,g_2g_3l)+\sigma(g_2g_3,l)\sigma(g_1,g_2g_3l)}}}}\ket{g_1g_2g_3l,g_1g_2g_3r}\bra{g_1g_2g_3l,g_1g_2g_3r}\notag\\
    ={}&\sum_{l,r}\frac{{\color{blue}{\exp \left\{\ii (-)^{\tilde\rho(g_1g_2)}\alpha(g_3,l)\sigma(g_1,g_2g_3l) \right\}}}}{{\color{red}{\exp \left\{\ii (-)^{\tilde\rho(g_1g_2)}\alpha(g_3,l)\sigma(g_1,g_2g_3l)  \right\}}}}\frac{{\color{cyan}{(-)^{\sigma(g_3,l)\sigma(g_1,g_2g_3l)}}}}{{\color{violet}{(-)^{\sigma(g_3,l)\sigma(g_1,g_2g_3l)}}}}\ket{g_1g_2g_3l,g_1g_2g_3r}\bra{g_1g_2g_3l,g_1g_2g_3r}
    =1
\end{align}
where we use three cocycle condition defined in Eq.~\eqref{eq:omegap_3-cocycle}, and then merge terms with the same color by two cocycle condition.

Similarly, $\lambda^{(1)}$ reads
\begin{align}
     \lambda^{(1)}(g_1,g_2) = \sum_{l,r}\frac{\omega^{\prime}(g_1,g_2,g_2^{-1}g_1^{-1}r)}{(-)^{\sigma(g_2,g_2^{-1}g_1^{-1}r)\sigma(g_1,g_1^{-1}r)}}\exp\left\{- \ii (-)^{\tilde\rho(g_1)}\alpha(g_2,g_2^{-1}g_1^{-1}r)\sigma(g_1,g_1^{-1}r) \right\}\ket{l,r}\bra{l,r}
     \label{eq:u1f_lambda_1_mpo_form}
\end{align}

Now let us check that Eq.~\eqref{eq:wg_from_mpo}, \eqref{eq:u1f_lambda_2_mpo_form} and Eq.~\eqref{eq:u1f_lambda_1_mpo_form} satisfy the second line in Eq.~\eqref{eq:group_relation_internal_legs}, where we rewrite here for convenience:
\begin{equation}
    W_{\beta_1}(g_1)\cdot\act{K^{s(g_1)}} {W}_{\beta_2}(g_2)=D_{(\beta_1,\beta_2)}(g_1,g_2)\cdot\exp[\ii\alpha(g_1,g_2)\cdot n_{f}]\cdot W(g_1g_2)\,,
    \label{eq:u1f_wg_fusion_app}
\end{equation}
The left-hand side equals
\begin{align}
   \io{2}{1}{g_1}\io{1}{1}{g_1}W(g_1)\io{2}{2}{g_2}\io{1}{2}{g_2}\act{K^{s(g_1)}}{W}(g_2)=\io{2}{1}{g_1}\io{1}{1}{g_1}\act{g_1}{\io{2}{2}{g_2}}\act{g_1}{\io{1}{2}{g_2}}W(g_1)\act{K^{s(g_1)}}{W(g_2)}\,,
\end{align}
while the right-hand side is 
\begin{align}
    &\lambda^{(2)}(g_1,g_2)\tikzmark{io}{\act{g_1}{\io{2}{2}{g_2}}}\io{2}{1}{g_1}\io{1}{1}{g_1}\tikzmark{end}{}\act{g_1}{\io{1}{2}{g_2}}\lambda^{(1)}(g_1,g_2)\ee^{\ii\alpha(g_1,g_2)n_f}W(g_1g_2)\nonumber\\
    =&\io{2}{1}{g_1}\io{1}{1}{g_1}\act{g_1}{\io{2}{2}{g_2}}\act{g_1}{\io{1}{2}{g_2}}(-)^{\act{g_1}{n}_{\lambda}^{(2)}(g_2)[n^{(2)}_{\lambda}(g_1)+n^{(1)}_{\lambda}(g_1)]}\lambda^{(2)}(g_1,g_2)\lambda^{(1)}(g_1,g_2)\ee^{\ii\alpha(g_1,g_2)n_f}W(g_1g_2)\,.
\begin{tikzpicture}[remember picture,overlay,>=stealth]
    \draw[->,blue] (io.north) -- ++(0,5pt)  -| (end.north);
\end{tikzpicture}
\end{align}
Therefore, we should check the following equation:
\begin{align}
    W(g_1)\act{K^{s(g_1)}}{W(g_2)} = (-)^{\act{g_1}{n}_{\lambda}^{(2)}(g_2)[n^{(2)}_{\lambda}(g_1)+n^{(1)}_{\lambda}(g_1)]}\cdot \lambda^{(2)}(g_1,g_2)\lambda^{(1)}(g_1,g_2)\ee^{\ii\alpha(g_1,g_2)n_{f}}W(g_1g_2)\,.
    \label{eq:u1f_W_1_2_12_coeff}
\end{align}
Putting in the definition of $W(g)$ and $\lambda^{(a)}$, we obtain the left hand side of Eq.~\eqref{eq:u1f_W_1_2_12_coeff} as
\begin{align}
    \sum_{l,r}\frac{(-)^{\sigma(g_1,g_2r)\sigma(g_2r,r^{-1}l)}}{\omega(g_1,g_2r,r^{-1}l)}\frac{(-)^{\sigma(g_2,r)\sigma(r,r^{-1}l)}}{[\omega(g_2,r,r^{-1}l)]^{(-)^{s(g_1)}}}\ket{g_1g_2l,g_1g_2,r}\bra{l,r}\,,
    \label{eq:u1f_W_1_2_12_coeff_lhs}
\end{align}
and the right hand side with the following lengthy expression
\begin{align}
    &\sum_{l,r}(-)^{\sigma(g_2,l)[\sigma(g_1,g_2l)+\sigma(g_1,g_2r)]}\cdot \frac{(-)^{\sigma(g_2,l)\sigma(g_1,g_2l)}}{\omega^{\prime}(g_1,g_2,l)}\cdot \omega^{\prime}(g_1,g_2,r)(-)^{\sigma(g_2,r)\sigma(g_1,g_2r)}\cdot \frac{(-)^{\sigma(g_1g_2,r)\sigma(r,r^{-1}l)}}{\omega(g_1g_2,r,r^{-1}l)}\\
    &\times\exp \left\{ \ii(-)^{\tilde\rho(g_1)}[\alpha(g_2,l)\sigma(g_1,g_2l)-\alpha(g_2,r)\sigma(g_1,g_2r)] \right\}\exp \left\{-\ii\alpha(g_1,g_2)\sigma(g_1g_2r,r^{-1}l) \right\}\ket{g_1g_2l,g_1g_2r}\bra{l,r}\nonumber\,.
\end{align}
We now use the relation between $\omega$ and $\omega^{\prime}$ in Eq.~\eqref{eq:omega_omegaprime}, where the detailed derivations are presented in Sec.VI of SM\footnotemark[10]:
The right-hand side of Eq.~\eqref{eq:u1f_W_1_2_12_coeff} now becomes
\begin{align}
    &\sum_{l,r}(-)^{[\sigma(g_2,l)+\sigma(g_2,r)]\sigma(g_1,g_2r)+\sigma(g_1g_2,r)\sigma(r,r^{-1}l)}\cdot\frac{\omega(g_1,g_2,r)}{\omega(g_1,g_2,l)\omega(g_1g_2,r,r^{-1}l)} \cdot \ee^{\ii \alpha(g_1,g_2)[\sigma(g_1g_2,l)-\sigma(g_1g_2,r)-\sigma(g_1g_2r,r^{-1}l)] }\notag\\
    &\cdot \ket{g_1g_2l,g_1g_2r}\bra{l,r}\nonumber\\
    ={}&\sum_{l,r}(-)^{[\sigma(g_2,l)+\sigma(g_2,r)]\sigma(g_1,g_2r)+\sigma(g_1g_2,r)\sigma(r,r^{-1}l)}\cdot \frac{\omega(g_1,g_2,r)\omega(g_1,g_2r,r^{-1}l)[\omega(g_2,r,r^{-1}l)]^{(-)^{s(g_1)}}}{\omega(g_1,g_2,l)\omega(g_1g_2,r,r^{-1}l)}\nonumber\\
    &\cdot \omega^{-1}(g_1,g_2r,r^{-1}l)[\omega^{-1}(g_2,r,r^{-1}l)]^{(-)^{s(g_1)}}\cdot \exp\left\{-\ii(-)^{\rho(g_1g_2)}\alpha(g_1,g_2)\sigma(r,r^{-1}l) \right\}\cdot \ket{g_1g_2l,g_1g_2r}\bra{l,r}\notag\\
    ={}&\sum_{l,r}\frac{(-)^{[\sigma(g_2,l)+\sigma(g_2,r)]\sigma(g_1,g_2r)}(-)^{[\sigma(g_1g_2,r)+\sigma(g_1,g_2)]\sigma(r,r^{-1}l)}}{\omega(g_1,g_2r,r^{-1}l)[\omega(g_2,r,r^{-1}l)]^{(-)^{s(g_1)}}}\ket{g_1g_2l,g_1g_2r}\bra{l,r}\nonumber\\
    ={}&\sum_{l,r}\frac{(-)^{\sigma(g_1,g_2r)\left [\sigma(g_2,l)+\sigma(g_2,r)+\sigma(r,r^{-1}l)\right ]}(-)^{\sigma(g_2,r)\sigma(r,r^{-1}l)}}{\omega(g_1,g_2r,r^{-1}l)[\omega(g_2,r,r^{-1}l)]^{(-)^{s(g_1)}}}\ket{g_1g_2l,g_1g_2r}\bra{l,r}\nonumber\\
    ={}& \sum_{l,r}\frac{(-)^{\sigma(g_1,g_2r)\sigma(g_2r,r^{-1}l)}(-)^{\sigma(g_2,r)\sigma(r,r^{-1}l)}}{\omega(g_1,g_2r,r^{-1}l)[\omega(g_2,r,r^{-1}l)]^{(-)^{s(g_1)}}}\ket{g_1g_2l,g_1g_2r}\bra{l,r}\,,
\end{align}
where we apply three cocycle condition for $\omega$ to get the second equation.
Comparing the above equation with Eq.~\eqref{eq:u1f_W_1_2_12_coeff_lhs}, we thus prove Eq.~\eqref{eq:u1f_W_1_2_12_coeff}. 

So, Eq.~\eqref{eq:wg_from_mpo} is indeed a solution for tensor equations with given fermionic three cocycle $(\sigma,\omega)$.

\subsection{$v-$sublattice}
The construction for $W(g)$ on internal legs of $v-$sublattice is similar to that on internal legs of $u-$sublattice presented in the previous part.
We also ignore internal leg index $(v\alpha)$ for simplicity.

Fermion charge number operator reads
\begin{align}
    n_f = \sum_{l,r}\sigma(l,l^{-1}r)\ket{l,r}\bra{l,r}
    \label{eq:nf_v_action}
\end{align}
where the sign difference comes from the opposite charges on different sublattices, and the exchange between $l$ and $r$ comes from the different definitions of local Hilbert space for different sublattices.

Internal leg $g$-action
\begin{align}
    W(g)=\sum_{l,r}\frac{\omega(g,l,l^{-1}r)}{(-)^{\sigma(g,r)\sigma(l,l^{-1}r)}}\ket{gl,gr}\bra{l,r} \,.
\label{eq:wg_from_mpo_v}
\end{align}

By acting $W(g)$ on $n_f$, we obtain
\begin{align}
    \act{g}{n}_f &= \sum_{l,r} \sigma(g^{-1}l,l^{-1}r) \ket{l,r}\bra{l,r}\nonumber\\
                 &= \sum_{l,r} (-)^{\rho(g)}\left( \sigma(l,l^{-1}r) + \sigma(g,g^{-1}l) -\sigma(g,g^{-1}r) \right) \ket{l,r}\bra{l,r}\,.
\end{align}
Therefore we can assign
\begin{align}
    n^{(2)}_{\lambda}(g) = \sum_{l,r} (-)^{\rho(g)}\sigma(g,g^{-1}l)\ket{l,r}\bra{l,r}\,,\quad
    n^{(1)}_{\lambda}(g) = \sum_{l,r} (-)^{\rho(g)+1}\sigma(g,g^{-1}r)\ket{l,r}\bra{l,r}\,,
    \label{eq:nlambda_from_mpo_v}
\end{align}
which shares the same form as those for the $u-$sublattice.

$\lambda(g_1,g_2)$'s for the $v$-sublattice can be taken as the same form as those for the $u-$sublattice, and Eq.~\eqref{eq:u1f_lam_omega_2_app} will also be satisfied.
To see this, we point out that that for Eq.~\eqref{eq:u1f_lam_omega_2_app}, the only subtlety between $u-$ and $v-$sublattice is $g_1-$action on $\lambda^{(2)}(g_2,g_3)$.
However, it is easy to check that such subtlety will not make any difference for these two sublattices, and therefore Eq.~\eqref{eq:u1f_lam_omega_2} works for $v-$sublattice as well.

Again we want to check that $W(g)$'s and $\lambda^{(a)}$'s proposed here satisfy Eq.~\eqref{eq:u1f_wg_fusion_app}.
We shall repeat what we have done in the last subsection. 
Fusion of two symmetry elements gives
\begin{align}
    W_{\beta_1}(g_1)\cdot\act{K^{s(g_1)}}{W}_{\beta_2}(g_2) &= \io{2}{1}{g_1}\io{1}{1}{g_1}\act{g_1}{\io{2}{2}{g_2}}\act{g_1}{\io{1}{2}{g_2}}W(g_1)\act{K^{s(g_1)}}{W(g_2)}
\end{align}
where
\begin{align}
    W(g_1)\cdot \act{K^{s(g_1)}}{W(g_2)}&= \sum_{l,r}\frac{\omega(g_1,g_2l,l^{-1}r)}{(-)^{\sigma(g_1,g_2r)\sigma(g_2l,l^{-1}r)}}\cdot \frac{[\omega(g_2,l,l^{-1}r)]^{(-)^{s(g_1)}}}{(-)^{\sigma(g_2,r)\sigma(l,l^{-1}r)}}\ket{g_1g_2l,g_1g_2r}\bra{l,r}\,.
    \label{eq:u1f_W_1_2_12_coeff_lhs_v}
\end{align}
For the right-hand side of Eq.~\eqref{eq:u1f_wg_fusion_app}, we have
\begin{align}
    &D_{(\beta_1,\beta_2)}(g_1,g_2)\cdot\exp\{ \ii\alpha(g_1,g_2)\cdot n_{f} \}\cdot W(g_1g_2)\nonumber\\
    =&\io{2}{1}{g_1}\io{1}{1}{g_1}\act{g_1}{\io{2}{2}{g_2}}\act{g_1}{\io{1}{2}{g_2}}\cdot (-)^{\act{g_1}{n}_{\lambda}^{(2)}(g_2)[n^{(2)}_{\lambda}(g_1)+n^{(1)}_{\lambda}(g_1)]}\cdot \lambda^{(2)}(g_1,g_2)\lambda^{(1)}(g_1,g_2)\cdot \ee^{\ii\alpha(g_1,g_2)n_f}\cdot W(g_1g_2)
\end{align}
in which
\begin{align}
    &(-)^{\act{g_1}{n}_{\lambda}^{(2)}(g_2)[n^{(2)}_{\lambda}(g_1)+n^{(1)}_{\lambda}(g_1)]}\cdot \lambda^{(2)}(g_1,g_2)\lambda^{(1)}(g_1,g_2)\cdot \ee^{\ii\alpha(g_1,g_2)n_f}\cdot W(g_1g_2)\nonumber\\
    ={}& \sum_{l,r}(-)^{\sigma(g_2,l)[\sigma(g_1,g_2l)+\sigma(g_1,g_2r)]}\cdot \frac{(-)^{\sigma(g_2,l)\sigma(g_1,g_2l)}}{\omega^{\prime}(g_1,g_2,l)}\cdot \omega^{\prime}(g_1,g_2,r)\cdot (-)^{\sigma(g_2,r)\sigma(g_1,g_2r)}\cdot \frac{\omega(g_1g_2,l,l^{-1}r)}{(-)^{\sigma(g_1g_2,r)\sigma(l,l^{-1}r)}}\\
    &\times\exp \left\{ \ii(-)^{\tilde\rho(g_1)}[\alpha(g_2,l)\sigma(g_1,g_2l)-\alpha(g_2,r)\sigma(g_1,g_2r)] \right\}\cdot \exp \left[\ii\alpha(g_1,g_2)\sigma(g_1g_2l,l^{-1}r) \right]\cdot \ket{g_1g_2l,g_1g_2,r}\bra{l,r}\nonumber\\
    ={}&\sum_{l,r} \frac{(-)^{\sigma(g_2,l)[\sigma(g_1,g_2l)+\sigma(g_1,g_2r)]}}{(-)^{\sigma(g_1g_2,r)\sigma(l,l^{-1}r)}}\cdot \frac{\omega(g_1g_2,l,l^{-1}r)\omega(g_1,g_2,r)}{\omega(g_1,g_2,l)\omega(g_1,g_2l,l^{-1}r)[\omega(g_2,l,l^{-1}r)]^{(-)^{s(g_1)}}}\nonumber\\
    &\times\exp \left\{ -\ii(-)^{\tilde{\rho}(g_1g_2)}\alpha(g_1,g_2)\sigma(l,l^{-1}r)\right\}\cdot \omega(g_1,g_2l,l^{-1}r)[\omega(g_2,l,l^{-1}r)]^{(-)^{s(g_1)}}\ket{g_1g_2l,g_1g_2,r}\bra{l,r}\\
    ={}&\sum_{l,r} \frac{(-)^{\sigma(g_2,l)[\sigma(g_1,g_2l)+\sigma(g_1,g_2r)]}}{(-)^{\sigma(l,l^{-1}r)[\sigma(g_1g_2,r)+\sigma(g_1,g_2)]}}\cdot \omega(g_1,g_2l,l^{-1}r)[\omega(g_2,l,l^{-1}r)]^{(-)^{s(g_1)}}\ket{g_1g_2l,g_1g_2,r}\bra{l,r}\nonumber\\
    =&\sum_{l,r} \frac{\omega(g_1,g_2l,l^{-1}r)}{(-)^{\sigma(g_1,g_2r)\sigma(g_2l,l^{-1}r)}}\cdot \frac{[\omega(g_2,l,l^{-1}r)]^{(-)^{s(g_1)}}}{(-)^{\sigma(g_2,r)\sigma(l,l^{-1}r)}}\ket{g_1g_2l,g_1g_2r}\bra{l,r}\,,
\end{align}
which equals Eq.~\eqref{eq:u1f_W_1_2_12_coeff_lhs_v}.
So, we conclude that Eq.~\eqref{eq:u1f_wg_fusion_app} is also satisfied on $v-$sublattice.

\section{Details about $\Omega_{a/b}(g_1,g_2)$}\label{app:Omega_ab}
\subsection{Explicit constructions of $\Omega_{a/b}$}
In this section let us derive the explicit form of $\Omega_a(g_1,g_2)$ and $\Omega_b(g_1,g_2)$.
The input of the derivation is only the tensor equations that we explained and introduced in Section \ref{sec:tensor_equation}.
For a symmetry action $\forall g\in G_b$, the MPO form of this symmetry operation on a segment of edge $M$ ($1,\,2,\,\cdots,\, l$) is given by:
\begin{equation}
    U_M(g) = \pe\cdot d_{\beta}^{-1}\fTr \left[\Tr_\beta \left[w_{_{\beta,l+1}}(g)\otimes_f w_{_{\beta,L}}(g)\BOF{j=1}{l}W_{\beta,j}(g) \right] \right]K^{s(g)}\cdot\pe\,,
\end{equation}
in which
\begin{align}
    W_{\beta,j}(g) = I_{\beta,j}(g)\cdot W_j(g);~w_{_{\beta,L}}(g) = \iota_{\beta,\frac{1}{2}}^{(1)}(g)\cdot w_L(g);~w_{_{\beta,l+1}}(g) = \iota_{\beta,l+\frac{1}{2}}^{(2)}(g)\cdot w_{l+1}(g)\,.
\end{align}
The fusion of two symmetry actions on the same segment $M$ is then
\begin{align}
    U_M(g_1)\cdot U_M(g_2)=\pe\cdot d_{\beta}^{-1}\fTr \Bigg[ \Tr_{\beta}\Bigg[&(w_{_{\beta_1,l+1}}(g_1) \act{K^{s(g_1)}}{w}_{_{\beta_2,l+1}}(g_2))\otimes_f(w_{_{\beta_1,L}}(g_1) \act{K^{s(g_1)}}{w}_{_{\beta_2,L}}(g_2))\nonumber\\
    &\BOF{j=1}{l}W_{\beta_1,j}(g_1)\act{K^{s(g_1)}}{W}_{\beta_2,j}(g_2) \Bigg] \Bigg]K^{s(g_1g_2)} \cdot\pe\,.
\end{align}
Locally, a fusion of two symmetry actions gives the IGG $D_{(\beta_1,\beta_2),j}(g_1,g_2)$, that introduced in section \ref{sec:tensor_equation} Eq.~\eqref{eq:u1f_wg_fusion}
\begin{equation}
    W_{\beta_1,j}(g_1)K^{s(g_1)}\cdot W_{\beta_2,j}(g_2)K^{s(g_2)}=D_{(\beta_1,\beta_2),j}(g_1,g_2)\cdot\exp\{ \ii\alpha(g_1,g_2)\cdot n_{f,j} \}\cdot W_{j}(g_1g_2)K^{s(g_1g_2)}\,,
\end{equation}
such fusion is then
\begin{align}
    U_M(g_1)\cdot U_M(g_2)=\pe\cdot d_{\beta}^{-1}\fTr \Bigg[ \Tr_{\beta}\Bigg[&(w_{_{\beta_1,l+1}}(g_1) \act{K^{s(g_1)}}{w}_{_{\beta_2,l+1}}(g_2))\otimes_f(w_{_{\beta_1,L}}(g_1) \act{K^{s(g_1)}}{w}_{_{\beta_2,L}}(g_2))\nonumber\\
    &\BOF{j=1}{l}D_{(\beta_1,\beta_2),j}(g_1,g_2)\cdot\exp\{\ii\alpha(g_1,g_2)n_{_{f,j}}\}\cdot W_j(g_1,g_2) \Bigg] \Bigg]K^{s(g_1g_2)} \cdot\pe\,.
\end{align}
By separating $D_{(\beta_1,\beta_2)}(g_1,g_2)$'s into plaquette IGG $\Lambda$'s, most of the plaquette terms will be absorbed by $\pe$, and consequently we can rewrite the such segment fusion as
\begin{align}
    U_M(g_1)\cdot U_M(g_2)=\pe\cdot d_{\beta}^{-1}\fTr \Bigg[& \Tr_{\beta}\Bigg[(w_{_{\beta_1,l+1}}(g_1) \act{K^{s(g_1)}}{w}_{_{\beta_2,l+1}}(g_2))\otimes_f(w_{_{\beta_1,L}}(g_1) \act{K^{s(g_1)}}{w}_{_{\beta_2,L}}(g_2))\Lambda^{(2)}_{(\beta_1,\beta_2),\frac{1}{2}}(g_1,g_2)\nonumber\\
    &\Lambda^{(1)}_{(\beta_1,\beta_2),l+\frac{1}{2}}(g_1,g_2)\exp\{\ii\alpha\,n_{_{f,M}} \}\cdot W_j(g_1,g_2)  \Bigg] \Bigg]K^{s(g_1g_2)} \cdot\pe\,,
\end{align}
with $n_{_{f,M}} = \sum_{i=1}^{l}n_{_{f,i}}$.
For a symmetry group element $(g,\alpha)\in G_f$, its symmetry action on a segment of the edge is given by
\begin{equation}
    \exp\{\ii \alpha\, (n_{_{f,L}} + n_{_{f,l+1}})\}  
    \exp\{\ii \alpha\, n_{_{f,M}}\}\cdot U_M(g)\nonumber
    = \exp\{\ii \alpha\,n_{_{f,\overline{M}}}\}\cdot U_{M}(g)\,,
\end{equation}
since we want to ask that $(g,0)\cdot(e,\alpha) = (g,(-)^{\tilde{\rho}(g)}\alpha)$ also holds on a segment, such that
\begin{align}
    \left(U_M(g)K^{s(g)}\right)\cdot \exp\{\ii \alpha\, n_{_{f,\overline{M}}}\} \cdot \left(U_M(g)K^{s(g)}\right)^{-1} =\exp \{\ii (-)^{\tilde{\rho}(g)}\alpha\, n_{_{f,\overline{M}}}\}\,.
\end{align}
The fusion is then related to $U_M(g_1g_2)$ as
\begin{align}
    &U_M(g_1)\cdot U_M(g_2)=\pe\cdot d_{\beta}^{-1}\fTr \Bigg[ \Tr_{\beta}\Bigg[(w_{_{\beta_1,l+1}}(g_1) \act{K^{s(g_1)}}{w}_{_{\beta_2,l+1}}(g_2))\otimes_f(w_{_{\beta_1,L}}(g_1) \act{K^{s(g_1)}}{w}_{_{\beta_2,L}}(g_2))\Lambda^{(2)}_{(\beta_1,\beta_2),\frac{1}{2}}(g_1,g_2)\nonumber\\
    &\Lambda^{(1)}_{(\beta_1,\beta_2),l+\frac{1}{2}}(g_1,g_2)w^{-1}_L(g_1g_2) w^{-1}_{l+1}(g_1g_2)\ee^{-\ii\alpha(g_1,g_2)\,(n_{_{f,L}}+n_{_{f,l+1}})}\ee^{\ii\alpha(g_1,g_2)n_{_{f,\overline{M}}}}\cdot U_M(g_1g_2)  \Bigg] \Bigg]K^{s(g_1g_2)} \cdot\pe\,.
\end{align}

Finally we need to deal with additional boundary operators $w_L$'s and $w_{l+1}$'s. For $w_L$'s, we have
\begin{align}
    P_{\frac{1}{2}}\cdot w_{_{\beta_1,L}}(g_1) \act{K^{s(g_1)}}{w}_{_{\beta_2,L}}(g_2) =& P_{\frac{1}{2}}\cdot  \iota_{\beta_1,\frac{1}{2}}^{(1)}(g_1) w_L(g_1)~ \iota_{\beta_2,\frac{1}{2}}^{(1)}(g_2) \act{K^{s(g_1)}}{w_L(g_2)}\nonumber\\
    =& P_{\frac{1}{2}}\cdot \iota_{\beta_1,\frac{1}{2}}^{(1)}(g_1) \act{g_1}{\iota_{\beta_2,\frac{1}{2}}^{(1)}(g_2)}w_L(g_1) \act{K^{s(g_1)}}{w_L(g_2)}\,,
\end{align}
which can be perfectly absorbed together with $\Lambda^{(2)}_{(\beta_1,\beta_2),\frac{1}{2}}(g_1,g_2)$,
\begin{align}
    & P_{\frac{1}{2}}\cdot \iota_{\beta_1,\frac{1}{2}}^{(1)}(g_1) \act{g_1}{\iota_{\beta_2,\frac{1}{2}}^{(1)}(g_2)}\left( w_L(g_1) \act{K^{s(g_1)}}{w_L(g_2)}\right)\Lambda^{(2)}_{(\beta_1,\beta_2),\frac{1}{2}}(g_{1},g_{2}) \nonumber\\
    =& P_{\frac{1}{2}}\cdot  \left(w_L(g_1) \act{K^{s(g_1)}}{w_L(g_2)} \right)\lambda^{(2)}_{\frac{1}{2}}(g_1,g_2)\,.
\end{align}
Thus we can conclude that $\Omega_b(g_1,g_2)$ is
\begin{equation}
    \Omega_b(g_1,g_2) = \pe \cdot w_L(g_1) \act{K^{s(g_1)}}{w_L(g_2)}  w^{-1}_L(g_1g_2) \ee^{-\ii\alpha(g_1,g_2)\,n_{_{f,L}}}\,\lambda^{(2)}_{\frac{1}{2}}(g_1,g_2)  \cdot \pe
\end{equation}
Similarly, we can repeat what we have done for $w_{l+1}$'s, and we got
\begin{equation}
    \Omega_a(g_1,g_2) = \pe \cdot (-)^{\act{g_1}{n}_{\lambda,l+\frac{1}{2}}^{(2)}(g_2)\,{n}_{\lambda,l+\frac{1}{2}}^{(2)}(g_1)} w_{l+1}(g_1) \act{K^{s(g_1)}}{w}_{l+1}(g_2) w^{-1}_{l+1}(g_1g_2)\ee^{-\ii\alpha(g_1,g_2)\,n_{_{f,l+1}}}\,\lambda^{(1)}_{l+\frac{1}{2}}(g_1,g_2) \cdot \pe\,.
\end{equation}

\subsection{Fermion number of $\Omega_{a/b}$ }\label{subapp:2-cocycle_defects_fpeps}
We already know 
\begin{align*}
    [w_L(g),n_{_{f,L}}]_{\rho(g)} = n_{\lambda,\frac{1}{2}}^{(1)}(g)\cdot w_L(g)\,,
\end{align*}
then the commutation between $w^{-1}_L(g)$ and $n_{_{f,L}}$ should be
\begin{align}
    [w^{-1}_L(g),n_{f,L}]_{\rho(g)} = -(-)^{\rho(g)}\act{g^{-1}}{n}_{\lambda,\frac{1}{2}}^{(1)}(g)\cdot w^{-1}_L(g) \,,
\end{align}
such that
\begin{align*}
    w^{-1}_L(g)w_L(g) n_{_{f,L}} w^{-1}_L(g)w_L(g) = n_{_{f,L}}
\end{align*}
is satisfied.
The commutation between $\Omega_b(g_1,g_2)$ and $n_{_{f,L}}$ is given through
\begin{align}
    &w_L(g_1)w_L(g_2)w^{-1}_L(g_1g_2)\cdot n_{_{f,L}}\nonumber\\
    =&w_L(g_1)w_L(g_2)\cdot \left( (-)^{\rho(g_1g_2)}\left(n_{_{f,L}} -\act{g_2^{-1}g_1^{-1}}{n}_{\lambda,\frac{1}{2}}^{(1)}(g)\right)\right)\cdot w^{-1}_L(g_1g_2) \nonumber\\
    =&w_L(g_1) \cdot \left((-)^{\rho(g_1g_2)}\left((-)^{\rho(g_2)}n_{_{f,L}} + n_{\lambda,\frac{1}{2}}^{(1)}(g_2) - \act{g_1^{-1}}{n}^{(1)}_{\lambda,\frac{1}{2}}(g_1g_2) \right) \right)\cdot w_L(g_2)w^{-1}_L(g_1g_2)\nonumber\\
    =&(-)^{\rho(g_1g_2)}\left((-)^{\rho(g_1g_2)}n_{_{f,L}} + (-)^{\rho(g_2)}n_{\lambda,\frac{1}{2}}^{(1)}(g_1) + \act{g_1}{n}_{\lambda,\frac{1}{2}}^{(1)}(g_2) - n^{(1)}_{\lambda,\frac{1}{2}}(g_1g_2)  \right)\cdot w_L(g_1)w_L(g_2)w^{-1}_L(g_1g_2)\nonumber\\
    =&(n_{_{f,L}}-\sigma(g_1,g_2))\cdot w_L(g_1)w_L(g_2)w^{-1}_L(g_1g_2)\nonumber\\
    \Rightarrow&[w_L(g_1)w_L(g_2)w^{-1}_L(g_1g_2),n_{_{f,L}}] = -\sigma(g_1,g_2)w_L(g_1)w_L(g_2)w^{-1}_L(g_1g_2)\nonumber\\
    \Rightarrow& [\Omega_b(g_1,g_2),n_{_{f,L}}]=-\sigma(g_1,g_2)\Omega_b(g_1,g_2)\,.
    \label{eq:Omega_ab_fermion_number_app}
\end{align}
In the equations above, condition given by Eq.~\eqref{eq:u1f_n_lam_fusion} is used.
Similarly for $\Omega_a(g_1,g_2)$ we have
\begin{align}
    [\Omega_a(g_1,g_2),n_{_{f,L}}] = +\sigma(g_1,g_2)\Omega_a(g_1,g_2)\,.
\end{align}
Thus we recover the fermion number condition in Eq.~\eqref{eq:Omega_ab_fermion_number}.

\subsection{3-cocycle data $\omega(g_1,g_2,g_3)$}\label{subapp:3-cocycle_defects_fpeps}
To obtain the condition of 3-cocycle $\omega$ (Eq.~\eqref{eq:omegaab_twist_two-cocycle}), a fusion of three symmetry actions to the same segment is needed: $\forall g_{1,2,3}\in G_b$
\begin{align}
    &U_M(g_1)K^{s(g_1)}\cdot U_M(g_2)K^{s(g_2)}\cdot U_M(g_3)K^{s(g_3)}\nonumber\\
    = &\Omega(g_1,g_2)\cdot\Omega(g_1g_2,g_3)\cdot \ee^{\ii \left(\alpha(g_1,g_2)+\alpha(g_1g_2,g_3)\right)n_{_{f,M}}}\cdot U_M(g_1g_2g_3)K^{s(g_1g_2g_3)}\label{eq:UM_fusion_1}\,,
\end{align}
while we can also fuse $g_2$ and $g_3$ first:
\begin{align}
    &U_M(g_1)K^{s(g_1)}\cdot U_M(g_2)K^{s(g_2)}\cdot U_M(g_3)K^{s(g_3)}\nonumber\\
    =& \act{U_M(g_1)K^{s(g_1)}}{\Omega(g_2,g_3)}\cdot\Omega(g_1,g_2g_3)\cdot \ee^{\ii \left((-)^{\tilde{\rho}(g_1)}\alpha(g_2,g_3) + \alpha(g_1,g_2g_3) \right)n_{_{f,M}} }\cdot U_M(g_1g_2g_3)K^{s(g_1g_2g_3)}\,.
    \label{eq:UM_fusion_2}
\end{align}
Equating Eq.~\eqref{eq:UM_fusion_1} and \eqref{eq:UM_fusion_2}, we obtain a condition that $\Omega$ should satisfy:
\begin{align}
    \Omega(g_1,g_2)\cdot \Omega(g_1g_2,g_3)\cdot \pe = \act{U_M(g_1)K^{s(g_1)}}{\Omega(g_2,g_3)}\cdot \Omega(g_1,g_2g_3)\cdot\pe\,.
\end{align}
As we have already seen that $\Omega$'s can be decomposed into boundary terms, $\Omega_a$ and $\Omega_b$ at the two ends of the segment, then we have the condition that $\Omega_a$ and $\Omega_b$ should satisfy:
\begin{align}
    \Omega_b(g_1,g_2)\Omega_b(g_1g_2,g_3)=\frac{1}{\omega(g_1,g_2,g_3)}&{}^{U_M(g_1)K^{s(g_1)}}\Omega_b(g_2,g_3)\Omega_b(g_1,g_2g_3)\label{eq:u1f_Omega_b_fusion_app}\\ 
    (-)^{\sigma(g_1,g_2)\sigma(g_1g_2,g_3)}\Omega_a(g_1,g_2)\Omega_a(g_1g_2,g_3)\frac{1}{\omega(g_1,g_2,g_3)}&={}^{U_M(g_1)K^{s(g_1)}}\Omega_a(g_2,g_3)\Omega_a(g_1,g_2g_3)(-)^{\sigma(g_2,g_3)\sigma(g_1,g_2g_3)}\nonumber.
\end{align}
We can now check Eq.~\eqref{eq:omegaab_twist_two-cocycle} directly using the explicit form of $\Omega_{a/b}$. Here only the case of $\Omega_b$ is checked in details, while $\Omega_a$'s case can be checked in a quite similar way. 
The left hand side of the equation above is 
\begin{align}
    &\Omega_b(g_1,g_2)\Omega_b(g_1g_2,g_3)\nonumber\\
    =&\pe \cdot w_L(g_1) \act{K^{s(g_1)}}{w_L(g_2)}  w^{-1}_L(g_1g_2)  \tikzmark{ep}{\boxed{\ee^{-\ii \alpha(g_1,g_2) n_{_{f,L}}}}} \, \lambda^{(2)}_{\frac{1}{2}}(g_1,g_2)w_L(g_1g_2) \act{K^{s(g_1g_2)}}{w_L(g_3)}  w^{-1}_L(g_1g_2g_3)\tikzmark{end}{}\nonumber\\
    &  \ee^{-\ii \alpha(g_1g_2,g_3) n_{_{f,L}}} \, \lambda^{(2)}_{\frac{1}{2}}(g_1g_2,g_3)\cdot\pe\nonumber\\
    =&\pe \cdot w_L(g_1) \act{K^{s(g_1)}}{w_L(g_2)}\act{K^{s(g_1g_2)}}{w_L(g_3)} w^{-1}_L (g_1g_2g_3) \ee^{-\ii(\alpha(g_1,g_2)+\alpha(g_1g_2,g_3))\,n_{_{f,L}} } {\color{blue}{\ee^{-\ii\alpha(g_1,g_2)\,\sigma(g_1g_2,g_3)}}}\nonumber\\
    &\lambda^{(2)}_{\frac{1}{2}}(g_1,g_2) \lambda^{(2)}_{\frac{1}{2}}(g_1g_2,g_3)\cdot\pe\,,
    \label{eq:u1f_Omega_ab_expansion_1}
\begin{tikzpicture}[remember picture,overlay,>=stealth]
    \draw[->,blue] (ep.north) -- ++(0,5pt)  -| (end.north);
\end{tikzpicture}
\end{align}
where the additional exponential factor in blue color is due to rearranging terms according to the blue arrow in the equation above, while such factor can be read out from the relation Eq.~\eqref{eq:Omega_ab_fermion_number_app}.
The right hand side is a little more complicated:
\begin{align}
    &\act{U_M(g_1)K^{s(g_1)}}{\Omega}_b(g_2,g_3)\Omega_b(g_1,g_2g_3)\nonumber\\
    =&\pe\cdot \left(w_L(g_1)\tikzmark{t1}{\boxed{w_{l+1}(g_1)W_M(g_1)K^{s(g_1)}}}\right)\cdot \left(w_L(g_2)\act{K^{s(g_2)}}{w}_L(g_3)w^{-1}_L(g_2g_3) \lambda^{(2)}_{\frac{1}{2}}(g_2,g_3)\right)\cdot \tikzmark{t2}{}  \nonumber\\
    &\nonumber\\
    &\left(K^{s(g_1)}W_M^{-1}(g_1)w_{l+1}^{-1}(g_1)w_L^{-1}(g_1)\right)\cdot\tikzmark{h1}{\boxed{\ee^{-\ii (-)^{\tilde{\rho}(g_1)}\alpha(g_2,g_3)n_{_{f,L}}} \ee^{-\ii(-)^{s(g_1)}\alpha(g_2,g_3)n_{\lambda,\frac{1}{2}}^{(1)}(g_1)}}}\cdot \Big(w_L(g_1)w_L(g_2g_3)w^{-1}_L(g_1g_2g_3)\tikzmark{h2}{}\nonumber\\
    &\ee^{-\ii\alpha(g_1,g_2g_3)\,n_{_{f,L}}}\lambda^{(2)}_{\frac{1}{2}}(g_1,g_2g_3) \Big)\cdot\pe\nonumber\\
    =&\pe\cdot w_L(g_1)\act{K^{s(g_1)}}{w_L(g_2)}\act{K^{s(g_1g_2)}}{w_L(g_3)}\act{g_1}{\lambda^{(2)}_{\frac{1}{2}}(g_2,g_3)}{\color{blue}{(-)^{n^{(2)}_{\lambda,\frac{1}{2}}(g_1)\sigma(g_2,g_3)}}} w^{-1}_L(g_1g_2g_3)\ee^{-\ii(-)^{\tilde{\rho}(g_1)}\alpha(g_2,g_3)n_{_{f,L}}}\nonumber\\
    &{\color{red}{ \ee^{-\ii(-)^{\tilde\rho(g_1)}\alpha(g_2,g_3)\sigma(g_1,g_2,g_3)}}} \ee^{-\ii(-)^{s(g_1)}\alpha(g_2,g_3)n_{\lambda,\frac{1}{2}}^{(1)}(g_1)}\ee^{-\ii\alpha(g_1,g_2g_3)n_{f_{f,L}}}\lambda^{(2)}_{\frac{1}{2}}(g_1,g_2,g_3)\cdot\pe\nonumber\\
    =&\pe\cdot \ee^{\ii(-)^{s(g_1)}\alpha(g_2,g_3)n_{\lambda,\frac{1}{2}}^{(2)}(g_1)}  w_L(g_1)\act{K^{s(g_1)}}{w_L(g_2)}\act{K^{s(g_1g_2)}}{w_L(g_3)} w^{-1}_L(g_1g_2g_3)\ee^{-\ii(-)^{\tilde{\rho}(g_1)}\alpha(g_2,g_3)n_{_{f,L}}} \ee^{-\ii\alpha(g_1,g_2g_3)n_{f_{f,L}}}\nonumber\\
    &(-)^{n^{(2)}_{\lambda,\frac{1}{2}}(g_1)\sigma(g_2,g_3)}\ee^{\ii(-)^{\tilde\rho(g_1)}\alpha(g_2,g_3)\sigma(g_1,g_2,g_3)}\act{g_1}{\lambda}^{(2)}_{\frac{1}{2}}(g_2,g_3) \lambda^{(2)}_{\frac{1}{2}}(g_1,g_2g_3)\cdot\pe\,.
    \label{eq:u1f_Omega_ab_expansion_2}
\begin{tikzpicture}[remember picture,overlay,>=stealth]
        \draw[->,blue] (t1.south) -- ++(0,-5pt)  -| (t2.north);
        \draw[->,red] (h1.north) -- ++(0,5pt)  -| (h2.north);
\end{tikzpicture}
\end{align}
Now we can conclude the relation between $\omega$ and $\lambda$'s by comparing Eq.~\eqref{eq:u1f_Omega_b_fusion_app}, \eqref{eq:u1f_Omega_ab_expansion_1}, and  \eqref{eq:u1f_Omega_ab_expansion_2}:
\begin{align}
    &\ee^{-\ii\alpha(g_1,g_2)\sigma(g_1g_2,g_3)}\lambda^{(2)}_{\frac{1}{2}}(g_1,g_2) \lambda^{(2)}_{\frac{1}{2}}(g_1g_2,g_3)\nonumber\\
     = &\frac{\ee^{\ii(-)^{s(g_1)}\alpha(g_2,g_3)n_{\lambda,\frac{1}{2}}^{(2)}(g_1)}}{\omega(g_1,g_2,g_3)}(-)^{n^{(2)}_{\lambda,\frac{1}{2}}(g_1)\sigma(g_2,g_3)}\left(\ee^{-\ii(-)^{\tilde\rho(g_1)}\alpha(g_2,g_3)\sigma(g_1,g_2,g_3)}\right)\act{g_1}{\lambda}^{(2)}_{\frac{1}{2}}(g_2,g_3) \lambda^{(2)}_{\frac{1}{2}}(g_1,g_2g_3)\,.
     \label{eq:u1f_lambda_omega_relation}
\end{align}
Now we can compare Eq.~\eqref{eq:u1f_lambda_omega_relation} and \eqref{eq:u1f_lambda_omega_prime_relation}, then the relation between $\omega$ and $\omega^{\prime}$ can be read out:
\begin{align}
    \omega(g_1,g_2,g_3) = \omega^{\prime}(g_1,g_2,g_3)\exp\big[ -\ii(-)^{\tilde\rho(g_1)}\alpha(g_2,g_3)\sigma(g_1,g_2g_3)+\ii\alpha(g_1,g_2)\sigma(g_1g_2,g_3) \big]\,.
    \label{}
\end{align}
We have already known the 3-cocycle condition that $\omega^{\prime}$ satisfies, then the calculation of 3-cocycle condition for $\omega$ are simplified by calculating those exponential factors:
\begin{align}
    &\frac{\omega(g_1,g_2,g_3)\omega(g_1,g_2g_3,g_4)[\omega(g_2,g_3,g_4)]^{(-)^{s(g_1)}}}{\omega(g_1g_2,g_3,g_4)\omega(g_1,g_2,g_3g_4)}\nonumber\\
    =&\frac{\omega^{\prime}(g_1,g_2,g_3)\omega^{\prime}(g_1,g_2g_3,g_4)[\omega^{\prime}(g_2,g_3,g_4)]^{(-)^{s(g_1)}}}{\omega^{\prime}(g_1g_2,g_3,g_4)\omega^{\prime}(g_1,g_2,g_3g_4)}\times \ee^{{\color{red}{\ii\alpha(g_1,g_2)\sigma(g_1g_2,g_3)}}{\color{blue}{-\ii(-)^{\tilde\rho(g_1)}\alpha(g_2,g_3)\sigma(g_1,g_2g_3)}}} \nonumber\\
    &\times \ee^{{\color{violet}{\ii\alpha(g_1,g_2g_3)\sigma(g_1g_2g_3,g_4)}}{\color{cyan}{-\ii(-)^{\tilde\rho(g_1)}\alpha(g_2g_3,g_4)\sigma(g_1,g_2g_3g_4)}} }\times \ee^{{\color{blue}{\ii(-)^{s(g_1)}\alpha(g_2,g_3)\sigma(g_2g_3,g_4)}}{\color{teal}{-\ii(-)^{\tilde\rho(g_2)+s(g_1)}\alpha(g_3,g_4)\sigma(g_2,g_3g_4)}} }\nonumber\\
    &\times \ee^{{\color{violet}{-\ii\alpha(g_1g_2,g_3)\sigma(g_1g_2g_3,g_4)}}{\color{teal}{+\ii(-)^{\tilde\rho(g_1g_2)}\alpha(g_3,g_4)\sigma(g_1g_2,g_3g_4)}}}\times \ee^{{\color{red}{-\ii\alpha(g_1,g_2)\sigma(g_1g_2,g_3g_4)}}{\color{cyan}{+\ii(-)^{\tilde\rho(g_1)}\alpha(g_2,g_3g_4)\sigma(g_1,g_2g_3g_4)}}}\,.
\end{align}
Merging terms that are in the same color together, we have
\begin{align}
    &\frac{\omega(g_1,g_2,g_3)\omega(g_1,g_2g_3,g_4)[\omega(g_2,g_3,g_4)]^{(-)^{s(g_1)}}}{\omega(g_1g_2,g_3,g_4)\omega(g_1,g_2,g_3g_4)}\nonumber\\
    =&(-)^{\sigma(g_1,g_2)\sigma(g_3,g_4)}~\ee^{\ii (-)^{\tilde\rho(g_1g_2)} \alpha(g_3,g_4)\sigma(g_1,g_2) }\times \ee^{{\color{red}{\ii\alpha(g_1,g_2)\left[(-)^{\rho(g_1g_2)}\sigma(g_3,g_4)-\sigma(g_1g_2g_3,g_4)  \right]}}}\nonumber\\
    &\times \ee^{{\color{blue}{\ii(-)^{\tilde\rho(g_1)}\alpha(g_2,g_3)\left[\sigma(g_1g_2g_3,g_4)-\sigma(g_1,g_2g_3g_4)  \right]   }}} \times \ee^{{\color{violet}{\ii\sigma(g_1g_2g_3,g_4)\left[\alpha(g_1,g_2)-(-)^{\tilde\rho(g_1)}\alpha(g_2,g_3)  \right]  }}}\nonumber\\
    &\times \ee^{{\color{cyan}{\ii(-)^{\tilde\rho(g_1)}\sigma(g_1,g_2g_3g_4)\left[\alpha(g_2,g_3)-(-)^{\tilde\rho(g_2)}\alpha(g_3,g_4)  \right]  }}}\times \ee^{{\color{teal}{\ii(-)^{\tilde\rho(g_1g_2)}\alpha(g_3,g_4)\left[\sigma(g_1,g_2g_3g_4)-\sigma(g_1,g_2)  \right]  }}}\nonumber\\
    =&(-)^{\sigma(g_1,g_2)\sigma(g_3,g_4)}~\ee^{\ii (-)^{\rho(g_1g_2)} \alpha(g_1,g_2)\sigma(g_3,g_4) }\,,
\end{align}
thus we recover 3-cocycle condition that we obtained from the anomalous symmetry actions on edge in Section \ref{sec:classification} from tensor equations.

\section{Details about Kasteleyn orientation}\label{app:kasteleyn}
\subsection{Kasteleyn orientation and symmetric FPEPS}\label{subapp:kasteleyn_sym}
In this section, we define rules to extract an orientation from orders of symmetry actions on internal legs and prove that the physical wavefunction is $g$-symmetric only when the right hand side of Eq.~(\ref{eq:honeycomb_to_fisher_kasteleyn}) is a Kasteleyn orientation.
According to Eq.~(\ref{eq:u1f_g_act_nf}), the commutator between $W(g)$ and parity $F\equiv\exp[\ii\pi n_f]$ reads
\begin{align}
    W(g)\cdot F\cdot W^{-1}(g)=D(g)\cdot F
    \label{}
\end{align}
where $D(g)=\exp[\ii\pi n_D(g)]$, whose eigenvalues equal $\pm1$. 
Thus, $D(g)=P_e(g)-P_o(g)$, where $P_{e/o}$ are projectors to subspace with eigenvalue $\pm1$.

With these projectors, we can decompose site and bond tensors to orthogonal sectors, e.g. $P_{e,x}(g)\otimes_f P_{o,y}(g)\otimes_f P_{o,z}(g)\cdot \hat{T}$.
By inserting $P_e(g)+P_o(g)$ on all internal legs of the tensor network wavefunction, the wavefunction equals summation of tensor contractions for  different sectors.
However, as $D(g)$ is an IGG element, local tensors vanish if acted by odd number of $P_o(g)$'s.
In other words, action of $P_o(g)$ must form loops on the network.
Given a loop configuration $d$, by acting $P_o(g)$'s on internal legs along loops and $P_e(g)$'s on internal legs within domains, we obtain site tensors $\hat{T}^d$'s and bond tensors $\hat{B}^d$'s.
By contracting internal legs of $\hat{T}^d$'s and $\hat{B}^d$'s, we get physical state $\ket{\psi_d}$, and physical wavefunction can be expressed as $\ket{\Psi}=\sum_d\ket{\psi_d}$.

In the following, we will prove that $\forall d,~g\circ\ket{\psi_d}=\ket{\psi_d}$ if orders of $W(g)$ action gives Kasteleyn orientation, and thus $\ket{\Psi}$ is invariant under $g$.

We start from configuration $d$ without loop.
In this case, we only get $W_e(g)$'s when acting $g$, which are parity even and free to permute. 
$\ket{\psi_d}$ is apparently invariant under $g$.

For configuration $d$ with a single loop, the situation is similar to Section~\ref{subsec:kasteleyn_1d}. 
Let the number of internal ket legs to be $2L$, internal legs along this loop are labeled counter-clockwise by $l$, where site tensors sit between $l=2k-1$ and $l=2k$. 
Site tensors are then named as $\hat{T}_{2k-1,2k}^d$, while bond tensors as $\hat{B}_{2k,2k+1}^d$.
As all tensors are parity even, we rearrange tensors along the loop together in the following way:
\begin{align}
    \ket{\psi_d}={}&\fTr[\cdots\otimes_f \hat{B}_{2,3}^d\otimes_f \hat{B}_{3,4}^d\otimes_f\cdots\otimes_f \hat{B}_{2L,1}^d\otimes_f \notag\\
    &\hat{T}_{2L-1,2L}^d\otimes_f \hat{T}^d_{2L-3,2L-2}\otimes_f\cdots \otimes_f \hat{T}_{1,2}^d\otimes_f\cdots]
    \label{eq:psid_tn}
\end{align}
We now act $g$ on $\ket{\psi_d}$, and according to Eq.~\eqref{eq:site_g_action} and \eqref{eq:bond_g_action}, it equals action of $W(g)$'s on internal legs of $\hat{T}^d$'s and $\hat{B}^d$'s.
We define $W_{e/o}(g)\equiv P_{e/o}\cdot W(g)$, which is parity even/odd sector of $W(\TT)$.
From definition of $\ket{\psi_d}$, $W(g)$'s act as $W_o(g)$'s on internal legs along the loop, while acting as $W_e(g)$'s on internal legs away from the loop.
Fermion signs come from permuting $W_o(g)$'s on the loop.
We arrange the order of $W_o(g)$'s contraction according to Eq.~(\ref{eq:psid_tn}) as
    \begin{align}
        \fTr\bigg\{ & \Big( (-1)^{s_{2,3}}\cdot W_{o,2}(g)\otimes_f W_{o,3}(g) \Big) \otimes_f\cdots\otimes_f \Big( (-1)^{s_{2L,1}}\cdot W_{o,2L}(g)\otimes_f W_{o,1}(g) \Big)\bigotimes\subf \notag\\
        &\bigg[ \Big( (-1)^{s_{1,2}}\cdot W_{o,1}(g)\otimes_f W_{o,2}(g) \Big) \otimes_f\cdots\otimes_f \Big( (-1)^{s_{2L-1,2L}}\cdot W_{o,2L-1}(g)\otimes_f W_{o,2L}(g) \Big) \bigg]^{-1} \bigg\}\notag\\
        ={}&(-1)^{1+\sum_{l}s_{l,l+1}}=1
        \label{eq:psid_wo}
    \end{align}
where $s_{l,l+1}=0/1$ if the arrow at $(l,l+1)$ is along/against the direction of the loop (counter-clockwise/clockwise direction).
The last line in Eq.~\eqref{eq:psid_wo} is from the definition of Kasteleyn orientation.
So, for $d$ with a single loop, $W(g)$'s and $[W(g)]^{-1}$'s cancels, and $\ket{\psi_d}$ is $g$-symmetric.
In contrast, if the orientation is not Kasteleyn, one can always find a loop configuration $d$, such that the last line of Eq.~(\ref{eq:psid_wo}) gives $-1$.
So, $\ket{\Psi}$ breaks $g$ symmetry if Kasteleyn orientation is not satisfied.

For configuration $d$ with more than one loops, we can arrange all tensors belonging to one loop together, and repeat the above calculation for every loop.
Thus, such $\ket{\psi_d}$ is also $g$-symmetric.
In conclusion, $\ket{\Psi}$ is $g$-symmetric if and only if the orientation extracted from $W(g)$'s is a Kasteleyn orientation.

As we mentioned in Section~\ref{subsec:kasteleyn_1d}, by flipping arrows on all edges connecting to certain vertices, one gets another Kasteleyn orientation.
In the tensor language, flipping arrows for edges connecting to vertex $(s\alpha)$ corresponds to modifying $W_{(s\alpha)}(g)$ to $D_{(s\alpha)}(g)\cdot W_{(s\alpha)}(g)$:
    \begin{align*}
        \left[ D_{(sx)}(g)\cdot W_{(sx)}(g)\otimes_f W_{(sy)}(g) \otimes_f W_{(sz)}(g) \right]^{-1} \cdot \hat{T}_s  
        = \left[  W_{(sy)}(g) \otimes_f W_{(sz)}(g)\otimes_f W_{(sx)} \right]^{-1} \cdot \hat{T}_s  
        ~~~\Rightarrow \adjincludegraphics[valign=c,scale=2]{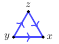}
    \end{align*}
Similar logic works for bond tensors.
Following these rules for arrows, there is one-to-one correspondence between gauge transformation $W(g)$ and Kasteleyn orientation.

The above argument can be easily generalized to any trivalent lattice, where each vertex connects three bonds, and thus there is at most one loop travelling through a vertex.
We now consider generic graphs, where more than one loops may meet at sites.

Let us present rules for extract orientations in generic lattices.
Similar as Eq.~(\ref{eq:honeycomb_to_fisher_kasteleyn}), we first construct a new planar graph, where a site connecting $n$-bonds in the original lattice splits to $n$ vertices in the new graph, and each vertex is labeled by the corresponding internal leg index.
Each pair of these $n$ vertices are connected by new edges.
Given $g$-action on internal legs, arrows on edges of the new planar graph follows similar rules presented in Eq.~(\ref{eq:site_arrow_rule}) and Eq.~(\ref{eq:bond_g_action}). 
For example, consider site tensor $s$ with four internal legs, the above rule reads
    \begin{align*}
        U(g)K^{s(g)} \cdot \hat{T}_s  
        = \left[  W_{(sa)}(g) \otimes_f W_{(sb)}(g)\otimes_f W_{(sc)}\otimes_f W_{(sd)} \right]^{-1} \cdot \hat{T}_s  
        ~~~\Rightarrow \adjincludegraphics[valign=c,scale=2]{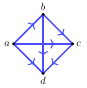}
    \end{align*}

To proceed, let us focus on a particular choice of Kasteleyn orientation.
We number the $n$ vertices from $1$ to $n$ clockwise, and let the arrow points from $i$ to $j$ if $i<j$.
It is easy to verify that any loop within these $n$ vertices matches the condition for Kasteleyn orientation.
Arrows on bond tensors are chosen to satisfy conditions for Kasteleyn orientation with larger loops.

For loop configurations without loop crossing, using similar argument presented in the honeycomb lattice case, we conclude that $W(g)$'s and $[W(g)]^{-1}$'s cancels without additional fermion sign.

\begin{figure}[htpb]
    \centering
    \includegraphics[scale=2]{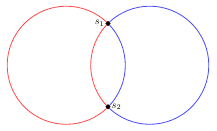}
    \caption{Two domain wall loops~(red and blue) intersect at site $s_1$ and $s_2$.
    Here, $s_{1,2}$ are not ``true crossing points'' of these two loops: they are separate at these two points.}.
    \label{fig:two_loops}
\end{figure}

As shown in Fig.~\ref{fig:two_loops}, we consider configuration $d$ where two loops~(colored blue and red) intersect at site $s_1$ and $s_2$.
Loops are chosen such that internal legs belonging to one loop are neighbours.
Thus, there are no ``true crossing points'' between these two loops.

By inserting $P_o$'s on internal legs at these two loops, and $P_e$'s on other internal legs, we obtain tensors $\hat{T}^d$'s and $\hat{B}^d$'s, and wavefunction $\ket{\psi_d}$ from contracting $\hat{T}^d$'s and $\hat{B}^d$'s.
Let the number of internal ket legs of the red/blue loop to be $2L_1/2L_2$.
We label internal legs along these two loops by index $l$, where $1\le l\le 2L_1$ labels internal legs along the red loop and $2L_1+1\le l\le 2L_1+2L_2$ for the blue loop.
Due to the intersecting sites $s_{1/2}$, Eq.~(\ref{eq:psid_tn}) do not directly apply to the case here.
However, at any intersection point, two $W_o^{-1}(g)$'s belonging to one loop can be always moved together without sign, and in this way, Eq.~(\ref{eq:psid_wo}) still applies for every single loop.
The above argument can be easily generalized to any loop configurations.
We then conclude that for a $g$-symmetric FPEPS, the planar graph extracted from $W(g)$'s satisfies Kasteleyn orientation.

\subsection{Symmetry on the open boundary from Kasteleyn orientation}\label{subapp:kasteleyn_bdry}

We now consider a finite 2D FPEPS on region $A$, which is cut from an infinite FPEPS, as shown in Fig.~\ref{fig:kasteleyn_domain_wall}. The aim of this section is to show the validity of Eqn.~(\ref{eq:u1f_sym_edge_w}).
By contracting all internal legs within $A$,  we obtain a large tensor $\hat{T}_A$:
\begin{align}
    \hat{T}_A = \fTr_{A} [\mathbb{B}\otimes_f \mathbb{T}]\,,
\end{align}

We now act $g$ on physical legs of $\hat{T}_A$, which give $W(g)$'s and $W^{-1}(g)$'s on all bulk and boundary virtual legs of $\hat{T}_A$ according to Eq.~\eqref{eq:site_g_action} and \eqref{eq:bond_g_action}. The parity of $W(g)$'s are determined by $D(g)$'s which depend on the string configuration. Now consider a configuration $d$ that there is only one string with ends $i$ and $j$,  according to Eqn.~(\ref{eq:u1f_sym_edge_w}), the boundary symmetry action $W^{d}_{\partial A}$ is
\begin{align}
    W^{d}_{\partial A}(g)=(-1)^{\xi(j)-\xi(i)}\WW_e \otimes_{f} W_{i}(g)\otimes_{f} W_{j}(g)
    \label{eqn:app_sym_act_boundary_kasteleyn}
\end{align}
Where $\WW_e$ is the tensor product of parity even $W(g)$'s acting other boundary internal legs. Now we turn to the $1$D string $L=\{l_1,l_2,\dots,l_n\}$ with ends $l_1=i$ and $l_n=j$. As parity even $W(g)$'s and $W(g)^{-1}$'s acting on other bulk internal legs already cancel, we can read the symmetry equation according to Eqn.~(\ref{eq:kasteleyn_g_on_open_chain}):
\begin{align}
    U_{A}(g)\cdot\hat{T}^{d}_{A}&=(-1)^{\sum s(l_i,l_{i+1})}\fTr[\WW_e\otimes_f W^{-1}_{l_3}(g)\otimes_{f} W^{-1}_{l_2}(g)\otimes_f\dots \otimes_{f}W^{-1}_{l_{n-1}}\otimes_{f}W^{-1}_{l_{n-2}}\otimes_{f}W_{l_1}(g)\otimes_{f}\dots\otimes_{f}W_{l_n}(g)\otimes_f \hat{T}^{d}_{A}]\nonumber\\
    &=(-1)^{\sum s(l_i,l_{i+1})}\fTr[ \WW_e \otimes_f W_{l_1}(g)\otimes_{f}W_{l_n}(g)\otimes_f \hat{T}^{d}_{A}]\nonumber\\
    &=\fTr[ W_{\partial A}^{d}(g)\otimes_f \hat{T}^{d}_{A}]
\end{align}
Where $\hat{T}^{d}_{A}$ is the tensor $\hat{T}_{A}$ with the specific domain configuration $d$.
We know from Kasteleyn orientation that $[\xi(j)-\xi(i)]+\sum s(l_i,l_{i+1})=0 \mod 2$, thus Eqn.~(\ref{eqn:app_sym_act_boundary_kasteleyn}) is valid, which means Eqn.~(\ref{eq:u1f_sym_edge_w}) is valid for one open string case. The verification of multi-string case is straightforward as bulk $W(g)$'s and $W^{-1}(g)$'s from different strings are always separable as shown in Fig.\ref{fig:two_loops}. Thus for multiple strings case, the $W^{d}_{\partial A}$ is
\begin{align}
    W^{d}_{\partial A}(g)=(-1)^{\sum_{\alpha} \xi(l^{(\alpha)}_{n})-\xi(l^{(\alpha)}_{1})} \WW_e \bigotimes_{\alpha}{}_{f} W^{(\alpha)}_{l_{1}}(g)\otimes_{f} W^{(\alpha)}_{l_{n}}(g) 
\end{align}
Here $\alpha$ denotes different strings. Thus, Eqn.~(\ref{eq:u1f_sym_edge_w}) is valid for all configurations.

\end{document}